\documentclass[aps,prl,amsmath,amssymb,floatfix,reprint,citeautoscript,noeprint,superscriptaddress,twocolumn]{revtex4-2}
\usepackage{dsfont}
\usepackage[english]{babel}
\usepackage[dvipsnames]{xcolor}

\usepackage{graphicx}
\usepackage{subcaption}

\usepackage{amsmath,amssymb,bm}
\usepackage[version=3]{mhchem}
\usepackage{verbatim}
\usepackage{multirow}
\usepackage{dcolumn}
\usepackage{nicefrac}
\usepackage{siunitx}
\usepackage{booktabs}
\usepackage{transparent}
\usepackage{tcolorbox}
\usepackage{relsize}
\usepackage[justification=centerlast, font={sf}, labelfont={bf}]{caption}
\usepackage{rotating}
\usepackage[colorlinks,allcolors=black,citecolor=blue,urlcolor=blue]{hyperref}
\hypersetup{
    citecolor=Turquoise,
    colorlinks=true,
    linkcolor=black,    
    urlcolor=Turquoise
    }

\newcommand{\tbf}[1]{\textbf{#1}}

\newcommand{\mbf}[1]{\boldsymbol{\mathit{#1}}}
\newcommand{\mrm}[1]{\mathrm{#1}}

\newcommand \imi { \mathrm{i} }
\newcommand \e[1] { \mathrm{e}^{#1} }
\newcommand \dd[1]  { \!\!\textrm d{#1} \,}   

\begin{document}

\title{Molecular interpretability of the bulk electrochemical impedance of concentrated electrolytes}

\author{Connie J. Fairchild}
\affiliation{Yusuf Hamied Department of Chemistry, University of
  Cambridge, Lensfield Road, Cambridge, CB2 1EW, United Kingdom}

\author{Stephen J. Cox}
\affiliation{Department of Chemistry, Durham University, South Road,
  Durham, DH1 3LE, United Kingdom}

\author{Benjamin Rotenberg}
\affiliation{Sorbonne Universit\'{e}, CNRS, Physical Chemistry of Electrolytes and
Interfacial Nanosystems (PHENIX), 4 place Jussieu, Paris, France}
\affiliation{R\'{e}seau sur le Stockage Electrochimique de l’Energie (RS2E), FR CNRS 3459, 80039 Amiens Cedex, France}

\author{Thomas Sayer}
\email{thomas.e.sayer@durham.ac.uk}
\affiliation{Department of Chemistry, Durham University, South Road,
  Durham, DH1 3LE, United Kingdom}

\date{\today}

\begin{abstract}
Electrochemical impedance spectroscopy (EIS) is a widely used technique to understand time-dependent response and relaxation under applied voltage. While these spectra contain a wealth of information, major gaps in our understanding can hinder our ability to interpret EIS spectra in terms of microscopic chemical mechanisms. We propose an alternative approach to common empirical fitting procedures for describing the contribution of the bulk electrolyte to the EIS spectrum. This new approach is rooted in determining the moments of the frequency-dependent conductivity, with molecular interpretability provided by a generalized Langevin equation description of an effective single particle dynamics; the `itinerant oscillator' (IO) model. In contrast to a Debye--Falkenhagen description, the IO model makes no assumptions regarding the concentration of the electrolyte, a fact we demonstrate by analysing molecular dynamics simulations of a room-temperature ionic liquid. By analysing the memory function from simulation within the framework provided by the IO model, we reveal the importance of capturing the separation of timescales within the memory function for describing the temperature dependent $\beta$-relaxation process. We go on to show how our impedance model directly reports on this distribution of timescales while retaining the simplicity of commonly employed workflows.
\end{abstract}

\maketitle
\setcounter{secnumdepth}{2}  
\section{Introduction}
Understanding the structure and dynamics of electrolyte solutions and ionic liquids is a central challenge in physical chemistry; these determine effective solute-solute interactions, the capacity to do work, and rates of dissipation. Yet, despite over a century of research, a comprehensive understanding remains elusive, especially at high ionic strengths \cite{perkin_is_2013, gebbie_ionic_2013, lee_are_2015, weingartner_understanding_2008, nordness_ion_2020}. From a theoretical perspective, difficulties arise due to the long-ranged nature of the Coulomb interaction between the ions of the fluid; the condition of local electroneutrality imposes additional length scales that are lacking in simple fluids, most notably the Debye length (or, more generally, multiple screening lengths) and the Bjerrum length \cite{gebbie_long_2017, kornyshev_double-layer_2007, bazant_double_2011}. The presence of such additional length scales in turn gives rise to additional time scales, which become relevant to understanding a system's nonequilibrium relaxation.

In the context of energy storage devices, an important measure of a system's time-dependent linear response is the frequency-dependent impedance,
\begin{equation}
\label{eqn:Z-def}
Z(\omega) = \frac{V(\omega)}{I(\omega)},
\end{equation}
where $V(\omega)$ and $I(\omega)$ are the system's frequency-dependent electric potential and current, respectively. While several experimental techniques exist for determining $Z(\omega)$, Eq.~\ref{eqn:Z-def} underpins the physical principle for characterizing the overall opposition to a current: the system is perturbed (e.g., with an alternating applied potential), and the resulting response (e.g., the alternating current) is measured \cite{wang_electrochemical_2021}. Although obtaining $Z(\omega)$ in such a manner is conceptually straightforward, attributing features of the impedance to specific microscopic processes is a complex task \cite{usler_impedance_2026}. Yet, our ability to rationally design ionic fluids for next-generation devices, both in terms of energy storage capability and output, depends powerfully on the understanding derived from a molecular-level interpretation of impedance spectra.

In this paper, we aim to reformulate the nonequilibrium relaxation of ionic fluids in terms of parameters that are, at least in principle, directly related to a system's microscopic correlations. By recasting the impedance of a bulk system in terms of moments of the frequency-dependent conductivity, we explicitly show that commonly-used equivalent circuit models (i.e., the Cole--Cole and Cole--Davidson models \cite{cole_dispersion_1941, davidson_dielectric_1950, davidson_dielectric_1951}) cannot faithfully describe the impedance of bulk ionic fluids; this fact is borne out by results from molecular simulations of a room temperature ionic liquid. 

In a bid to provide a relatively simple analytical theory that gives insight into the molecular-level processes that determine the conductivity moments, and which treats high and low ionic strengths on an equal footing, we present an analysis of the conductivity in the context of an ``itinerant oscillator'' (IO) model, which is rooted in a generalized Langevin equation (GLE) framework. The IO model reduces the equations of motion to an open, two-body problem of a tagged ion and its correlated counterion cloud (or cage). In this way we are able to construct a closed functional form for the frequency-dependent memory of the system. While such a simple model is limited, we will show it captures the most salient aspects of bulk conductivity across the full range of frequencies. Moreover, the extracted microscopic parameters vary in response to changing conditions (e.g., temperature) in an intuitive way, and thus provide insight into microscopic processes that determine impedance.

\subsection{Bulk Impedance in an Electrochemical Device}

\begin{figure}[!ht]
     \centering

    \includegraphics[width=0.48\textwidth]{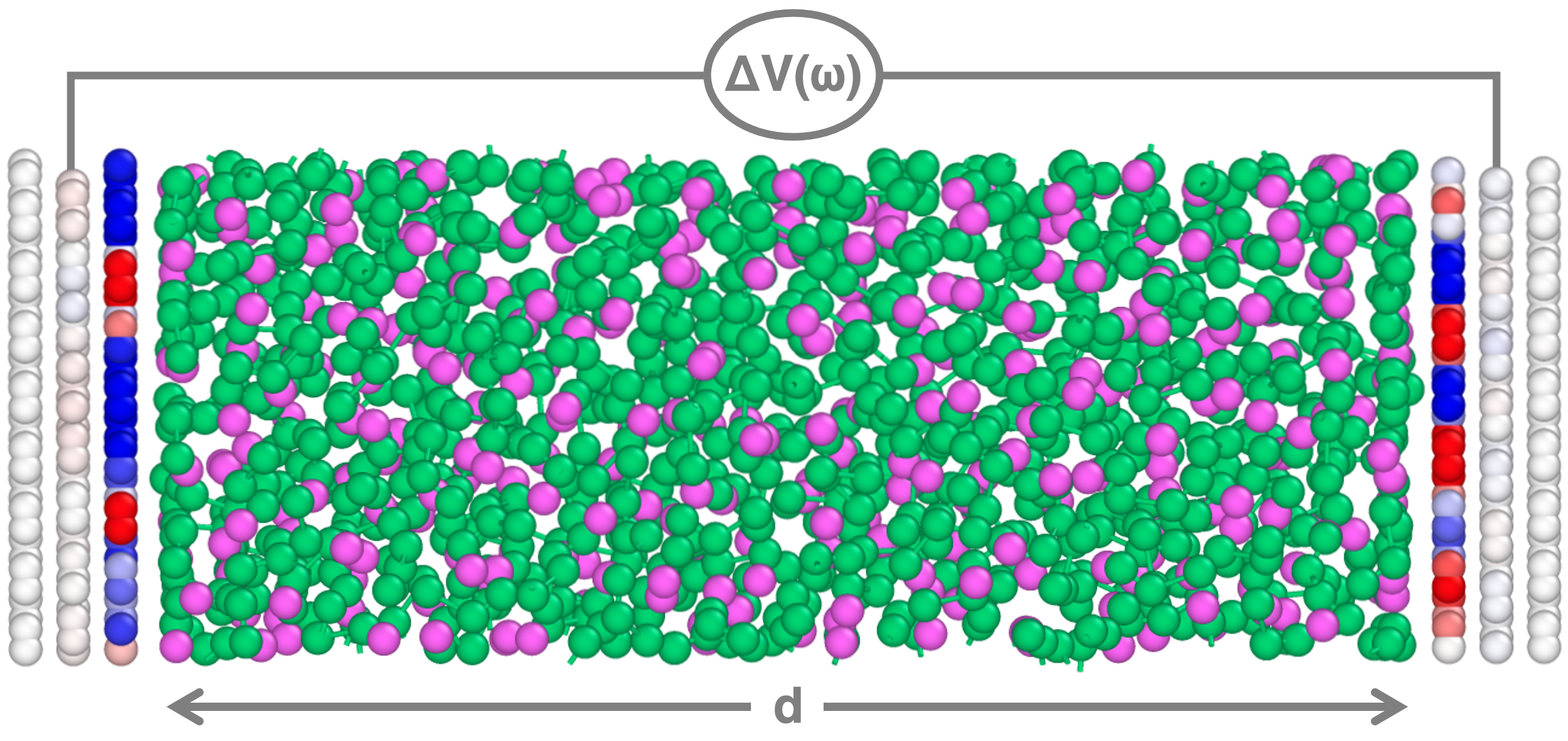}

    \caption{\label{fig:setup} Snapshot from a typical set-up for confined electrolyte simulations within a constant potential ensemble, where each electrode's charge (blue--red colormap) fluctuates in response to the electrolyte trajectory, facilitating the calculation in Eq.~\ref{eqn:Z-FDT}. In this paper we look to quantify the \textit{bulk impedance}. This is the impedance of the above nanocapacitor if the electrolyte were to behave bulk-like throughout and the interfaces had no effect.}
\end{figure}

A prototypical system of interest is shown schematically in Fig.~\ref{fig:setup}. Here, an ionic fluid (we use this term to encompass both electrolytes and ionic liquids) is confined between two electrodes such that the potential drop across the device, $\Delta V(\omega)$, is controlled by external means \cite{tee_fully_2022, dufils_simulating_2019}. The instantaneous charge on one of the electrodes, $Q(t)$, varies in time; this is not only due to the frequency-dependence of $\Delta V(\omega)$ directly, but also to the fluctuating polarization of the ionic fluid. As shown by Pireddu and Rotenberg, the impedance of such a system can be obtained from a fluctuation--dissipation relation \cite{pireddu_frequency-dependent_2023},
\begin{equation}\label{eqn:Z-FDT}
\begin{split}
  \frac{1}{Z(\omega)} = \beta\bigg[\imi&\omega\langle(\delta Q)^2\rangle 
  \\&+\omega^2\int_0^\infty\!\mrm{d}t\,\langle\delta Q(t)\delta Q(0)\rangle\exp(-\imi\omega t)\bigg].
\end{split}
\end{equation}
Equation~\ref{eqn:Z-FDT} provides a natural starting point to relate $Z(\omega)$ to the microscopic behaviour of the system with molecular simulations. For example, Ref.~\cite{pireddu_impedance_2024} found that, with appropriate placement of the dielectric boundary \footnote{As determined from the covariance of the local polarization with the total dipole of the fluid}, the differential capacitance (determined from the slope of ${Z^{-1}(\omega\to 0)}$) was well described by the standard dielectric continuum expression. Moreover, upon decomposing $Z(\omega)$ into ``bulk'' and ``interfacial'' contributions (see below) the bulk-like response of the confined fluid was confirmed to hold across a range of frequencies. While these observations from molecular simulations provide insight that would be difficult to obtain solely from experiments, the associated computational cost makes it challenging to generalize across a range of conditions and systems.


Our ultimate aim is to construct a simple analytical model that connects the microscopic parameters of an ionic fluid to its macroscopic impedance. In this work, we will tackle just one half of this challenge by focussing on contributions to the overall impedance that arise from the ionic fluid's \textit{bulk dynamics}. Specifically, in the case of the system geometry depicted in Fig.~\ref{fig:setup}, the ``bulk impedance'' can be expressed in terms of the frequency-dependent bulk conductivity, $\sigma (\omega)$, 
\begin{equation}
\label{eqn:Zbulk}
Z_{\rm bulk}(\omega ) = \frac{d}{A_{\rm el}}\frac{1}{\sigma (\omega)}.
\end{equation}
In Eq.~\ref{eqn:Zbulk}, the geometric factor $d/A_{\rm el}$ is simply the ratio of the distance between the two dividing surfaces that encompass the fluid and the surface area of an electrode. See Refs.~\cite{jcox_dielectric_2022} and~\cite{dos_santos_dielectric_2018} for different examples of choosing $d$ appropriately. Note that, as $\sigma (\omega)$ is a material property, Eq.~\ref{eqn:Zbulk} makes explicit that $Z_{\rm bulk}(\omega)$ is not only an extensive quantity, but that it also depends upon the shape of the system. Clearly, the microscopic dynamics that describe $Z_{\rm bulk}(\omega)$ are encoded in $\sigma(\omega)$. Our aim can therefore be recast as developing a microscopic theory for the bulk conductivity. 

\subsection{Microscopic Correlations in Bulk Impedance}\label{sec:bulk_impedance}

We first describe the minimal equivalent circuit (EC) model for the bulk impedance, that of a parallel resistor-capacitor (RC) circuit,
\begin{equation}
  \label{eqn:ZRC}        
  Z_{\rm bulk}^{\rm (RC)}(\omega) = \frac{R}{1+\imi\omega\tau_{\rm RC}},
\end{equation}
where $\tau_{\rm RC}$ is the RC time \cite{fletcher_universal_2014, zhang_electrochemical_2004, irvine_electroceramics_1990, hernandez_correct_2016, schwan_linear_1992, chassagne_compensating_2016}. Within the radius of convergence, the real and imaginary parts are given by
\begin{align}
  \operatorname{Re}[Z_{\rm bulk}^{\rm (RC)}(\omega)] &= R\sum_{n=0}^\infty (-1)^{n}  (\omega\tau_{\rm RC})^{2n} \label{eqn:ZRC-Re} \\
  \operatorname{Im}[Z_{\rm bulk}^{\rm (RC)}(\omega)] &= R\sum_{n=0}^\infty (-1)^{n+1}(\omega\tau_{\rm RC})^{2n+1} \label{eqn:ZRC-Im},
\end{align}
which makes explicit that $Z_{\rm bulk}^{\rm (RC)}(\omega)$ is governed by a single time scale, $\tau_{\rm RC}$; this places strict requirements on the form of $\sigma(\omega)$. Specifically, the frequency-dependent bulk conductivity is related to the Fourier-Laplace transform of the current fluctuations, which
for a system of $N$ particles in a volume $\Omega$, at temperature $T$, reads

\begin{equation}
\label{eqn:sigma-FDT}
  \sigma(\omega) = \frac{\beta}{3\Omega}\int_0^\infty\!\mrm{d}t\,\langle\mbf{J}(t)\cdot\mbf{J}(0)\rangle\exp(-\imi\omega t),
\end{equation}
with $\beta = 1/k_{\rm B}T$ ($k_{\rm B}$ is Boltzmann's constant), and
\begin{equation}\label{eqn:J_def}
  \mbf{J}(t) = \sum_i^N q_i\mbf{v}_i(t),
\end{equation}
where $q_i$ is the charge of particle $i$ and $\mbf{v}_{i}(t)$ is its velocity at time $t$ \cite{schroder_computation_2008}. The conductivity can be written as a series expansion around $\omega = 0$,
\begin{align}
  \operatorname{Re}[\sigma(\omega)] &= \sum_{n=0}^\infty \sigma_{2n}\omega^{2n}    \label{eqn:sigma-Re} \\
  \operatorname{Im}[\sigma(\omega)] &= \sum_{n=0}^\infty \sigma_{2n+1}\omega^{2n+1} \label{eqn:sigma-Im},
\end{align}
where
\begin{align}
    \sigma_{2n} &= \frac{(-1)^n}{2n!}\frac{\beta}{3 \Omega}\int_{0}^{\infty} \dd{t} \langle \mbf{J}(0)\cdot\mbf{J}(t)\rangle t^{2n} \label{eqn:sig even}
\\
    \sigma_{2n+1} &= \frac{(-1)^{n+1}}{(2n+1)!}\frac{\beta}{3 \Omega}\int_{0}^{\infty} \dd{t} \langle \mbf{J}(0)\cdot\mbf{J}(t)\rangle t^{2n+1} \label{eqn:sig odd}
\end{align}
After some straightforward but tedious algebra, we find that the real and imaginary parts of $Z_{\rm bulk}(\omega)$ are
\begin{widetext}
\begin{equation}\label{eqn:Re-Z expan}
    \operatorname{Re}[Z_{\mathrm{bulk}}(\omega)] = \frac{d}{A_{\rm el}\sigma_0}\Bigg[1
      - \left( \frac{\sigma_1^2}{\sigma_0^2} +\frac{\sigma_2}{\sigma_0} \right)\omega^2  + \left(  \frac{\sigma_1^4}{\sigma_0^4} + \frac{3\sigma_2\sigma_1^2}{\sigma_0^3} + \frac{\sigma_2^2 -2\sigma_3\sigma_1}{\sigma_0^2} -\frac{\sigma_4}{\sigma_0}\right)\omega^4  + \dots \Bigg],
\end{equation}
\begin{equation}\label{eqn:Im-Z expan}
     \operatorname{Im}[Z_{\mathrm{bulk}}(\omega)] = \frac{-d}{A_{\rm el}\sigma_0}\Bigg[\frac{\sigma_1}{\sigma_0}\omega
     - \left( \frac{\sigma_1^3}{\sigma_0^3} + \frac{2\sigma_2\sigma_1}{\sigma_0^2} - \frac{\sigma_3}{\sigma_0} \right)\omega^3  + \left( \frac{\sigma_1^5}{\sigma_0^5} + \frac{4\sigma_2\sigma_1^3}{\sigma_0^4} + \frac{3(\sigma_2^2\sigma_1-\sigma_3\sigma_1^2)}{\sigma_0^3} -\frac{2(\sigma_4\sigma_1+\sigma_3\sigma_2)}{\sigma_0^2} + \frac{\sigma_5}{\sigma_0}\right)\omega^5  + \dots \Bigg],
\end{equation}
\end{widetext}
in which higher order terms follow an obvious pattern where the coefficient of $\omega^n$ contains contributions from all conductivity moments $\sigma_0, \sigma_1, \ldots,\sigma_n$. Comparing Eqs.~\ref{eqn:Re-Z expan} and~\ref{eqn:Im-Z expan} with Eqs.~\ref{eqn:ZRC-Re} and~\ref{eqn:ZRC-Im}, it is clear that the RC equivalent circuit model truncates the conductivity at first order, $\sigma(\omega) = \sigma_0 + i\omega\sigma_1$, such that
\begin{equation}\label{eqn:tauRC} 
    \tau_{\rm RC} = \sigma_1/\sigma_0 .   
\end{equation}
In contrast, the full expressions provided by Eqs.~\ref{eqn:Re-Z expan} and~\ref{eqn:Im-Z expan}, while cumbersome, make clear that $Z_{\rm bulk}(\omega)$ is not, in general, characterized by a single timescale.

\begin{figure}[!b]
     \vspace{-20pt}
     \centering
     \begin{subfigure}[b]{0.45\textwidth}
        \centering
        \includegraphics[width=\textwidth]{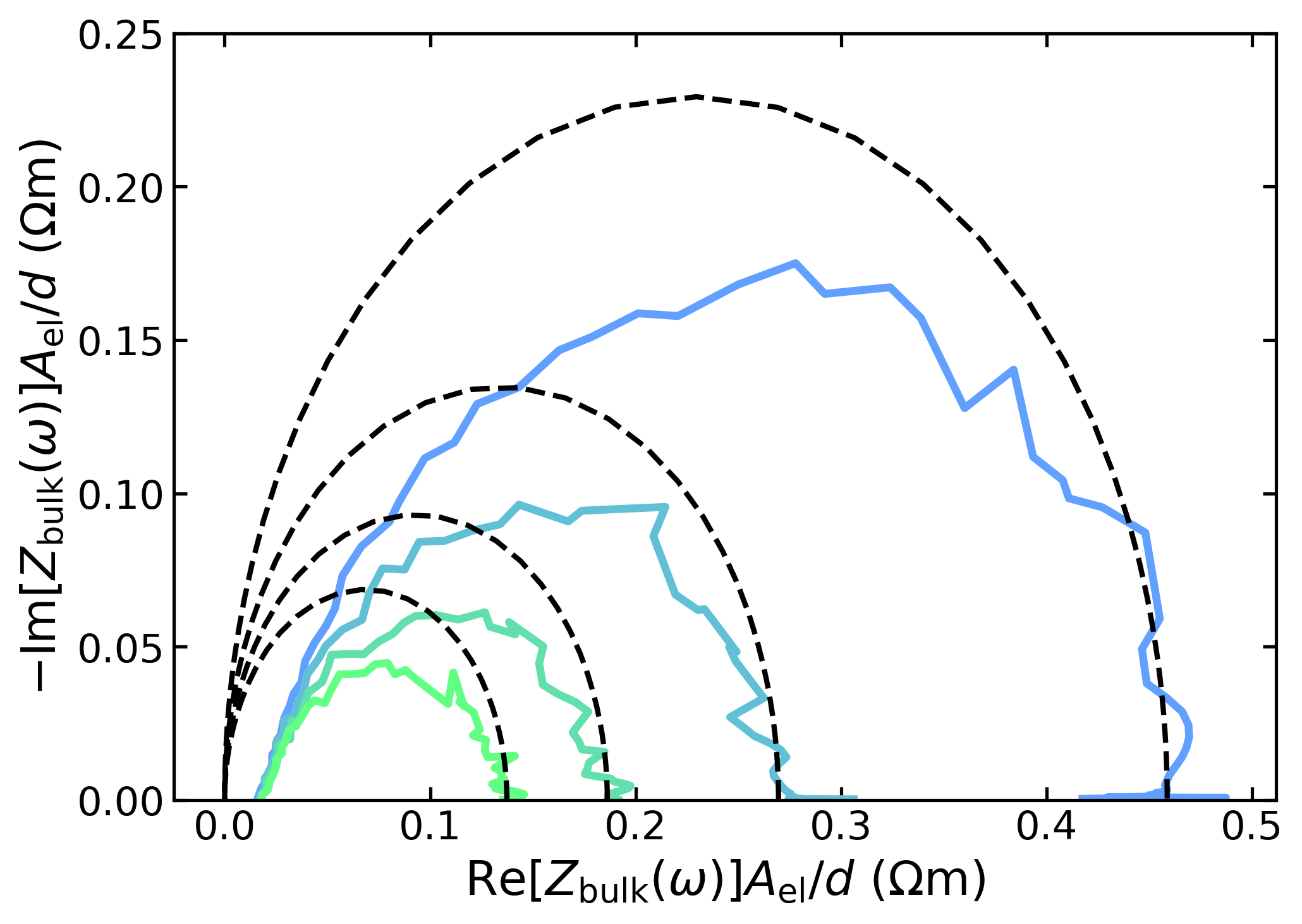}
        \vspace{-14pt}
        \caption{}
     \end{subfigure}
     \\
     \begin{subfigure}[b]{0.45\textwidth}
        \centering
        \includegraphics[width=\textwidth]{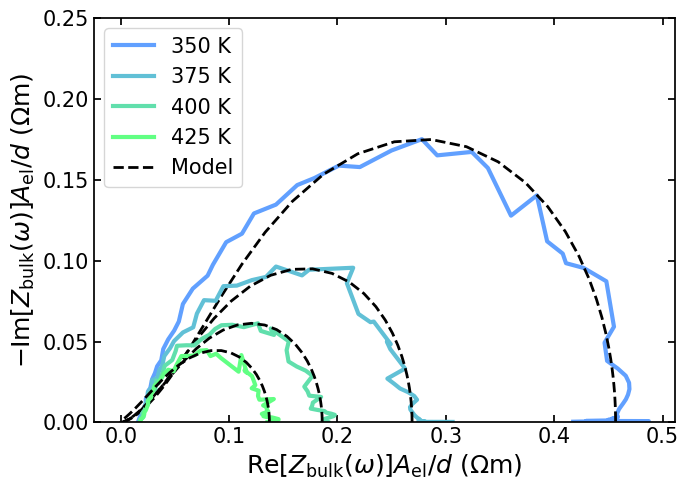}
        \vspace{-14pt}
        \caption{}
     \end{subfigure}
     \vspace{-8pt}
    \caption{\label{fig:Nyquist} Nyquist plots of $Z_{\rm bulk}(\omega)$ and
    their predicted model counterparts for Eq.~\ref{eqn:Zbulk}, with
    $\sigma(\omega)$ terminated at \textbf{(a)} first-order
    (RC circuit representation in Eq.~\ref{eqn:ZRC}) and \textbf{(b)} second-order (Eq.~\ref{eqn:Zbulk 2order}). 
    $\sigma_0$ and $\sigma_1$ were calculated from the simulation data using the Einstein-Helfand relation; $\sigma_2$ is extracted from the low-$\omega$ part of $\operatorname{Re}[\sigma(\omega)]$ using Eq.~\ref{eqn:sigma-Re} (Methods~\ref{app:einstein}).}
\end{figure}

The impact of conductivity moments beyond $\sigma_1$ can be directly assessed using results from molecular simulations. In particular, in this paper, we investigate a simple point-charge, coarse-grained model of the bulk ionic liquid (IL) \ce{[BMIM]+[BF4]^-}. Anticipating our results, in Fig.~\ref{fig:Nyquist}(a) we show $Z_{\rm bulk}(\omega)$ obtained by calculating $\sigma(\omega)$ from Eq.~\ref{eqn:sigma-FDT} in a box of 432 ion pairs at various temperatures above the glass transition. To directly connect with experimental workflows, we plot the impedance in the traditional Nyquist form (for the separate real and imaginary parts see Appendix~\ref{app:RCfit}). Alongside the simulation data, we also plot $Z^{\rm (RC)}_{\rm bulk}(\omega)$ with $R = d/A\sigma_0$ and $\tau_{\rm RC} = \sigma_1/\sigma_0$, where $\sigma_0$ and $\sigma_1$ have been obtained by the Einstein-Helfand method (see Methods~\ref{app:einstein}) \cite{schroder_computation_2008, picalek_molecular_2007}. While $Z^{\rm (RC)}_{\rm bulk}(\omega)$ qualitatively captures some of the temperature dependent trends of the simulation results, such as the decreasing maximum in $\operatorname{Im}[Z_{\rm bulk}(\omega)]$ with increasing temperature, the lack of flexibility results in a very poor quantitative description. In
Fig.~\ref{fig:Nyquist}(b), we show the same simulation data, alongside the approximate form,
\begin{equation}\label{eqn:Zbulk 2order}
  Z_{\rm bulk}(\omega) \approx \frac{d}{A_{\rm el}\sigma_0}\frac{1}{1 + \imi(\sigma_1/\sigma_0)\omega + (\sigma_2/\sigma_0)\omega^2},
\end{equation}
where we have expanded $\sigma(\omega)$ in Eq.~\ref{eqn:Zbulk} to second-order. While some quantitative differences remain, we see that including the effects of the microscopic correlations encoded in $\sigma_2$ significantly improves the overall description of the bulk impedence. In what follows, we will explain that this amounts to including effects of memory in the relaxation dynamics that are missing in the simple RC-circuit model.

The inadequacies of the single time scale RC-circuit model are already well known, with extensions such as the Cole--Cole equation,
\begin{equation}
  \label{eqn:ZCC}
  Z_{\rm CC}(\omega) = \frac{R}{1+(\imi\omega \tau)^\alpha},
\end{equation}
and Cole--Davidson equation,
\begin{equation}
  \label{eqn:ZCD}
  Z_{\rm CD}(\omega) = \frac{R}{(1+\imi\omega \tau)^\beta}
\end{equation}
commonly used to fit experimental data that deviate from the simple RC-circuit form of Eq.~\ref{eqn:ZRC} \cite{cole_dispersion_1941, davidson_dielectric_1950, davidson_dielectric_1951, friesen_impedance_2000, stoppa_interactions_2008}. For example, in Ref.~\cite{davidson_dielectric_1951} Davidson \textit{et al}. use their empirical dispersion model (Eq.~\ref{eqn:ZCD}) to describe the frequency-dependent behaviour of glycerol, a glassy, highly correlated liquid, whose asymmetric distribution does not agree with an RC-circuit form. Most commonly these functions are used to describe bulk systems whose correlation functions can be broadly described by a stretched exponential \cite{williams_non-symmetrical_1970}, however these expressions are also used to model the impedance response from transport within specific regions of an electrochemical cell itself \cite{lazanas_electrochemical_2023}. In practical usage, the $\alpha$ and $\beta$ of Eqs.~\ref{eqn:ZCC} and~\ref{eqn:ZCD} serve as additional fitting parameters to account for effects arising from a distribution of system timescales around a characteristic timescale $\tau$. In particular, the Cole--Davidson representation assumes that relaxation is well described by a distribution of multiple, uncorrelated RC processes \cite{beckmann_spectral_1988}. While such empirical equations are widely used, they are often invoked without establishing a firm connection to the underlying microscopic dynamics, which complicates interpretability \cite{iglesias_approach_2017, powles_interpretation_1951, shoar_abouzari_physical_2009, rezaei_niya_possible_2016, cordoba-torres_relationship_2015}. 

Hence, to make a principled comparison between Eqs.~\ref{eqn:Zbulk 2order},~\ref{eqn:ZCC},~and~\ref{eqn:ZCD} as function forms for fitting to empirical data, we desire a theory for $\sigma(\omega)$ which starts at the microscopic level.

The bulk of this article details how we construct an interpretable model for the frequency-dependent conductivity via a mean-field picture. Since we are studying ionic fluids, our model needs to be applicable at both low and high concentrations, in contrast to previous studies on dilute electrolytes \cite{wei_dielectric_1991,chandra_frequency_1993,chandra_frequency_2000,chandra_beyond_2000,dufreche_ionic_2002}. In Sec.~\ref{sec:bulk_sims} we will present our full simulation results and show they can be described by a GLE of the total current. Then, in Sec.~\ref{sec:analytical memory}, we will derive the connection between this GLE and the atomic equations of motion. While this will limit us to only describing single-particle correlations explicitly, we show that this is sufficient to produce a qualitative microscopic description of the fully interacting system. With this in hand, in Sec.~\ref{sec:new impedance} we will return to the total current to deduce precisely why including $\sigma_2$ in an impedance model produces the stark improvements seen between Fig.~\ref{fig:Nyquist}(a) and (b), ultimately proposing Eq.~\ref{eqn:Zbulk 2order} as a more informative alternative to the Cole-Davidson equation (Eq.~\ref{eqn:ZCD}) for modelling the impedance of glassy electrolytes.

\section{Conductivity and Memory Effects}\label{sec:bulk_sims}
\subsection{Generalized Langevin Framework}

A powerful approach for interrogating time correlation functions, such as that which appears in the integrand of the Green--Kubo formula for the conductivity (Eq.~\ref{eqn:sigma-FDT}), is to employ the Mori--Zwanzig framework \cite{zwanzig_ensemble_1960, fujisaka_continued_1987}. To expose the distinct timescales underlying the conductivity we can write down a GLE for the total current of the isotropic bulk fluid, $\mbf{J}(t)$,
\begin{equation}\label{eq:current_GLE}
    \frac{\partial}{\partial t}\bm{J}(t) = -\int_0^t \dd{t'} \kappa(t')\bm{J}(t-t')+\bm{f}^R(t),
\end{equation}
where $\kappa(t)$ is the isotropic, time-dependent friction kernel that represents the dissipation of energy out of the current-carrying degrees of freedom, and $\bm{f}^R$ is the (dimensionalised) random force that maintains detailed balance. Unlike in the simple, memoryless Langevin equation of Brownian motion, the friction appears as a time-dependent (decaying) response that takes into account the history of the current. Thanks to this, the non-Markovian expression Eq.~\ref{eq:current_GLE} is formally exact. Averaging this GLE at equilibrium gives the equation of motion for the current-current correlation function,
\begin{equation}\label{eqn:langevin in J}
    \frac{\partial}{\partial t} \langle\mbf{J}(t)\cdot\mbf{J}(0)\rangle = -\int^{t}_0 \dd{t'} \kappa(t')\langle\mbf{J}(t-t')\cdot\mbf{J}(0)\rangle.
\end{equation}
The Laplace transform of Eq.~\ref{eqn:langevin in J} can be compared to Eq.~\ref{eqn:sigma-FDT} in order to express the total conductivity in terms of $\kappa(t)$ \cite{boon_molecular_1991}:
\begin{equation}\label{eq:exact_friction_conductivity}
    \sigma(\omega)=\frac{\beta}{3\Omega}\langle\mbf{J}^2\rangle\frac{1}{\kappa(\omega)+\imi\omega}.
\end{equation}
Equation~\ref{eq:exact_friction_conductivity} states that obtaining a mathematical form for the friction kernel is equivalent to defining a complete model of the frequency-dependent conductivity. The use of a memory function is advantageous because it often assumes a simpler form than the full correlation function, as we will go on to show. It can also decay on a faster timescale. Together these two properties allow the Mori--Zwanzig analysis to reveal the minimal information needed to describe the observable of interest. 

\begin{figure*}[!ht]
     \centering
     \begin{subfigure}[b]{0.49\textwidth}
         \centering
         \includegraphics[width=\textwidth]{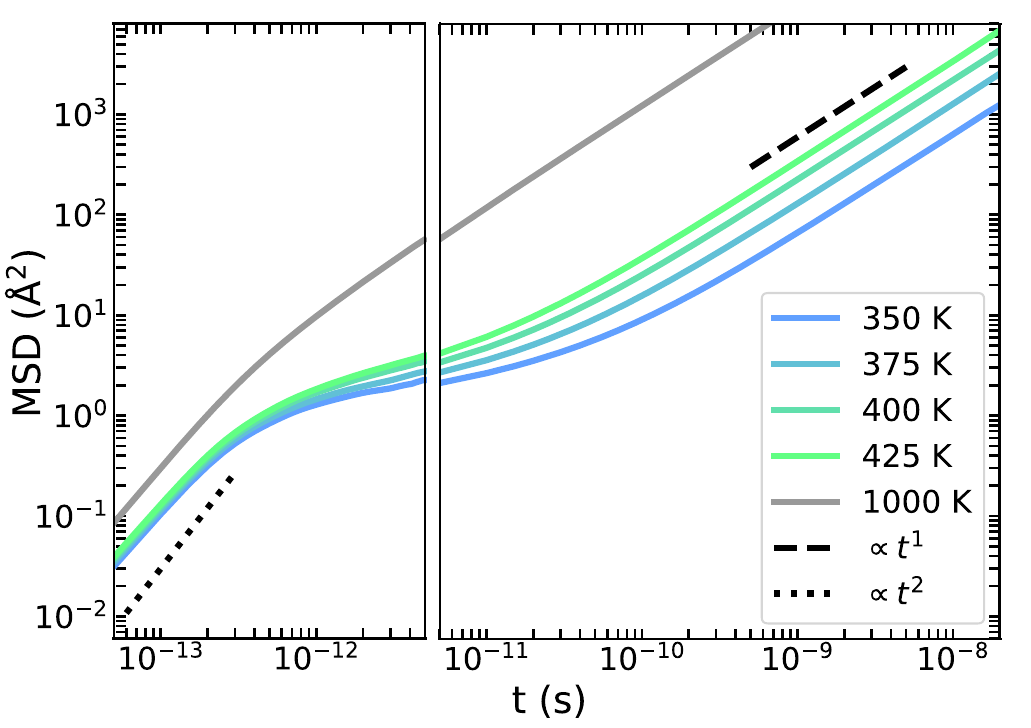}
         \caption{}
     \end{subfigure}
     \hfill
     \begin{subfigure}[b]{0.49\textwidth}
         \centering
         \includegraphics[width=\textwidth]{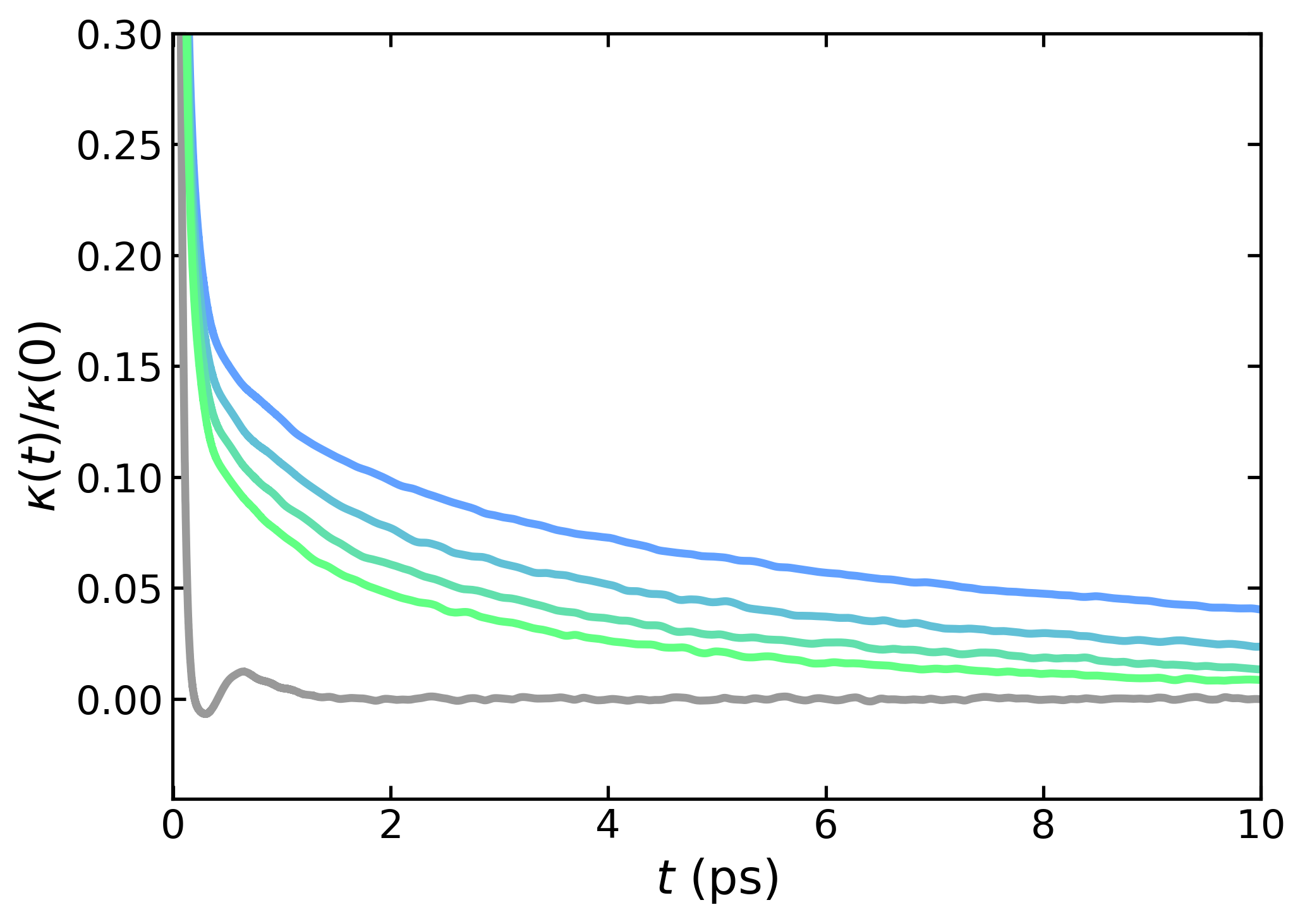}
         \caption{}
     \end{subfigure}
     \hfill
     \begin{subfigure}[b]{0.49\textwidth}
         \centering
         \includegraphics[width=\textwidth]{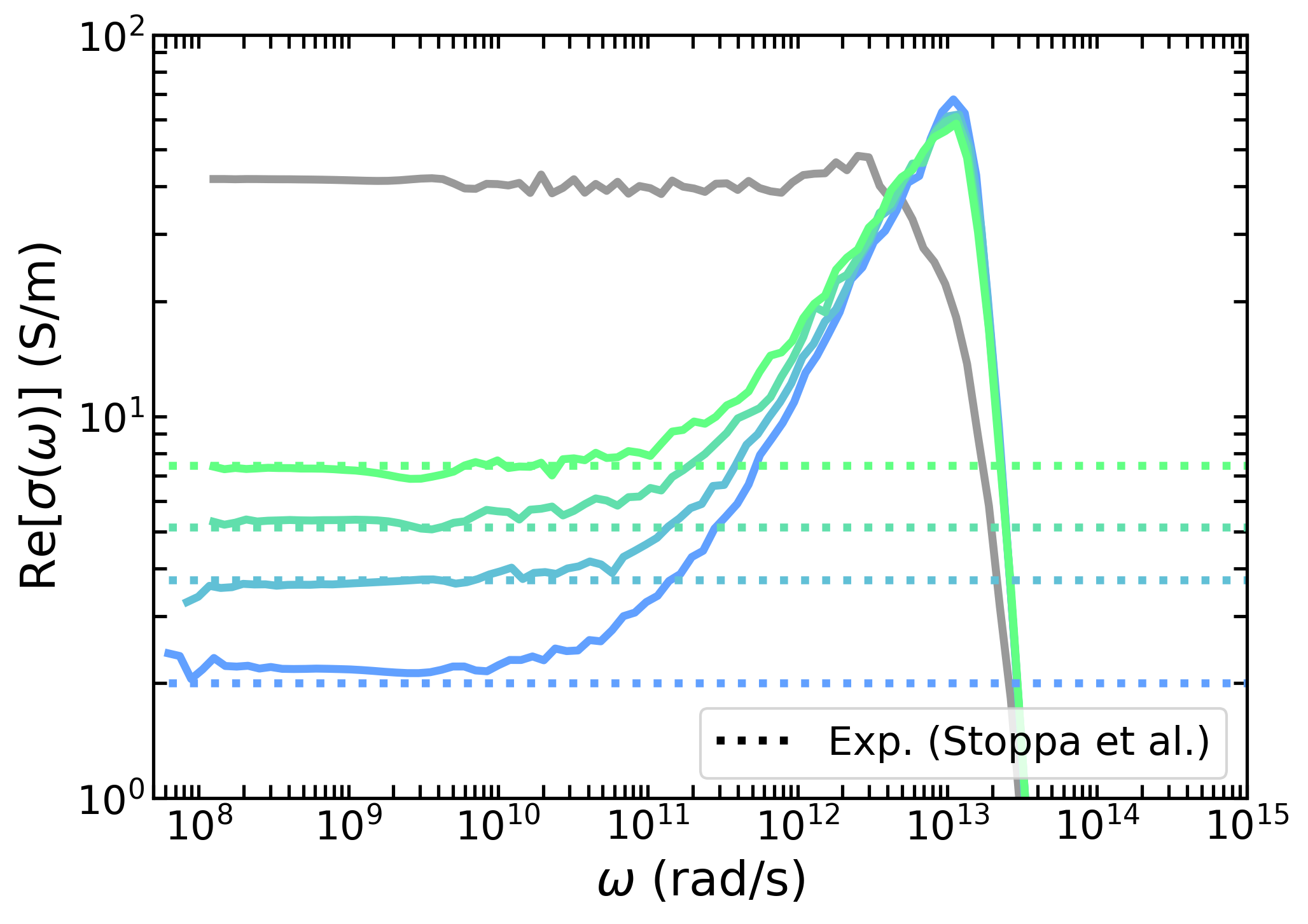}
         \caption{}
     \end{subfigure}
     \hfill
     \begin{subfigure}[b]{0.49\textwidth}
         \centering
         \includegraphics[width=\textwidth]{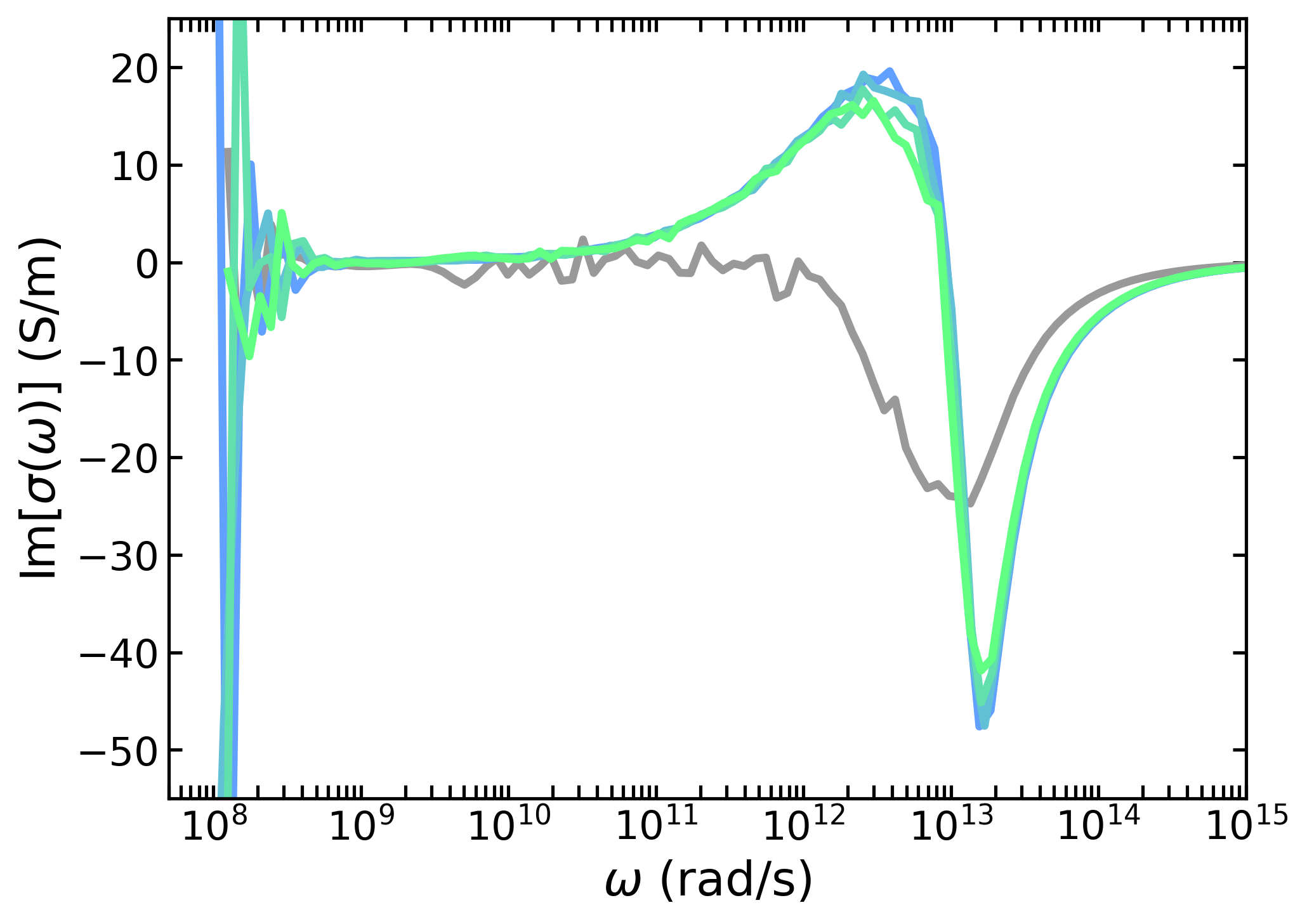}
         \caption{}
     \end{subfigure}
        \caption{\label{fig:bulkcond}
        MD data for bulk \ce{[BMIM]+[BF4]^-} over a range of temperatures. \textbf{(a)} Mean-squared displacement (MSD) of \ce{[BF4]-} anion, showing the subdiffusive regime expected for frustrated diffusion in room temperature ILs. The MSD for the \ce{[BMIM]+} cation shows the same features. At 1000 K we are at sufficiently high temperatures that there is a smooth transition from ballistic to diffusive behaviour. \textbf{(b)} Friction kernels found using the transfer tensor method. \textbf{(c)} Real-part of the bulk conductivity spectrum. The static conductivity found from simulation is compared to that found from experiment at 348.15~K, 378.15~K, 398.15~K and 428.15~K \cite{stoppa_conductivity_2010}. \textbf{(d)} Imaginary-part of the conductivity spectrum. Comparison between (a) and (c) shows that the timescales seen in the MSD (ballistic, sub-diffusive and diffusive regimes) are mirrored within the conductivity spectrum. For example, the lifetime of the subdiffusive regime in (a) aligns with the frequency range between the plateau and maximum in (c).}
\end{figure*}
\subsection{Bulk Simulations of \texorpdfstring{\ce{[BMIM]+[BF4]^-}}{[BMIM][BF4]}}

Ionic liquids are ``infinite concentration'' electrolytes \cite{perkin_is_2013} where the competition between long-range electrostatic interactions and short-ranged packing of molecules results in highly correlated dynamics \cite{hayes_structure_2015, pei_ionic_2022, ngo_thermal_2000}. Every species within the liquid is charged and contributes to the conductivity over the whole frequency range, and it is 
these broad electrical responses that motivate many of the potential applications of ILs. In particular, room temperature ILs have the potential to produce high energy density electric double layer capacitors due to their wide operating potential window, high ionic conductivity, and good thermal stability \cite{eftekhari_supercapacitors_2017, miao_ionic_2021, salanne_ionic_2018, ray_application_2021}.

In this work, we simulate the bulk ionic liquid \texorpdfstring{\ce{[BMIM]+[BF4]^-}}{[BMIM][BF4]} at several different temperatures above the glass transition (187.7~K) \cite{merlet_new_2012, shamim_glass_2010}. To access the long timescales required to extract the $\omega \rightarrow 0$ part of the conductivity, we employ a coarse-grained force field in which \texorpdfstring{\ce{[BMIM]+}}{[BMIM]} molecules comprise three Lennard--Jones centres, each carrying a partial charge, and \texorpdfstring{\ce{[BF4]^-}}{[BF4]} molecules comprise a single Lennard--Jones centre with a partial charge (each ion pair is electroneutral $z=\pm0.78$~e). Boxes of 432 ion pairs were propagated for at least $100$~ns in 5 parallel simulations to obtain converged statistics. Full simulation details are outlined in Methods~\ref{app:simdetails}. The memory function at each temperature is extracted by numerical inversion of the converged $\sigma(\omega)$ \cite{cerrillo_non-markovian_2014}, see Methods~\ref{app:TTM} for full details.

Figure~\ref{fig:bulkcond}(a) shows the mean-squared dispacement of the \ce{[BF4]-} anion throughout the simulation \cite{fong_onsager_2020}. The plateau seen at approximately 2~ps occurring between the ballistic and diffusive regimes is a signature of the frustrated diffusion process characteristic of ionic liquids; physically, this corresponds to the timescale over which ions are trapped within their coordination cage before undergoing the significant cooperative structural reorganisation required for diffusion \cite{rusciano_fickian_2022, de_souza_energy_2008}. This subdiffusive ``$\beta$-relaxation'' \cite{roy_dynamics_2010} timescale decreases with increasing temperature as the barrier to reorganisation is more easily overcome \cite{knorr_spectroscopic_2016, shiraishi_joharigoldstein_2023, karmakar_overview_2016, sha_dynamical_2019}. This strong temperature dependence is reflected in the decay of the memory function shown in Fig.~\ref{fig:bulkcond}(b); as the temperature increases, the extent of the memory, i.e. the lifetime of $\kappa(t)$, decreases.

In Fig.~\ref{fig:bulkcond}(c)~and~(d) we show the real and imaginary parts of $\sigma(\omega)$, respectively, obtained from the Green--Kubo relationship (Eq.~\ref{eqn:sigma-FDT}) alongside the windowing procedure outlined in Ref.~\cite{pireddu_impedance_2024}. Here, we observe that the temperature-dependence manifests in two ways. First, and most strikingly, the static conductivity increases significantly with temperature, as seen from the $\omega\to 0$ limit in Fig.~\ref{fig:bulkcond}(c). Second, the maximum in both the real and imaginary parts at $\omega\lesssim 10^{13}$ rad/s is seen to decrease with increasing temperature. Similarly, the minimum in the imaginary part at $\omega\gtrsim 10^{13}$ rad/s is seen to decrease with increasing temperature.

To confirm the subdiffusive origin of these features, in addition to the four operational temperatures $350$\,K--$425$\,K, we also ran simulations at 1000~K. We find that, at this elevated temperature, all objects do appear qualitatively different. In particular, the memory function has a much shorter lifetime, approaching a Markovian limit, and so the conductivity spectrum appears close to what would be predicted for transport with a single diffusive timescale. This is consistent with our description as at this temperature the energetic barrier to cage-breaking is easily overcome, allowing us to ignore the timescale corresponding to ion-cage dynamics.

In dilute electrolytes, the increase in $\operatorname{Re}[\sigma(\omega)]$ from the zero-frequency plateau seen in Fig.~\ref{fig:bulkcond}(c) is commonly attributed to the Debye-Falkenhagen effect. At low, but finite, frequencies, the field acts to distort the relative position of each ion with its counter-ion (Debye) cloud \cite{debye_dispersion_1928}. The restoring force associated with the resulting induced dipole acts to retard the motion of the central ion \cite{lesikar_debye-falkenhagen_1980, anderson_debye-falkenhagen_1994}. Upon increasing the frequency, the induced dipole becomes smaller, resulting in a reduced friction force and higher conductivity. At very high frequencies, the inertia of the ion itself resists a change in momentum; for the simple point-charge model that we investigate, as there are no vibrational or electronic degrees of freedom \cite{pireddu_impedance_2024}, this causes the conductivity to go to zero. The crucial insight from this dilute electrolyte picture is the notion of an ``atmospheric'' timescale ($\tau_{\rm atm}$) associated with the relaxation of the induced dipole, in addition to the diffusive timescale \cite{chandra_frequency_1993, ibuki_effect_1990, bonneau_frequency-dependent_2024}.

We argue that, in an ionic liquid, a similar physical picture can be imagined but with the counter-ion cloud replaced with the counter-ion cage, and the atmospheric timescale corresponding to that of $\beta$-relaxation. Our simulations support this assignment, as the frequency range for which this increase in $\operatorname{Re}[\sigma(\omega)]$ is occurring matches the lifetime of the subdiffusive regime of our IL. This picture suggests that in order to capture the microscopic mechanism that results in the maximum in $\operatorname{Re}[\sigma(\omega)]$ in our analytical model, we must look beyond a single-timescale description to properly capture the frequency-resolved transport properties of \ce{[BMIM]+[BF4]^-} \cite{janssen_mode-coupling_2018, boon_molecular_1991}.

\section{An Analytical Memory Function}\label{sec:analytical memory}
\subsection{The Microscopic Connection}

While the Mori--Zwanzig approach can simplify analysis, $\kappa(t)$ itself defines a collective property of many ions and does not lead directly to a clear atomistic interpretation. To establish a connection to the underlying microscopic dynamics, we also introduce a GLE for the dynamics of individual ions. Specifically, for a tagged ion $i$, the equation of motion for its velocity can be written as \cite{canales_generalized_1998},
\begin{equation}\label{eqn:langevin}
    \frac{\partial}{\partial t} \mbf{v}_i(t) = -\int^{t}_0 \dd{t'} \zeta(t')\mbf{v}_i(t-t') + \frac{\mbf{F}_i^{\mrm{R}}(t)}{m_i},
\end{equation}
where $\mbf{F}_i^\mrm{R}(t)$ is the random force acting upon it due to its interactions with the rest of the system, and $m_i$ is its mass. Similar to $\kappa(t)$, $\zeta(t)$ describes how energy is dissipated; for simplicity, we assume $\zeta(t)$ is identical for anions and cations (we test the validity of this approximation in  Appendix~\ref{app:cross_terms}). 

This atomistic non-Markovian expression, Eq.~\ref{eqn:langevin}, is closely related to the expression involving the full bulk current, Eq.~\ref{eqn:langevin in J}. To express the current autocorrelation in terms of individual ions, we average the current, Eq.~\ref{eqn:J_def}, against its initial value,

\begin{align}
    \langle \mbf{J}(t)\cdot\mbf{J}(0)\rangle &= \sum_{ij}q_iq_j \langle\mbf{v}_i(t)\cdot\mbf{v}_j(0)\rangle  \\
    &= \sum_i q_i^2\langle\mbf{v}_i(t)\cdot\mbf{v}_i(0)\rangle + \sum_{ij}{}' q_iq_j \langle\mbf{v}_i(t)\cdot\mbf{v}_j(0)\rangle,\label{eq:JJ_two_terms}
\end{align}
which is composed of `self' and `cross' terms such that in the second sum the prime denotes $i\neq j$. Taking the time derivative,
\begin{align}
   \frac{\partial}{\partial t}\langle\mbf{J}(t)\cdot\mbf{J}(0)\rangle &= -\sum_i q_i^2  \int_0^t\dd{t'}\zeta(t')\langle\mbf{v}_i(t-t')\cdot\mbf{v}_i(0)\rangle \nonumber \\
   &\hspace{11pt}-\sum_{ij}{}' q_iq_j\Big[\int_0^t\dd{t'}\zeta(t')\langle\mbf{v}_i(t-t')\cdot\mbf{v}_j(0)\rangle \nonumber\\
   &\hspace{50pt}+\frac{1}{m_i}\langle \mbf{F}^\mrm{R}_i(t)\cdot\mbf{v}_j(0) \rangle\Big],\label{eq:dJJ_two_terms}
\end{align}
provides an expression for the object of Eq.~\ref{eqn:langevin in J}. Here, averaging Eq.~\ref{eqn:langevin} against $\mbf{v}_j(0)$ gives only a single term for $i=j$ since correlations between the random force $\mbf{F}^\mrm{R}_i(t)$ and $\mbf{v}_i$ are strictly zero, but for $i\neq j$ these correlations are in general finite \footnote{This is because the random force remains in the complementary subspace of $\mbf{v}_i$ at all times, while the velocities of all the other ions $\mbf{v}_j$ start in the complementary subspace and only partial enter the projected space under time evolution.}. It is the presence of this $\langle\mbf{F}^\mrm{R}_i(t)\cdot\mbf{v}_j(0) \rangle$ term that prevents direct identification of $\zeta(t)=\kappa(t)$. Although the random force term could in principle be tackled separately \cite{carof_two_2014, jung_iterative_2017, daldrop_butane_2018}, it is in general very challenging to determine the form of $\mbf{F}^\mrm{R}_i(t)$, even numerically; in applications of the GLE to generate simulated trajectories, the random force is usually replaced with a random variable sampled from a model distribution function. 

\subsection{The Mean Field Picture}

To make progress, one can adopt a simplified one-body picture in which the double sum in Eq.~\ref{eq:dJJ_two_terms} is ignored. That is, we assume velocity correlations between different ions are negligibly small,
\begin{align}\label{eq:self_term_JJ}
     \frac{\partial}{\partial t}\langle\mbf{J}(t)\cdot\mbf{J}(0)\rangle &\approx -\int_0^t\dd{t'}\zeta(t')\sum_iq_i^2\langle\mbf{v}_i(t-t')\cdot\mbf{v}_i(0)\rangle.
\end{align} 
For our complex ions we obviously cannot neglect intramolecular correlations so we adopt a simplified picture where we replace the sum over atoms in the same molecule by a single charge on the central bead, $z = |\sum_{i \in 
\mathrm{mol}}q_i |$, and neglect rotational motion.
Under this approximation, $\zeta(t)=\kappa(t)$, and the frequency-dependent conductivity reduces to a generalized Nernst--Einstein equation,
\begin{align}
\label{eqn:sigma-FDT-single}
  \sigma_{\rm s}(\omega) &= \frac{\beta}{3\Omega}\int_0^\infty\!\mrm{d}t\,\left(\sum_{i}^{N_\mathrm{mol}} z^2\langle\mbf{v}_i(t-t')\cdot\mbf{v}_i(0)\rangle\right)\exp(-\imi\omega t). \\
  &= \rho z^2 \left(\frac{1}{m_+}+\frac{1}{m_-}\right)\frac{1}{\zeta(\omega)+\imi\omega}, \label{eqn:sigma_from_zeta}
\end{align}
where $\rho$ is the number density, $z$ is the absolute value of the ionic charge, and we have used the equipartition result $\sum^{N_{\rm mol}}_i\langle\mbf{v}_i^2\rangle = 3N_{\rm mol}/\beta m_i$ where $m_i$ is the mass of species~$\alpha$. In the hydrodynamic limit, one can take $\zeta(\omega)=\bar{\zeta}$ as a constant and recover the Nernst-Einstein formula itself. While the effects of cross-correlations are known to be significant as $\omega\to 0$, this effective single-ion dynamics ought to capture, at least qualitatively, the behaviour of $\sigma(\omega)$ at higher frequencies.

In the original work of Debye and Falkenhagen \cite{debye_dispersion_1928} the form of $\zeta(t)\approx\kappa(t)$ was derived from analytic considerations valid for an electrolyte in the limit of infinite dilution \cite{lesikar_debye-falkenhagen_1980, anderson_debye-falkenhagen_1994}. In this regime, where correlations arise purely from long-range electrostatics,
\begin{equation}\label{eq:debye_falkenhagen_kernel}
    \kappa_{\rm DF}(t) = \bar{\zeta}\frac{1+\sqrt{q}}{1+\sqrt{q(1+\imi \omega \tau_{\rm atm}) }},
\end{equation}
where $q=1/2$ for a symmetric binary electrolyte and $\tau_{\rm atm}$ is the new timescale arising from interaction of the tagged ion with its correlated, neutralizing Debye cloud,
\begin{equation}
    \tau_{\mrm{atm}}=\frac{1}{(D_++D_-)q\kappa_{\rm D}^2}.
\end{equation}
$D_+$ and $D_-$ are the self-diffusion coefficients of the cation and anion respectively, and $\kappa_{\rm D}$ is the inverse Debye screening length of the electrolyte.

Subsequently, researchers have sought to extend the Debye-Falkenhagen theory to finite ionic strength. A series of papers combining mode-coupling theory and simulation \cite{wei_dielectric_1991,chandra_frequency_1993,chandra_frequency_2000,chandra_beyond_2000,dufreche_ionic_2002}, as well as an analytical investigation \cite{ibuki_effect_1990} of different model forms for $\kappa(t)$, have treated aqueous and model-aqueous electrolytes of concentrations up to around 1~M. These studies have arrived at expressions for $\kappa(t)$ containing two terms which separate the ion-ion and ion-solvent friction. Yet, ionic liquids exist at nominal concentration greatly in excess of even these extended models, and where the identity of the `solvent' does not translate into well-defined, molecular degrees of freedom. 

In order to construct an interpretable analytical model that maintains notions of tagged ions and their immediate environments, we will move away from an analytic expansion for $\kappa(t)$ and instead work directly with the single-particle memory function $\zeta(t)$. Working with $\zeta(t)$ significantly reduces the complexity of the task, and all of the information we learn about the system through quantifying $\zeta(t)$ must also be contained within $\kappa(t)$ through Eq.~\ref{eq:dJJ_two_terms}. Therefore, in Section~\ref{sec:new impedance} we will return to a direct analysis of the simulation $\kappa(t)$ using what we have uncovered about the nature of $\zeta(t)$ to interpret the form of the full impedance spectrum. 

\begin{figure*}[!ht]
     \centering
     \begin{subfigure}[b]{0.49\textwidth}
         \centering
         \includegraphics[width=\textwidth]{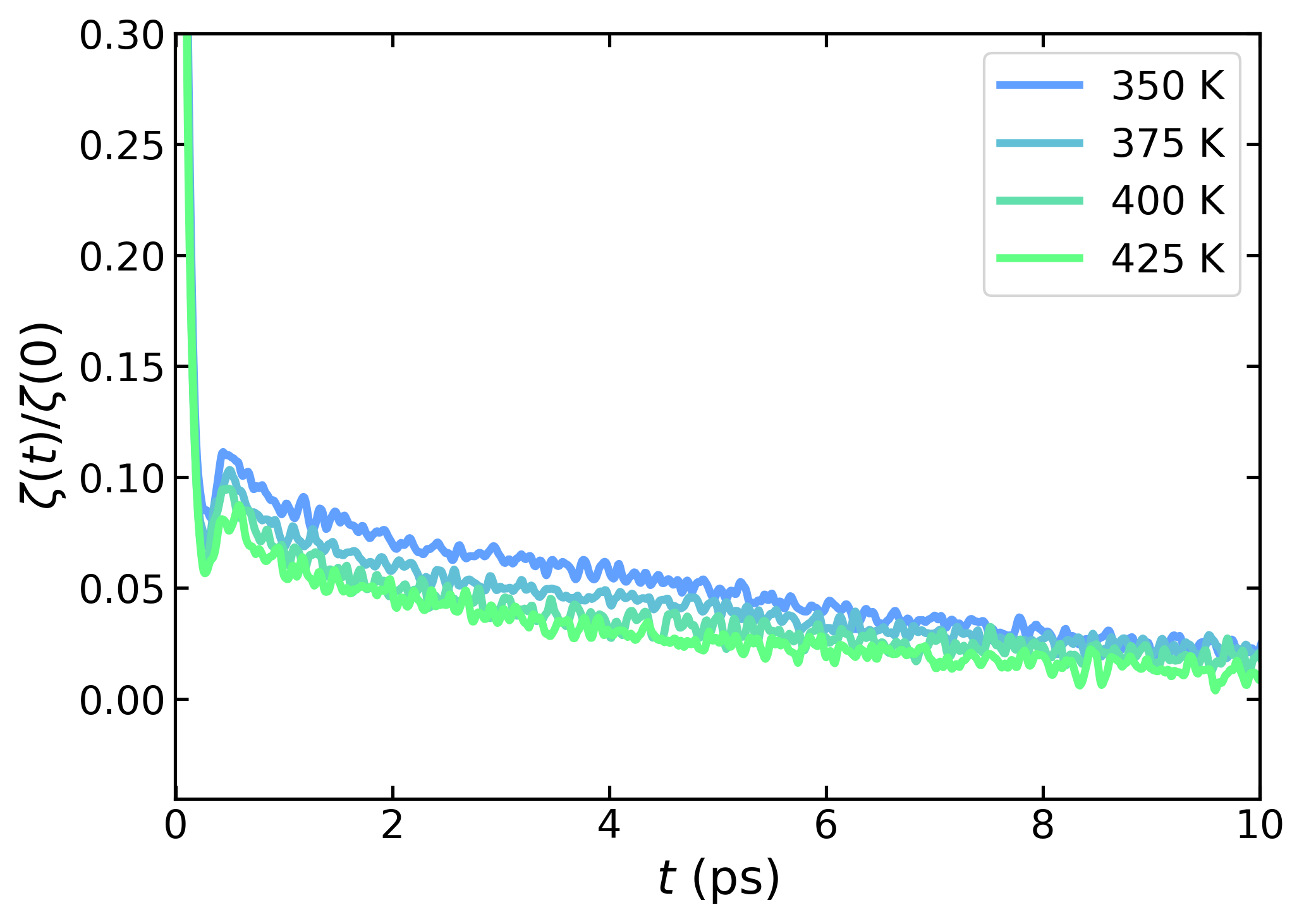}
         \caption{}
     \end{subfigure}
     \hfill
     \begin{subfigure}[b]{0.49\textwidth}
         \centering
         \includegraphics[width=\textwidth]{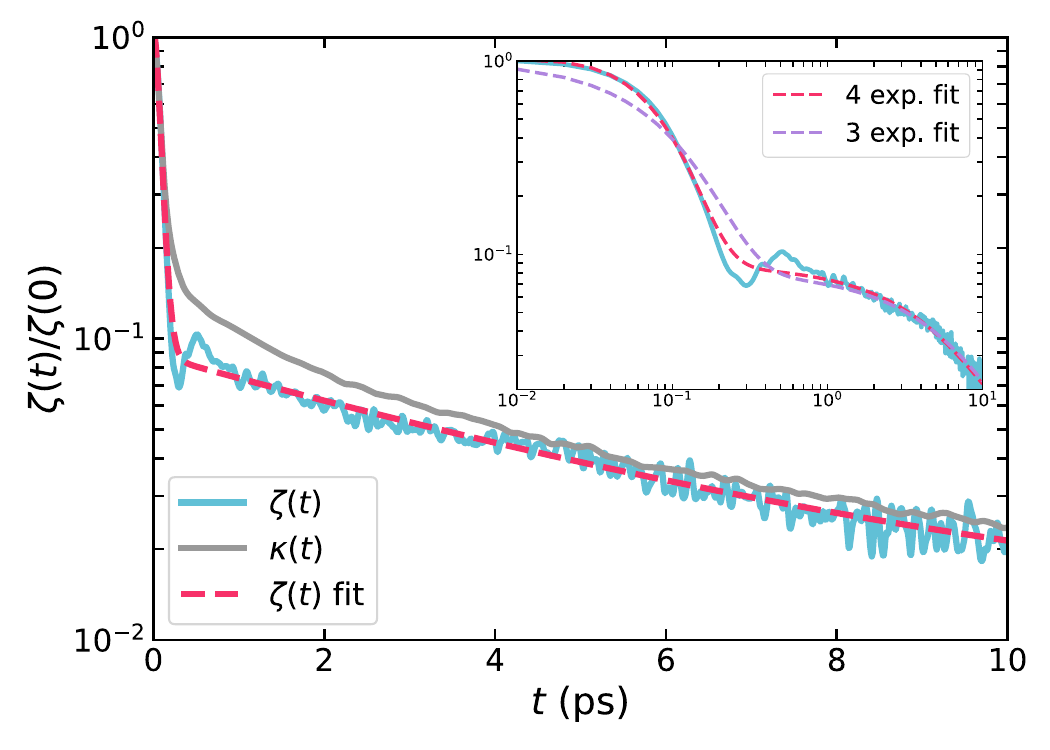}
         \caption{}
     \end{subfigure}
     \hfill
     \begin{subfigure}[b]{0.49\textwidth}
         \centering
         \includegraphics[width=\textwidth]{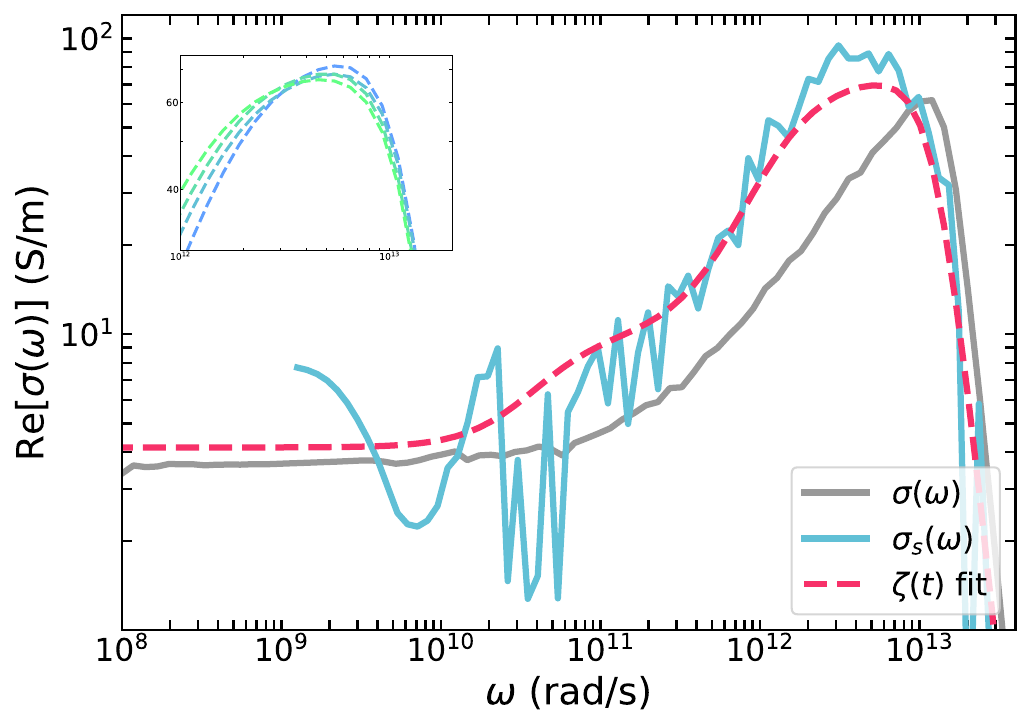}
         \caption{}
     \end{subfigure}
     \hfill
     \begin{subfigure}[b]{0.49\textwidth}
         \centering
         \includegraphics[width=\textwidth]{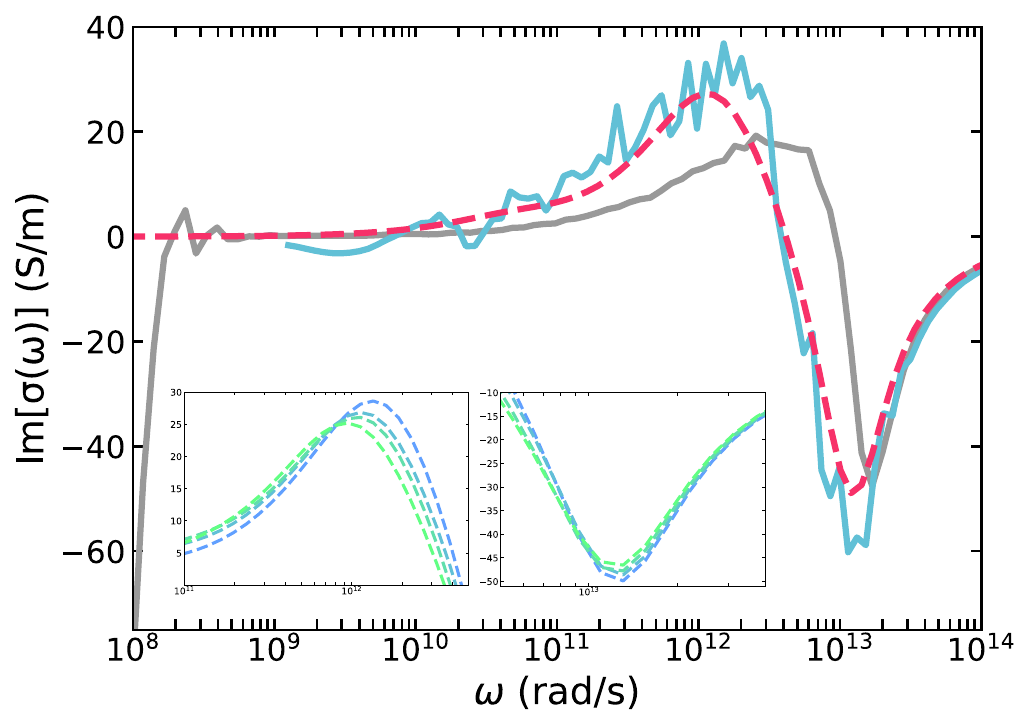}
         \caption{}
     \end{subfigure}
        \caption{\label{fig:zeta_data} Single tagged-ion MD data compared to the model fit Eq.~\ref{eqn:3exptime}. \textbf{(a)} Normalised single-particle memory function as defined in Eq.~\ref{eqn:langevin} at the four temperatures exhibiting subdiffusive behaviour. Direct comparison to Fig.~\ref{fig:bulkcond}(b) highlights the similar functional form of these two objects despite the increased complexity in the origin of $\kappa(t)$. \textbf{(b)} (inset) Four-exponential fit of the memory function at 375\,K (see Appendix~\ref{app:cross_terms}~and~\ref{app:singleioncond} for other temperature) produces quantitative agreement over both long and short timescales. When this fit is transformed to the single-particles conductivity (Eq.~\ref{eqn:sigma_from_zeta}), \textbf{(c)} and \textbf{(d)} show the good agreement with simulation results, and highlight that this form of the memory function is able to capture the key features seen in both the full ($\sigma$) and single-particle ($\sigma_s$) conductivity.}
\end{figure*}

The efficacy of this approach can be appreciated by comparing Fig~\ref{fig:zeta_data}(a) to~\ref{fig:bulkcond}(b), which highlights the similar functional forms of $\zeta(t)$ and $\kappa(t)$, alongside their similar behaviour with changing temperature. When $\zeta(t)$ found from simulation is transformed to the conductivity via Eq.~\ref{eqn:sigma-FDT-single}, we find that the features of interest within the total spectrum (such as the increase in $\operatorname{Re}[\sigma_{\rm max}(\omega)]$ with frequency) are very well captured by the single-particle memory, highlighting that collective effects are less important at high frequency. This allows us to assert that a single-particle analytical model is sufficient to understand the molecular origin of these spectral features. We now turn to a minimal model that gives rise to a friction kernel with form presented in Fig.~\ref{fig:zeta_data}(a).

\vspace{-8pt}
\subsection{The IO Model}

The Nernst-Einstein picture reduces the many-body dynamics to individual ions diffusing on some long, hydrodynamic timescale. This is the $\omega\rightarrow0$ limit of the conductivity \textit{in a mean field picture}. The Debye-Falkenhagen theory goes beyond this by considering individual ions to be `dressed' by their counter-ion clouds, thereby introducing the second timescale $\tau_{\rm atm}$. To treat an electrolyte in the `infinite-concentration' limit, we cannot resort to perturbative arguments; there will be more than just one additional timescale. We therefore seek a more general model that nonetheless retains the essential ideas of an ion-cloud picture.

To achieve this goal, we invoke the itinerant oscillator (IO) model as a microscopic starting point for introducing additional friction timescales \cite{sears_itinerant_1965, boon_molecular_1991, damle_itinerant-oscillator_1968-1}. The IO model reduces, by means of Mori--Zwanzig projection, the full dynamics of a complex fluid to a 2-body picture of a tagged particle and its immediate, correlated surroundings; both `quasiparticles' are coupled to each other harmonically, and with all other dynamics experienced as friction from the remaining environment. The correlated surroundings can be considered as the Poisson-Boltzmann cloud in a Debye-Falkenhagen picture, or instead as the counter-ion cage in an ionic liquid. 

The mathematical definition of the IO model comes from the Mori--Zwanzig projection that results in the following open, two-body system,
\begin{equation}\label{eqn:itinerant1}
\begin{split}
    \frac{\bm{F}_1(t)}{m_1} = \bm{\Dot{v}}_1(t)+\int^t_0\dd{t'}\gamma_1(t-t')\bm{v}_1(t') + ~~~~\\
    \omega_1^2\int^t_0\dd{t'}[\bm{v}_1(t')-\bm{v}_0(t')] ,
\end{split}
\end{equation}
\begin{equation}\label{eqn:itinerant0}
\begin{split}
    \frac{\bm{F}_0(t)}{m_0} = \bm{\Dot{v}}_0(t)+\int^t_0\dd{t'}\gamma_0(t-t')\bm{v}_0(t') + ~~~~\\
    \omega_0^2\int^t_0\dd{t'}[\bm{v}_0(t')-\bm{v}_1(t')],
\end{split}
\end{equation}
where the tagged ion is given subscript 1 and the correlated surroundings are denoted by the subscript 0. Equation~\ref{eqn:itinerant0} serves as a definition for the identity of the correlated surroundings as the (quasi)particle chosen to satisfy the harmonic coupling to the tagged ion, characterized by $\omega_n^2 =k/m_n$, where $m_n$ is the quasiparticle mass and $k$ is an effective spring constant. The time-dependent friction coefficients $\gamma_n(t)$ represent the two particles interacting with the remaining environment (other ions, impurities, solvent, etc.). Different chemical identities for the species in the ionic liquid will give different values for the coupling ($\omega_n^2$) and friction parameters ($\gamma_n(t)$), expressing changes in size, mass, shape, etc. The model parameters are also dependent on ensemble parameters such as the temperature, and we go on to show how they vary in an intuitive way in Sec.~\ref{sec:DF}.

The memory function of the ion alone is found by integrating out the motion of its correlated surroundings, reducing Eq.~\ref{eqn:itinerant1} to the one-body GLE of Eq.~\ref{eq:self_term_JJ} (see Appendix~\ref{app:itinerant} for the full derivation). The resulting memory function is \cite{boon_molecular_1991}
\begin{equation}\label{eqn:IOkernel1}
\zeta_{\mathrm{IO}}(\omega)=\gamma_1(\omega)+\frac{\omega_1^2}{\imi\omega+\frac{\omega_0^2}{\imi\omega+\gamma_0(\omega)}}.
\end{equation}
This is a continued fraction expression to at least second order in $\omega$. The flexibility in the expression stems from the as-yet indeterminate forms of $\gamma_n(\omega)$. The memory function prescribed by the IO model maintains microscopic interpretability: in an IL, the first term of Eq.~\ref{eqn:IOkernel1} describes the interaction between the central ion and its surroundings, while the timescales in the second term are determined solely by parameters defined by the cage; the prefactor that controls the degree to which these cage timescales contribute to the memory function is equal to the coupling frequency $\omega_1^2$.

This separation into two terms is analogous to the results from previous work on finite, low concentration electrolytes \cite{dufreche_ionic_2002, ibuki_effect_1990}. In these studies, the ion-solvent friction was found to be well captured by a delta function \cite{ibuki_effect_1990}, i.e., Markovian friction. In contrast, for an IL, the appropriate forms of $\gamma_n(t)$ are unclear. If we apply a similar Markovian simplification, we arrive at a form for $\zeta_{\mathrm{IO}}(t)$ comprising three functions: a short, delta-like ion-environment function, and two exponential decays associated with the ion-cloud coupling. This represents the simplest possible IO model, but there is no evidence yet to justify it. Therefore, we instead motivate a particular form of $\zeta_{\mathrm{IO}}(t)$ by comparing to atomistic simulation.

\subsection{The Debye-Falkenhagen Effect in Highly Concentrated Electrolytes}\label{sec:DF}

In Fig.~\ref{fig:zeta_data}(a) we show $\zeta(t)$ obtained from the velocity autocorrelation functions of the ions at different temperatures. Our first observation regarding the form of $\zeta(t)$ is that there exists a fast, well-separated timescale at all temperatures. This timescale is however not delta-like, especially at the lower temperatures where it is of the order of 100~fs. This rules out a Markovian approximation for $\gamma_1(t)$. The second observation also concerns this short-time behaviour; there is an oscillation at approximately 0.5~ps, which tells us that our model must contain either an oscillatory component or a decaying function with a negative prefactor in order to capture this transition between short and long-time behaviour. Our third observation is that, as is clear from a logarithmic plot seen in Fig.~\ref{fig:zeta_data}(b), there is a slow decay exceeding the apparent lifetime of the $\langle \bm{J}(0)\cdot\bm{J}(t) \rangle$ correlation function. This long time behaviour determines the static conductivity of the electrolyte.

Motivated by these observations, we fit $\zeta(t)$ in a basis of exponential functions. To fit the long-time decay we use the known limit of Eq.~\ref{eqn:sigma_from_zeta} to ensure that the model recovers the Nernst-Einstein static conductivity,
\begin{equation}\label{eq:nernst-einstein}
    \lim_{\omega\rightarrow 0}\sigma_s(\omega) = \sigma_{\rm NE} \equiv \beta \rho q^2 (D_+ + D_-),
\end{equation}
where $D_+$ and $D_-$ are the diffusion coefficients of the cation and anion respectively. We found that it was not possible to quantitatively capture this long-time limit with a single exponential function and therefore the long-time behaviour is modelled as a biexponential decay. To capture the short-time drop, the fastest decay is written as $\gamma_1(t) = \gamma_1\exp(-\alpha_1t)$. As seen in the inset of Fig.~\ref{fig:zeta_data}(b) this three-exponential fit is only semi-quantitative. Indeed, as a classical equilibrium correlation function, the expression should be even around $t=0$. To give enough flexibility for the linear contribution to be cancelled at short time, we introduce a second short-time exponential function, this time with a negative prefactor. The best-fit from these four exponentials still does not capture the oscillation around 0.5~ps, but is significantly more accurate at early times, as is shown in the Fig.~\ref{fig:zeta_data}(b)~inset. 

We now relate this model memory function to the single-particle conductivity using the generalized Nernst--Einstein expression, Eq.~\ref{eqn:sigma_from_zeta}. We present the real and imaginary parts of the model $\sigma_s(\omega)$ in Fig.~\ref{fig:zeta_data}(c)~and~(d), alongside the results directly from simulation. Not only does the four-exponential memory capture both the low and high-frequency behaviour qualitatively, it also reproduces the shoulder feature and the measured decrease in both $\operatorname{Re}[\sigma_s(\omega_\mathrm{max})]$ and $\operatorname{Im}[\sigma_s(\omega_\mathrm{max})]$ with temperature (see inset). The fact that the IO model captures the temperature dependence observed in the total conductivity spectra (Fig.~\ref{fig:bulkcond}(c)~and~(d)) strongly suggests that it contains enough of the essential physics of the system to understand the molecular mechanisms that govern relaxation in the ionic liquid. We have therefore found the minimal form of $\zeta(t)$ to be a linear combination of four exponential functions.

While we have established empirically that $\zeta(t)$ is well described by a four-exponential basis, the general expression for $\zeta_{\rm IO}(\omega)$ provides a means to understand how such a form arises. Specifically, if we assert that the ion-environment and cage-environment frictions $\gamma_1(t)$ and $\gamma_0(t)$ are single exponentials with characteristic timescales of $1/\alpha_1$ and $1/\alpha_0$, respectively, Eq.~\ref{eqn:IOkernel1} reduces to
\begin{equation}\label{eqn:3expfreq}
    \zeta_{\rm IO}(\omega) = \frac{\gamma_1}{\imi \omega + \alpha_1} + \frac{\omega_1^2}{\imi \omega +\frac{\omega_0^2}{\imi \omega + \frac{\gamma_0}{\imi \omega + \alpha_0}}},
\end{equation}
which in the time domain reads
\begin{equation}\label{eqn:3exptime}
    \zeta_{\rm IO}(t) = \gamma_1 \mathrm{e}^{-\alpha_1t} + \omega_1^2 \sum ^3_n A_n \mathrm{e}^{-s_n t}.
\end{equation}
Both $A_n$ and $s_n$ are defined solely by parameters associated with the ion cage ($\omega_0$, $\gamma_0$ and $\alpha_0$) and have complicated forms \cite{fujisaka_continued_1987, netz_barrier_crossing_2026}. The $s_n$ values are the roots of the cubic equation in the denominator of Eq.~\ref{eqn:3expfreq} and, as a result of its nested exponential form, we know that one of the $A_n$ prefactors must be negative. In the limit that the timescales $1/s_n$ are well separated they are known to take on a simpler form \cite{fujisaka_continued_1987}; applying this limit to the nested fraction of Eq.~\ref{eqn:3expfreq} yields a hierarchy of timescales: a fast timescale of approximately $1/\alpha_0$, an intermediate timescale of $\alpha_0/\gamma_0$ and a slow timescale of $\gamma_0/\omega_0^2 \alpha_0$,
\begin{equation}\label{eqn:3exptime params}
\begin{split}
    \zeta_{\rm IO}(t) \simeq& ~\gamma_1 \mathrm{e}^{-\alpha_1t} \\
    &+ \omega_1^2\left(-\vert A_1\vert\e{- \alpha_0t} + A_2\e{-\frac{\gamma_0}{\alpha_0}t} + A_3\e{-\frac{\omega_0^2\alpha_0}{\gamma_0}t} \right),
\end{split}
\end{equation}
where we have assigned the negative prefactor, $-\vert A_1\vert$, to the shortest timescale in the nested fraction to align the with the previous discussion surrounding the inset of Fig.~\ref{fig:zeta_data}(b).

\begin{figure}[!t]
     \begin{subfigure}[b]{0.235\textwidth}
         \centering
         \includegraphics[width=\columnwidth]{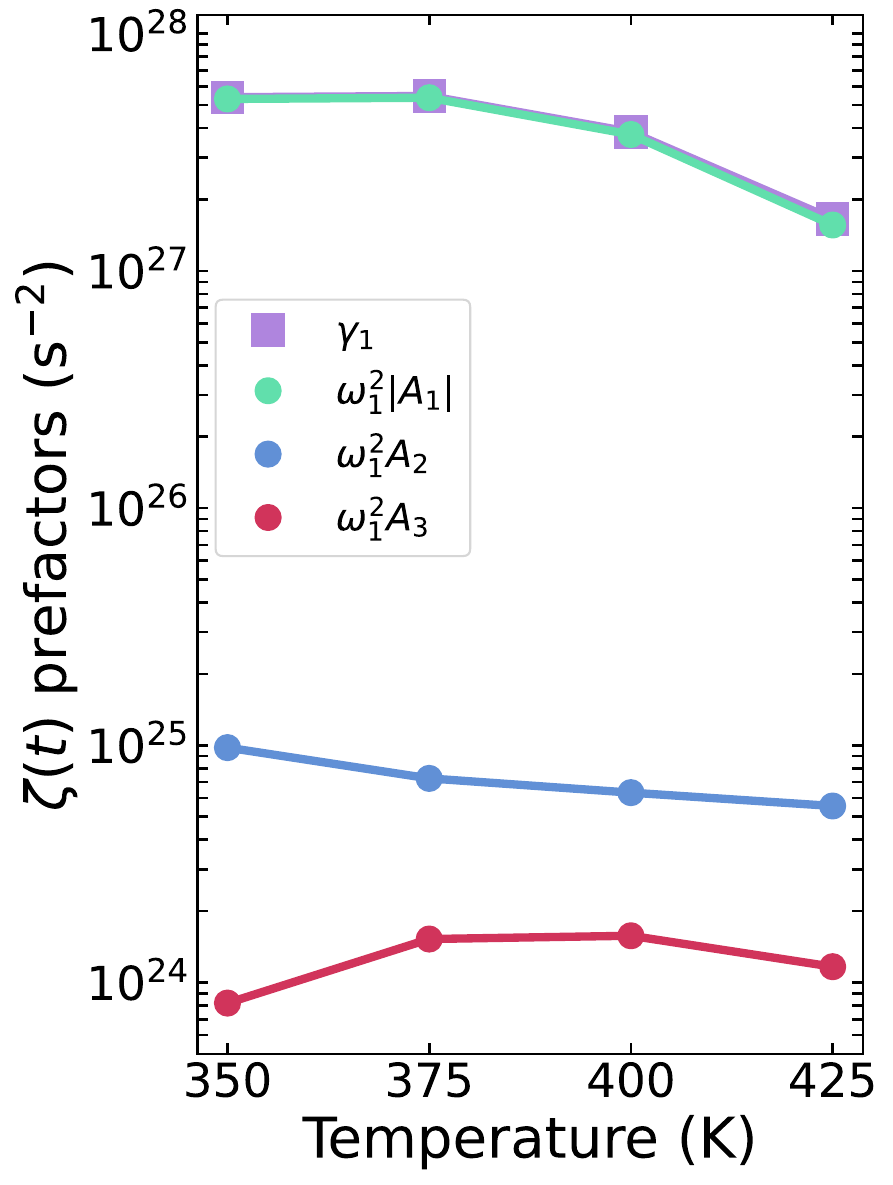}
         \caption{}
     \end{subfigure}
     \hfill
     \begin{subfigure}[b]{0.235\textwidth}
         \centering
         \includegraphics[width=\columnwidth]{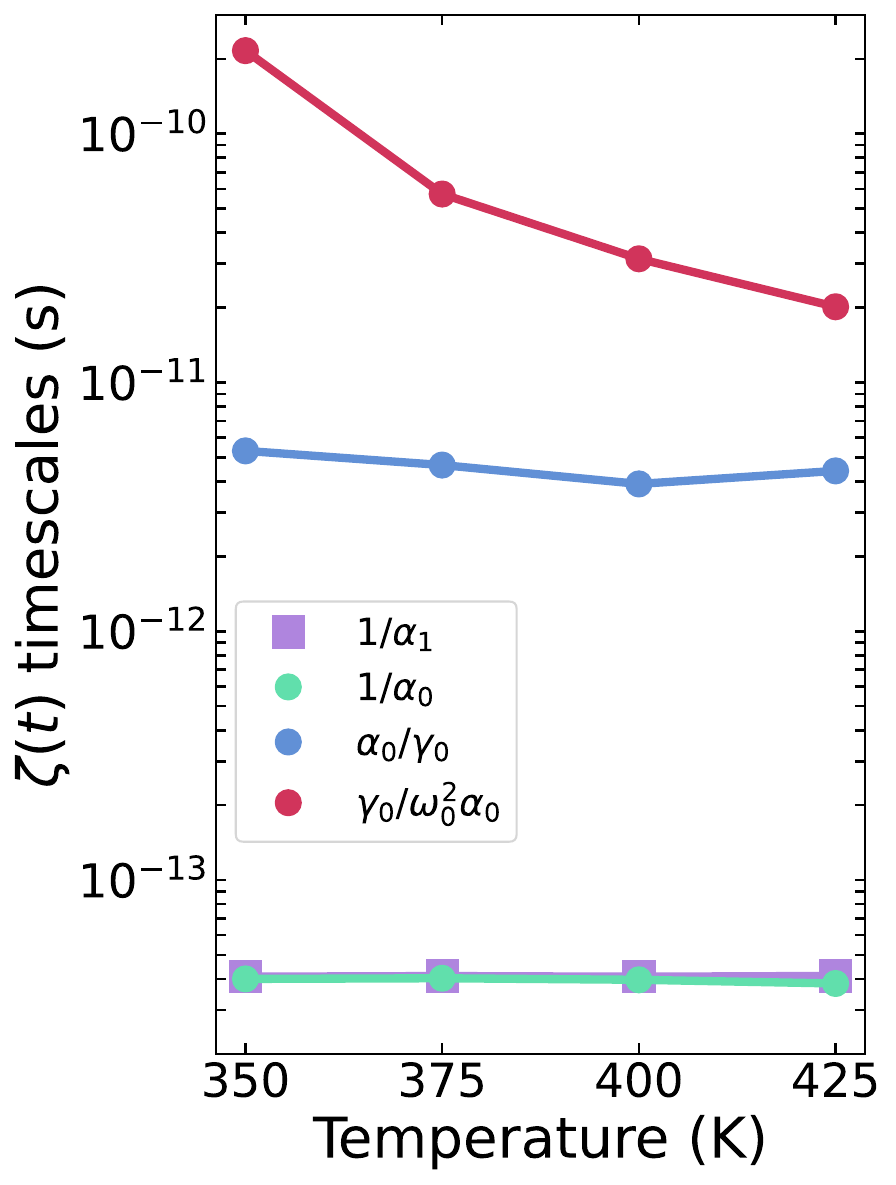}
         \caption{}
     \end{subfigure}
        \caption{\label{fig:parameters}
        \textbf{(a)} and \textbf{(b)} show how the microscopic parameters extract by a four-exponential fit to Eq.~\ref{eqn:3exptime params} vary in order to produce the temperature dependence of the memory and conductivity in Fig.~\ref{fig:zeta_data}. The fit is constrained to reproduce the Nernst-Einstein behaviour at zero frequency. Only the longest timescale, $\gamma_0/\omega_0^2\alpha_0$, is found to be temperature dependent.}
\end{figure}

Figure~\ref{fig:parameters}(a)~and~(b) shows how the constants in Eq.~\ref{eqn:3exptime params} vary with temperature. The fast cage-environment ($1/\alpha_0$) timescale takes on the negative prefactor $A_1$, acting to partially cancel out the dissipative contribution from the ion-environment friction. Indeed in Fig.~\ref{fig:parameters} both the prefactors and timescales for these processes have almost identical magnitudes at all temperatures, meaning there are only three \textit{distinct} physical timescales in the system, in contrast to the single timescale $\tau_{\rm atm}$ used to define the Debye-Falkenhagen memory function (Eq.~\ref{eq:debye_falkenhagen_kernel}). We note that this short timescale has very little temperature dependence, consistent with its interpretation as describing environmental collisions.

The results in Fig.~\ref{fig:parameters}(b) suggest that the temperature dependence of the conductivity spectrum is dominated by the ion-cage coupling terms ($\omega_n$), the impact of which is isolated to the longest decay mode of $\zeta_{\rm IO}(t)$. As the temperature increases from $350$~K to $425$~K, $\omega_0^2$ increases by an order of magnitude. This aligns with our previous analysis of Fig.~\ref{fig:bulkcond}(a) showing that the subdiffusive $\beta$-relaxation timescale in ILs quickly decreases with increasing temperature (as the energetic barrier to reorganization is more easily overcome).

Together, the results in Fig.~\ref{fig:zeta_data}~and~\ref{fig:parameters} show the success of a single-particle itinerant oscillator model in reproducing key spectral features alongside decoding temperature dependencies seen in both the single-particle and the full conductivity spectrum. Therefore we propose itinerant oscillator dynamics as a general model for the Debye-Falkenhagen effect in ionic liquids.

\section{A New Impedance Model}\label{sec:new impedance}

\begin{figure*}[!ht]
     \centering
     \begin{subfigure}[b]{0.45\textwidth}
         \centering
         \includegraphics[width=\textwidth]{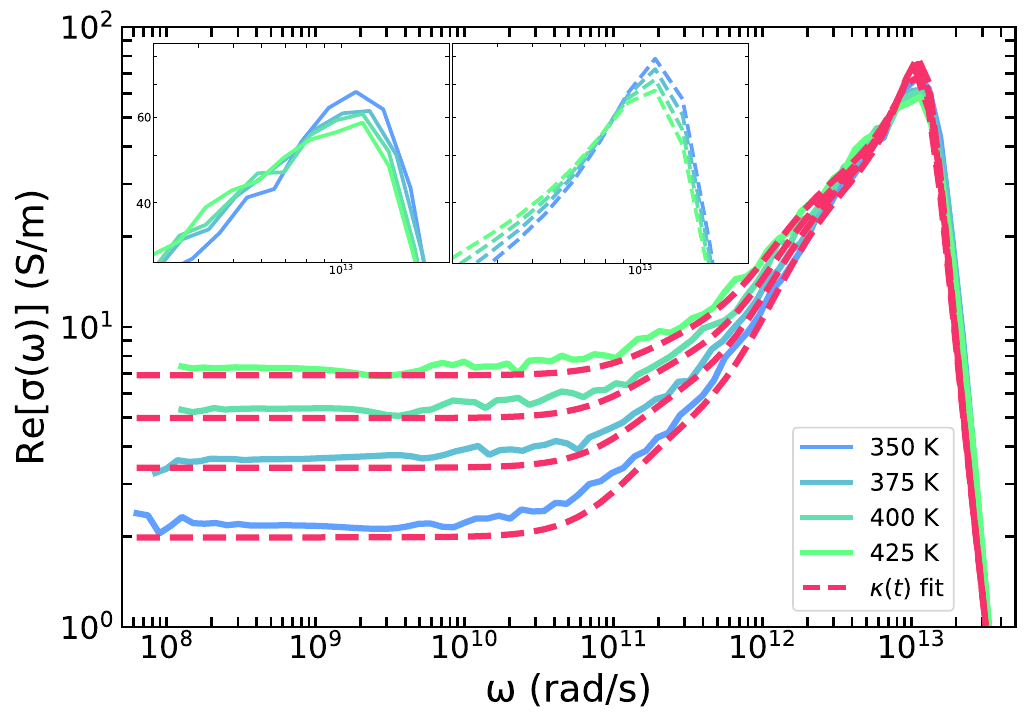}
         \caption{}
     \end{subfigure}
     \hfill
     \begin{subfigure}[b]{0.235\textwidth}
         \centering
         \includegraphics[width=\textwidth]{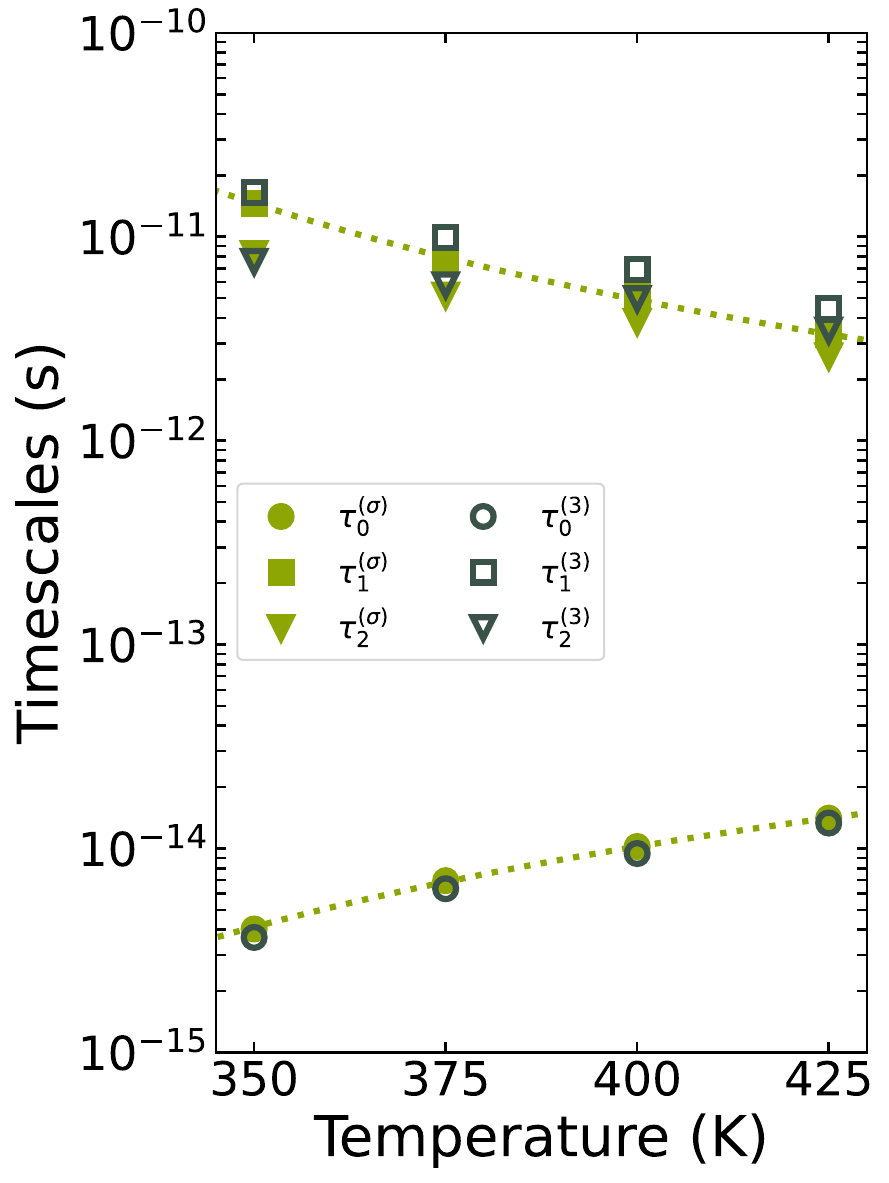}
         \caption{}
     \end{subfigure}
     \hfill
     \begin{subfigure}[b]{0.235\textwidth}
         \centering
         \includegraphics[width=\textwidth]{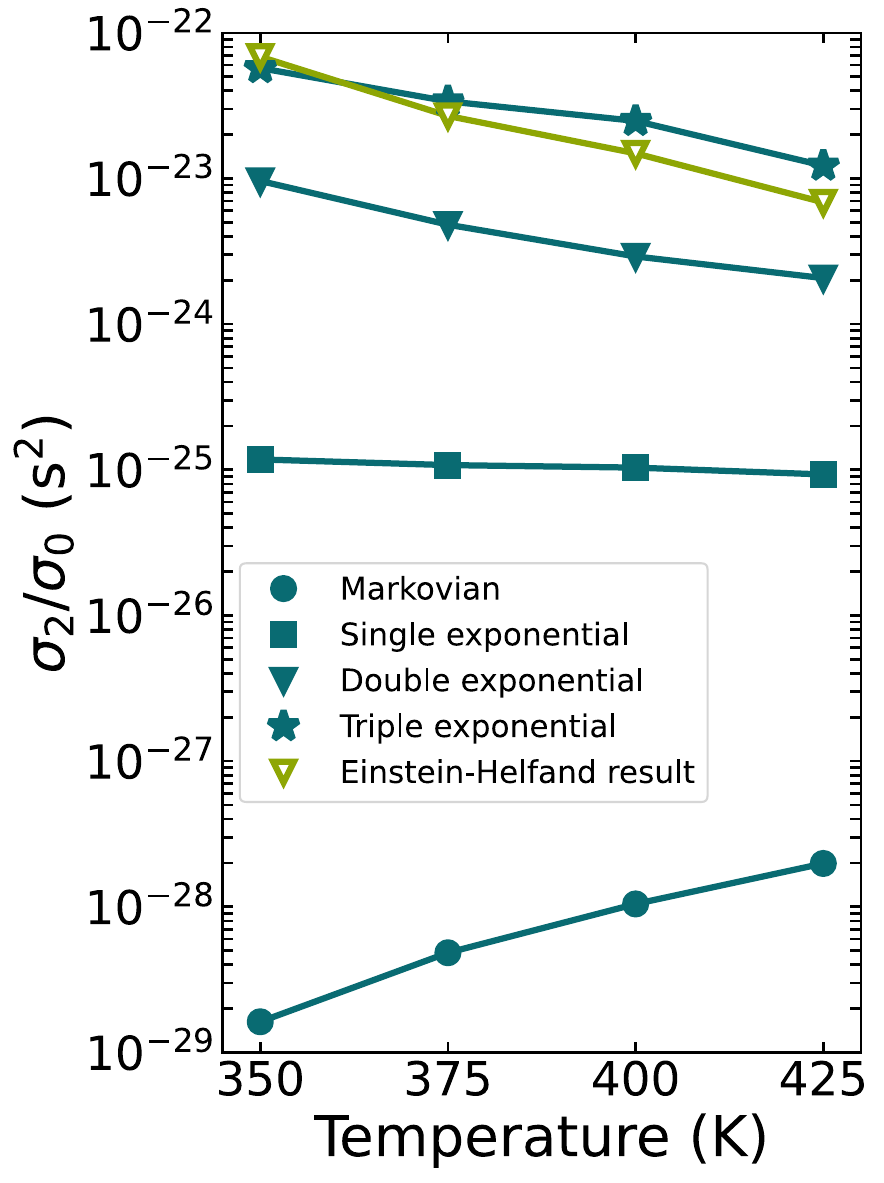}
         \caption{}
     \end{subfigure}
     \hfill
        \caption{\label{fig:timescales} 
        \textbf{(a)} When transformed to the conductivity, Eq.~\ref{eqn:kappaexp} captures the simulation results very well over all frequencies. Insets: Zoomed view of the maximum where left shows the simulation result and right shows the model reproducing the temperature dependence. \textbf{(b)} We have plotted values for $\tau_n$ found from two different methods to validate our three exponential approximation for $\kappa(t)$. $\tau_n^{(\sigma)}$ indicates the values were found using the Einstein-Helfand method to extract $\sigma_n$ and then Eqs.~\ref{eqn: 1exp sigma0},~\ref{eqn: 1exp sigma1}, and~\ref{eqn:sigma2} produce $\tau_n$. $\tau_n^{(3)}$ indicates the values were found from the $\kappa(t)$ fitting parameters using Eq.~\ref{eqn:tau0},~\ref{eqn:tau1}, and~\ref{eqn:tau2}. $\tau_0$ and $\tau_1$ can be well described using the Vogel-Fulcher-Tammann relationship. At 1000~K $\tau_0 \simeq 1.3\times10^{-13}$~s, whilst $\tau_1$ is pathologically found to be negative. Panel \textbf{(c)} highlights the importance of choosing a sufficiently complex memory function to correctly define $\sigma_2/\sigma_0$ for our system. For a single exponential ($\sigma_2/\sigma_0 = 2\tau_0^{(\sigma)}\tau_1^{(\sigma)} - (\tau_0^{(\sigma)})^2$) we cannot even correctly reproduce the temperature dependence found for $\sigma_2/\sigma_0$. Only when using a triple exponential memory function (Eq.~\ref{eqn:kappaexp}) do we include $\tau_2$ ($\sigma_2/\sigma_0 = 2\tau_0^{(\sigma)}\tau_1^{(\sigma)} - (\tau_0^{(\sigma)})^2 + (\tau_2^{(3)})^2$) and hence achieve the best quantitative agreement. The expressions for $\sigma_2/\sigma_0$ for multiple different memory functions can be found in Table~\ref{table:formulae}.}
\end{figure*}

We are now in a position to return to the impedance spectrum measured in experiment. Instead of fitting a functional form directly to the bulk impedance of Eq.~\ref{eqn:Zbulk}, which Eqs.~\ref{eqn:Re-Z expan}~and~\ref{eqn:Im-Z expan} showed are a complicated mixture of microscopically measurable quantities, we will work directly with $\sigma(\omega)$. Ultimately, this will provide a molecular level understanding of how the ``quadratic form'' arises in the denominator of $Z_{\rm bulk}(\omega)$ given by Eq.~\ref{eqn:Zbulk 2order}. In so doing, we will explain why it is to be preferred over the Cole--Davidson form.

\subsection{Understanding the Moments of the Conductivity}\label{sec:timescales}

Our multi-exponential analysis of the single-particle memory function revealed three \textit{distinct} timescales (Fig.~\ref{fig:parameters}(b)). Together with the expression linking $\kappa(t)$ to $\zeta(t)$ (Eq.~\ref{eq:dJJ_two_terms}), we propose a minimal form for the total memory function,
\begin{equation}\label{eqn:kappaexp}
    \kappa(t)= A\mathrm{e}^{-at} + B\mathrm{e}^{-bt} + C\mathrm{e}^{-ct}.
\end{equation}
See Appendix~\ref{app:minimalmemory} for further discussion comparing to a biexponential form. Due to the difficulty of relating the full conductivity to the single-particle memory $\zeta(t)$ (Eq.~\ref{eq:dJJ_two_terms}), we will not try to unravel the specific microscopic contributions to these 6~parameters. However, all of the information contained within the single-particle parameters of Eq.~\ref{eqn:3exptime} must also be present here. For example, the similarity in the forms of the total memory and single-particle memory means this fit to $\kappa(t)$ also produces a single short timescale and multiple longer decays, mirroring the results of Fig.~\ref{fig:parameters}(b). A fit to Eq.~\ref{eqn:kappaexp} produces excellent agreement with the directly computed full conductivity spectra shown in Fig.~\ref{fig:timescales}(a), including the temperature dependence of $\operatorname{Re}[\sigma_{\rm max}(\omega)]$ (see insets). Therefore, we argue that the qualitative physics of the full conductivity is the same as that investigated in the single-particle picture.

Now we have a suitable form for $\kappa(t)$, we can begin to understand what information is captured within the parameters of the second-order impedance spectrum Eq.~\ref{eqn:Zbulk 2order}. Specifically, we wish to understand how these parameters encode the additional physics at play beyond that of the RC-circuit model, Eq.~\ref{eqn:ZRC}.

We obtain an expression for the conductivity in terms of the memory function timescales in Eq.~\ref{eqn:kappaexp} through their frequency space relation (Eq.~\ref{eq:exact_friction_conductivity}). Equations~\ref{eqn:sigma-Re} and \ref{eqn:sigma-Im} then show how we can find the moments from this expression for the conductivity. We can therefore write the moments of the conductivity in terms of the timescales of the memory function. As the expressions that result from this procedure can be somewhat complicated, to help orient the reader, we first focus on a simple example where we model the total memory with a single exponential, $\kappa(t)=A\mrm{e}^{-at}$. We find 
\begin{equation}\label{eqn: 1exp sigma0}
        \sigma_0=\frac{\beta}{3\Omega}\langle\bm{J}^2\rangle\frac{a}{A} = \rho q^2\left(\frac{1}{m_+} + \frac{1}{m_-}\right)\tau_0
\end{equation}
\begin{equation}\label{eqn: 1exp sigma1}
        \frac{\sigma_1}{\sigma_0} = \tau_1 - \tau_0
\end{equation}
\begin{equation}\label{eqn: 1exp sigma2}
        \frac{\sigma_2}{\sigma_0} = 2\tau_0\tau_1 - \tau_0^2
\end{equation}
where we have introduced new parameters $\tau_0$ and $\tau_1$, which have units of time. In this simple example, $\tau_1 = 1/a$ is the decay timescale of $\kappa(t)$. $\tau_0$, the timescale associated with diffusion in the system, equals $a/A$, the inverse of $\tau_1$ multiplied by the friction strength, $A$. All higher moments can be written in terms of $\tau_0$ and $\tau_1$. These expressions show how changing the parameters in the memory function directly impact $\sigma_n$.

Our results indicate that \ce{[BMIM]+[BF4]^-} has a minimal memory function of a triple exponential decay of the form in Eq.~\ref{eqn:kappaexp}. For such a triexponential memory, the first two moments of the conductivity take the same form, Eq.~\ref{eqn: 1exp sigma0}~and~Eq.~\ref{eqn: 1exp sigma1} respectively. However, the second moment is modified to
    \begin{equation}\label{eqn:sigma2}
        \frac{\sigma_2}{\sigma_0}=2\tau_0\tau_1 - \tau_0^2 + \tau_2^2.
    \end{equation} 
Crucially, by introducing multiple timescales in our expression for $\kappa(t)$ we have introduced an additional timescale $\tau_2$ that appears only in $\sigma_2$. For this triexponential memory function the definition of the timescales $\tau_n$ are more involved, but we can still write down their exact form in terms of the $\kappa(t)$ parameters of  Eq.~\ref{eqn:kappaexp} (see Appendix~\ref{app:itinerant_cond}),
\begin{equation}\label{eqn:tau0}
        \tau_0 = \frac{1}{\left(\frac{A}{a} + \frac{B}{b} + \frac{C}{c}\right)},
\end{equation}
\begin{equation}\label{eqn:tau1}
        \tau_1 = \frac{\left(\frac{A}{a^2} + \frac{B}{b^2} + \frac{C}{c^2}\right)}{\left(\frac{A}{a} + \frac{B}{b} + \frac{C}{c}\right)},
\end{equation}
\begin{equation}\label{eqn:tau2}
        (\tau_2)^2 = \frac{\frac{AB}{ab}\left(\frac{1}{a} - \frac{1}{b}\right)^2 + \frac{AC}{ac}\left(\frac{1}{a} - \frac{1}{c}\right)^2 + \frac{BC}{bc}\left(\frac{1}{b} - \frac{1}{c}\right)^2}{\left(\frac{A}{a} + \frac{B}{b} + \frac{C}{c}\right)^2}.
\end{equation}
Our expressions show that for a memory function represented as a sum of $n$ exponentials, $1/\tau_0$ is a measure of the average decay rate of the memory function as a whole, $\tau_1$ is a weighted average decay timescale, and $\tau_2$ is a weighted variance in decay times. We can therefore conclude the importance of including $\sigma_2$ in a model for the impedance stems from the additional information provided by $\tau_2$. That is, it is only $\sigma_2$ that contains any information regarding the separation of different timescales contained within the memory function: $\sigma_0$ and $\sigma_1$ are just composed of averages.

In Fig.~\ref{fig:timescales}(b) we have plotted $\tau_n$ as a function of temperature. We show that both $\tau_1$ and $\tau_2$ decrease with increasing temperature. This result directly mirrors our finding for the single-particle model in Fig.~\ref{fig:parameters}(b) where we show that the longest two timescales converge at higher temperatures as a result of the temperature dependent $\beta$-relaxation process. Increasing temperature decreases both the average timescales ($\tau_1$) and the separation of timescales ($\tau_2$) in the memory function, again highlighting that the IO model contains sufficient physics to describe the entire memory function.

Figure~\ref{fig:timescales}(c) shows that this temperature dependent information is completely lost when we do not include the contribution from $\tau_2$ in $\sigma_2$. For this system, the $(\tau_2)^2$ term is the largest contribution to $\sigma_2$, as $\tau_2\approx\tau_1\gg\tau_0$ by three orders of magnitude. This regime only occurs due to the significant timescale separation seen in the memory function. Therefore we can now understand why including $\sigma_2$ in our impedance model produces such a stark improvement over the RC-circuit model. It is because no information regarding the separation of timescales in the memory function is accessible using $\sigma_0$ and $\sigma_1$ alone, which are the only moments that enter into the RC-circuit model.

\vspace{-6pt}
\subsection{Second-Order Impedance}\label{sec:impedance}

We now understand why an RC-circuit model is not able to describe the bulk impedance of our IL. This point can be developed to critique the use of generalized RC forms like Eq.~\ref{eqn:ZCC}~and~\ref{eqn:ZCD}. For example, consider the asymmetry seen in the Nyquist plot, Fig.~\ref{fig:Nyquist}. Asymmetry is characteristic of a highly correlated system and cannot be modelled by a simple RC form. Phenomenologically, moving to our approximate form for $Z_\mathrm{bulk}$ (Eq.~\ref{eqn:Zbulk 2order}) allows this asymmetry to be captured.  Given the link we established between $\sigma_2$ and $\kappa(t)$, we can therefore conclude that the asymmetry in our model is a result of well-separated timescales in Eq.~\ref{eqn:kappaexp}. Indeed if we artificially set $\tau_2=0$ (using Eq.~\ref{eqn: 1exp sigma2} for $\sigma_2$), the Nyquist plot in Fig.~\ref{fig:Nyquist}(b) no longer appears asymmetric. Contrastingly, in the literature this asymmetry for a glassy liquid is modelled using the Cole-Davidson equation, Eq.~\ref{eqn:ZCD}, which is derived starting from a distribution of many uncorrelated overlapping processes \cite{kudlik_dielectric_1999, stoppa_interactions_2008}. That is, the asymmetric profile is expressed as a sum of several symmetric, RC contributions using a model (Eq.~\ref{eqn:ZCD}) that only defines a single timescale. 

The key point is this: an RC model produces a poor fit to the data (Fig.~\ref{fig:Nyquist}(a)), and the single timescale extracted from this fit, or any of it's generalized forms (Eq.~\ref{eqn:ZCC}~and~\ref{eqn:ZCD}), would be unphysical. The microscopic implications of these models are clearly not fulfilled and therefore the value of $\tau_{\rm RC}$ found would hold no meaning. In contrast, we have shown fitting to the second-order expansion Eq.~\ref{eqn:Zbulk 2order} will produce a much better agreement and extracts moments which directly report on the microscopic timescales. That is, the agreement in Fig.~\ref{fig:timescales}(b) represents that values of $\sigma_0$, $\sigma_1$, and $\sigma_2$ which \textit{match} those computed with knowledge of the full, underlying microscopic dynamics (i.e. $\kappa(t)$ from MD simulations).

Put another way, the impedance arc of our simulated data, Fig.~\ref{fig:Nyquist}, corresponds to a distribution of correlation times $G(\tau)$ \cite{beckmann_spectral_1988},
\begin{equation}\label{eqn:DRT}
    Z(\omega) = \int_0^\infty \dd{\tau} \frac{G(\tau)}{1+\imi\omega\tau},
\end{equation}
that has multiple peaks that are well separated in time. Likewise, our second order impedance model also shows a $G(\tau)$ with several well-separated peaks (see Fig.~\ref{fig:DRT}(b)). In contrast, it has been shown that the Cole-Cole and Cole--Davidson expressions produce single peaks on the single timescale in a distribution of relaxation times (DRT) spectrum \cite{lasia_origin_2022, beckmann_spectral_1988}. Use of one of these expressions -- which may give a similar quality, or even superior, fit to the data -- would incorrectly lead one to argue for some kind of distribution of homogeneous environments which does not exist in reality. For a further discussion of the DRT results, see Appendix~\ref{app:DRT}.

Carrying forward this logic, a further expansion of $\sigma(\omega)$ to third or higher order will naturally increase the flexibility of Eq.~\ref{eqn:Zbulk 2order} and improve the possible fit of the impedance spectrum. However an interpretation for higher orders of $\sigma_n$ becomes analytically more challenging and computationally harder to access directly. As such, we stop at second-order to mirror the simplicity of the existing models (Eqs.~\ref{eqn:ZCC} and \ref{eqn:ZCD}) while highlighting that with this expression we can directly report on microscopic timescales. We note that when expanded up to second-order we have shown Eq.~\ref{eqn:Zbulk 2order} can reproduce the asymmetry in the complex plane normally modelled using the Cole-Davidson equation, however when expanded up to third order, this equation is also able to produce the symmetric suppression of the maximum in the imaginary plane normally modelled using the Cole-Cole equation \cite{lasia_origin_2022}.

\vspace{-8pt}
\section{Conclusion and Outlook}
\vspace{-4pt}
Decoding the information presented in an impedance spectrum is a significant challenge due to the complex signals produced by several simultaneous, correlated processes occurring throughout an electrochemical cell. In this work we focussed on understanding the contribution arising from the bulk electrolyte, producing a suitable-yet-simple model to begin improving our understanding of how microscopic correlations are borne out in these impedance signals. Importantly, this work provides a more physically meaningful analysis of the impedance than existing empirical equations.

When applying equivalent-circuit analyses, this work revealed and explained the importance of including a quadratic frequency term when modelling impedance: an $\omega^2$ term in the denominator is a prerequisite for including the contribution due to distinct relaxation timescales. This is particularly crucial for highly correlated electrolytes such as ILs, where our results recapitulate the well-known significant timescale separation, especially at lower temperatures. Including this quadratic term in the expression for the impedance allows us to model asymmetry in the Nyquist plot that was previously only considered using empirical equations. We have therefore provided a new, more physically intuitive, impedance model.

To justify our impedance model, we employed theory and simulation to establish the itinerant oscillator model as a quantitative tool for describing the microscopic mechanisms that give rise to the conductivity spectra of a prototypical ionic liquid. In particular, our investigation under the single-particle approximation, $\kappa(t)\approx\zeta(t)$, resulted in a deeper understanding of the temperature dependence of these conductivity spectra, attributing it to the changing timescale of the slowest process. This reproduced our qualitative understanding of the $\beta$-relaxation mechanism in a quantitative manner. Moreover, the IO model has many parallels to the existing dilute electrolyte models \cite{debye_dispersion_1928, lesikar_debye-falkenhagen_1980, anderson_debye-falkenhagen_1994, wei_dielectric_1991,chandra_frequency_1993,chandra_frequency_2000,chandra_beyond_2000,dufreche_ionic_2002, ibuki_effect_1990}, both depicting a central ion coupled to its immediate counter-ion surroundings, allowing it to remain consistent with the existing narrative for the origin of the Debye-Falkenhagen effect while simultaneously expanding the range of applicability to highly concentrated electrolytes.

The fundamental role of timescale separation also arises when studying electrolytes with GLE-type expressions derived using stochastic density functional theories (SDFTs) starting from the Dean-Kawasaki equation~\cite{illien_deankawasaki_2025}. Building on previous work introducing SDFT for ions in a polar solvent~\cite{illien_stochastic_2024} and comparing SDFT with Brownian Dynamics (BD) for the pure solvent~\cite{varghese_dynamic_2025}, a recent study of Varghese \textit{et al.} investigated a model electrolyte undergoing BD. The authors found an SDFT-based GLE well-described simulation data for the dynamic structure factors over a frequency range spanning six orders of magnitude \cite{varghese_solvent-induced_2026}. They observed the emergence of a two-step relaxation in the ion-ion dynamic structure factor brought on simply by varying the ratio of ionic and solvent timescales. This result further illustrates that beyond-RC-circuit impedance response is a general phenomenon and not restricted to our specific system: room temperature and pressure, the choice of ionic liquid, or even the extreme concentration regime.

With the ability to capture the bulk impedance, the next step is to separate out the impact of interfaces on the impedance of an electrochemical cell. While in previous work looking at the impedance of a salt-water system it was found that the electrolyte was bulk-like throughout \cite{pireddu_impedance_2024} -- only an additional capacitive term arose that was localised to the interfaces -- for an IL it is unlikely for this to be the case. Ionic liquids show unusual behaviour under confinement; their structure and dynamics can be drastically altered from the bulk behaviour \cite{fedorov_ionic_2014, borghi_ionic_2020, kritikos_molecular_2016}. Both simulation and experiment have shown that the interfacial region formed between an ionic liquid and a charged interface is not a single layer of co- and counter ions, as often seen in dilute electrolyte systems, but several layers that extend much further into the bulk \cite{hayes_at_2010, merlet_electric_2014}. Furthermore, confined IL simulations have shown that this adsorbed layer shows highly restricted dynamics, where the two-dimensional diffusion coefficient is significantly reduced near the surface, implying a large interfacial resistance \cite{ntim_role_2020, lhermerout_nanoconfined_2018, dufils_computational_2021}. 

Very recent advances in the experimental measurement of electrostatic screening in confined ILs have revealed that long-range screening is associated with extremely slow relaxation to equilibrium \cite{cross_short-range_2026}. This only raises more questions regarding the microscopic origin of this mysterious behaviour that spans several time decades. The work presented here, together with a simulation study of this IL under confinement, will help to quantify to what extent, and in what manner, interfaces disturb the dynamics of the bulk.

\textit{\textbf{Data availability.}}
Simulation data are available from the authors upon reasonable request.

\textit{\textbf{Acknowledgements.}} 
Via our membership of the UK's HEC Materials Chemistry Consortium, which is funded by EPSRC (EP/L000202), this work used the UK Materials and Molecular Modelling Hub for computational resources, which is partially funded by EPSRC (EP/T022213/1, EP/W032260/1 and EP/P020194/1).  S.J.C. is a Royal
Society University Research Fellow (Grant No. URF\textbackslash
R1\textbackslash 211144) at Durham University. This project received funding from the European Research Council under the European Union’s Horizon 2020 research and innovation program (Grant Agreement No. 863473). T.S.~is the recipient of an Early Career Fellowship from the Leverhulme Trust. 
\vspace{-20pt}

\section{Methods}\label{sec:methods}
\subsection{Simulation Details}\label{app:simdetails}
\ce{[BMIM]+[BF4]^-} was investigated using a coarse-grained force field parameterised by Merlet \textit{et al.} \cite{merlet_new_2012}. Bulk simulations were run with 432 ion pairs. The Lorentz-Berthelot mixing rules were used and the \ce{[BMIM]+} molecule geometry was fixed using the RATTLE algorithm \cite{andersen_rattle_1983}. All simulations were run using LAMMPS \cite{thompson_lammps_2022}.

Lennard-Jones interactions were implemented with cut and shift at 15~\AA. Coulombic interactions were cut off at 15~\AA, and long-range interactions were computed using a 3D particle-particle particle-mesh solver. All simulations were run with a timestep of 1~fs. A Nos\'{e}-Hoover thermostat fixed the temperature, with a chain length of 3 and time constant of 10~ps, and a barostat with a time constant of 20~ps was used to fix the pressure. 

The system was initially equilibrated under NPT for 20~ns at ambient pressure for five temperatures, 350~K, 375~K, 400~K, 425~K and 1000~K. The cell-length parameters we found to equilbrate to 51.89~\AA, 52.17~\AA, 52.46~\AA, 52.75~\AA, and 63.08~\AA  respectively. Then, five independent NVT replicas were run at each temperature for at least 100~ns for data collection.

The single-particle data presented in Fig.~\ref{fig:zeta_data} of the Main Text was found by directly extracting out the single-particle velocity autocorrelation functions for the anion and central bead of the cation for a 5~ns trajectory and averaging over at least 80 replicas. These data were then transformed to the single-particle memory function as outlined in the next section.

All conductivity spectra from simulation data were calculated via either Eq.~\ref{eqn:sigma-FDT-single} or Eq.~\ref{eqn:sigma-FDT} using the method outlined in Ref.~\cite{pireddu_impedance_2024}. All conductivity spectra calculated using a model for the memory function, $\zeta(t)$ or $\kappa(t)$, were calculated using Eq.~\ref{eqn:sigma_from_zeta} and Eq.~\ref{eq:exact_friction_conductivity} respectively.

\subsection{Transfer Tensor Method}\label{app:TTM}
As detailed in Boon and Yip \cite{boon_molecular_1991}, extraction of the friction kernel from reference (simulation or experimental) data is traditionally performed by taking second derivatives of the time correlation function and parameterizing the auxiliary kernels. The friction kernel is then found by an iterative convolution from these auxiliaries. However obtaining accurate results when taking second derivatives of noisy data is not feasible, which limits the quantitative interpretation of the friction kernel.

What is more, in these data the current-current correlation function decays from its maximal value at $t=0$ in around 2~ps to graphical accuracy. However, the memory function is (depending on temperature) still at around 5\% of its initial value at this time, and decays on a much longer timescale. Whether the current-current correlation is truly equilibrated (\textit{vis-a-vis} the long-time tail) or is taking some small but finite value (as in recent calculations performed on the dispersive Holstein model) we cannot directly observe due to the noise background. In either case, we can say that the auxiliary kernel approach to obtaining the friction kernel at these times is likely to fail to produce an accurate output.

Recently, discrete approaches to the Volterra problem were popularized in the chemical physics community by workers in the field of quantum dynamics. The transfer tensor method (TTM) obtains an equivalent description of the dynamics in discrete time \cite{cerrillo_non-markovian_2014},
\begin{equation}\label{eq:TTM}
    C(N\Delta t) = \sum_{m=1}^N T_{N-m} C(m\Delta t),
\end{equation}
where $\Delta t$ can in principle be much larger than the fundamental time step $\delta t$, here 1 fs. From a reference $C(t)$ the TTM at resolution $\Delta t$ is obtained iteratively provided $C^{-1}(0)$ exists. Therefore, no derivatives are required in order to parameterize this memory object. To make a link to the friction kernel we can taking a finite difference,
\begin{equation}\label{eq:limit}
    \frac{C_{N} - C_{N-1}}{\Delta t} = \frac{T_1 - 1}{\Delta t}C_{N-1} + \sum_{m=2} \delta t \frac{T_{N-m}}{\Delta t^2} C(m\Delta t),
\end{equation}
identifying $T_\mathrm{i>1}/\Delta t^2$ as $\zeta(t)$ in the limit $\Delta t \rightarrow 0$. For the current-current correlation the first term on the right-hand side (the ``free-streaming term'') is formally zero in the limit of infinitesimal resolution, but can be non-zero at finite resolution. Although the transfer matrix and the friction kernel are quantitatively different outside of the infinitesimal limit, both Eq.~\ref{eqn:langevin} and Eq.~\ref{eq:TTM} return the same numerical value at their respective levels of resolution.

\begin{figure*}[!ht]
     \centering
     \begin{subfigure}[b]{0.49\textwidth}
         \centering
         \includegraphics[width=\textwidth]{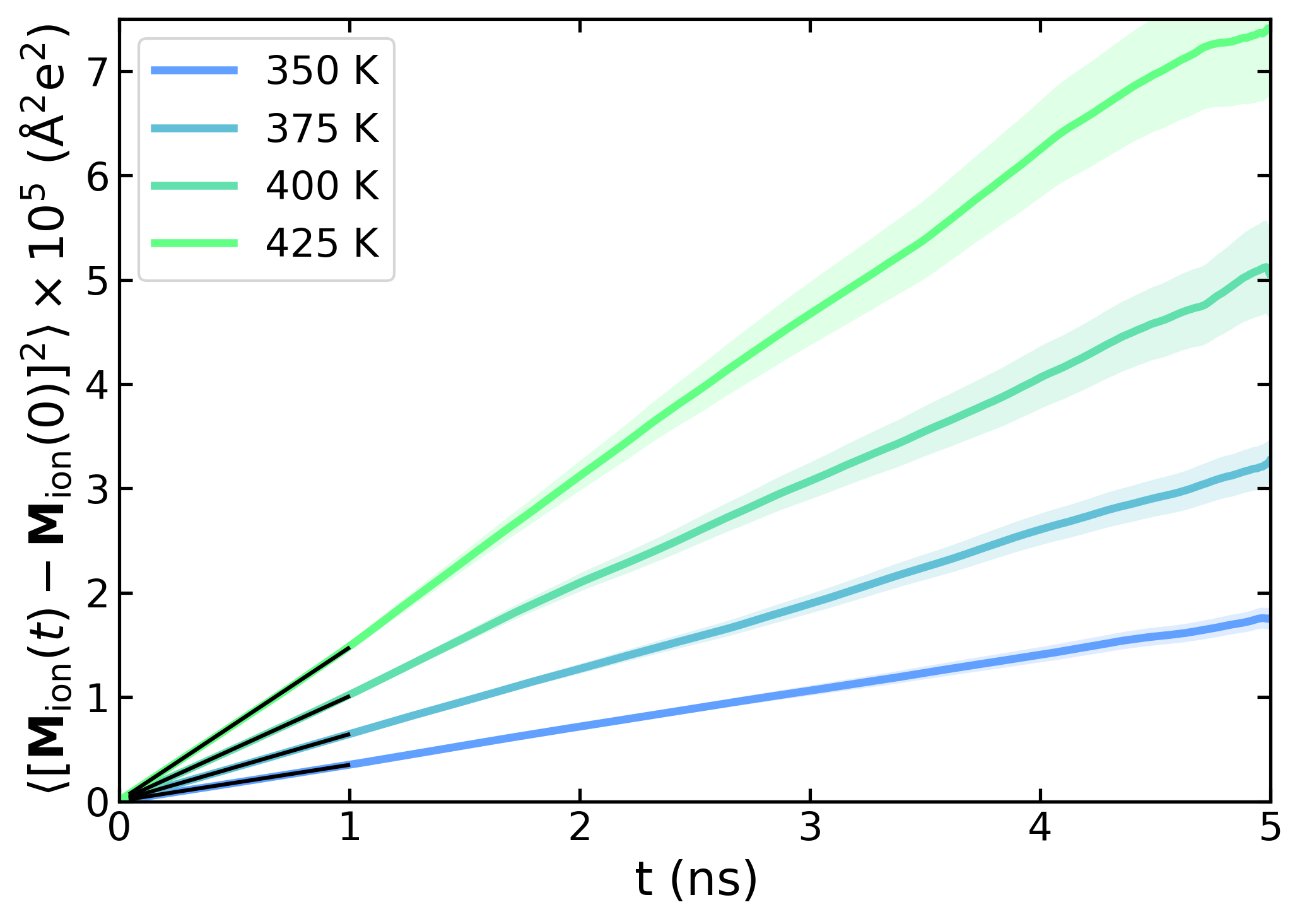}
         \caption{}
     \end{subfigure}
     \hfill
     \begin{subfigure}[b]{0.49\textwidth}
         \centering
         \includegraphics[width=\textwidth]{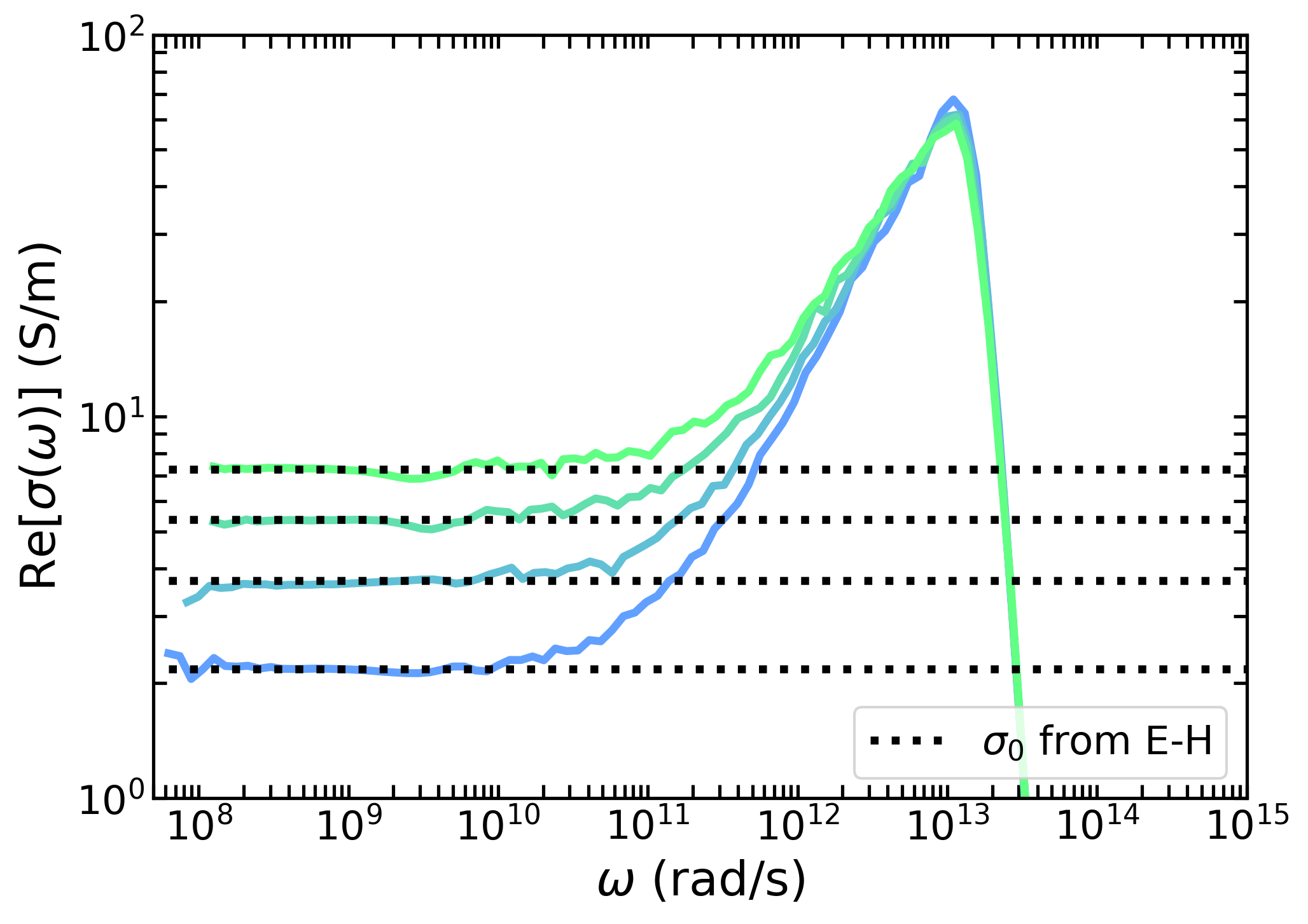}
         \caption{}
     \end{subfigure}
     \hfill
     \begin{subfigure}[b]{0.49\textwidth}
         \centering
         \includegraphics[width=\textwidth]{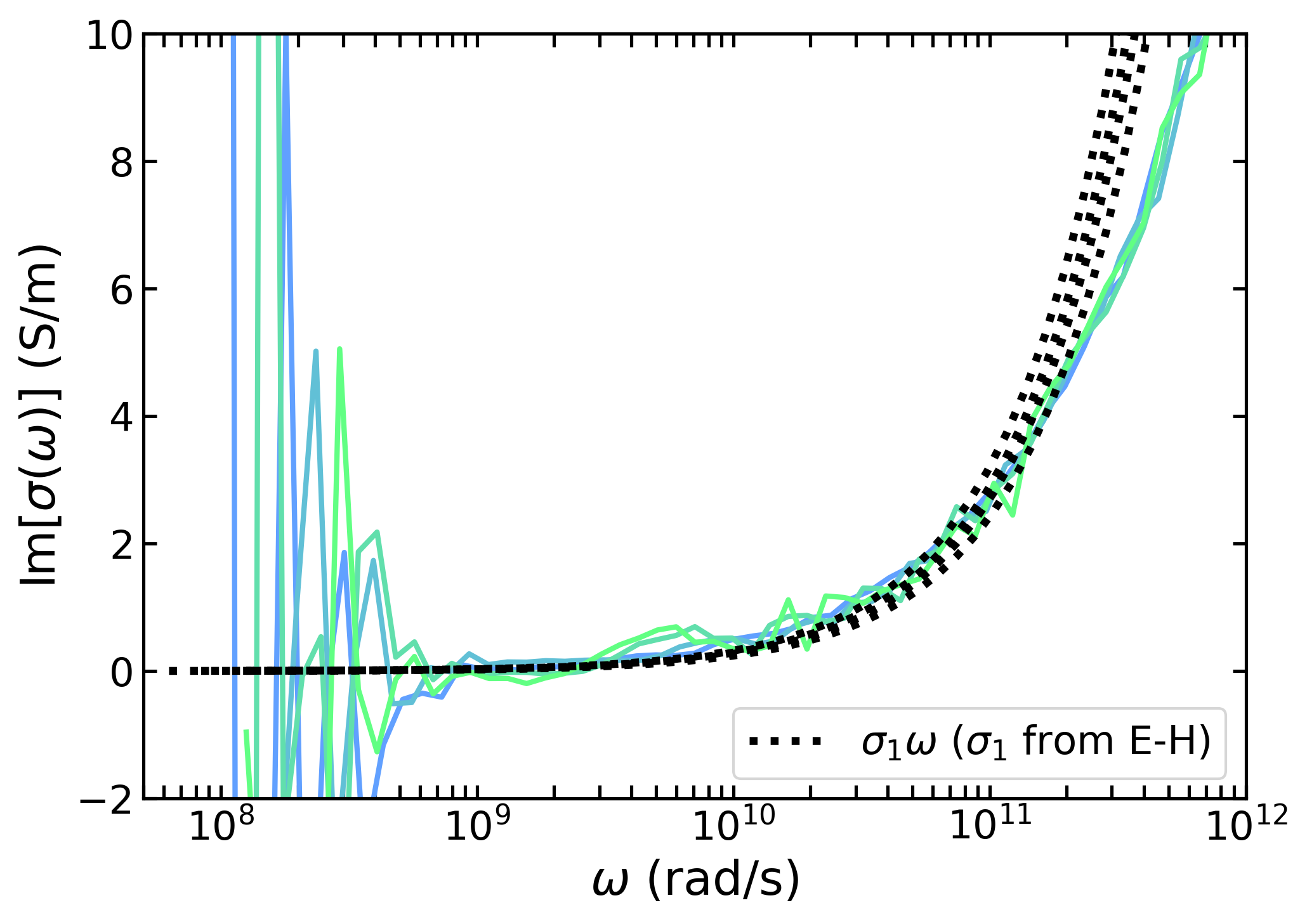}
         \caption{}
     \end{subfigure}
     \hfill
     \begin{subfigure}[b]{0.49\textwidth}
         \centering
         \includegraphics[width=\textwidth]{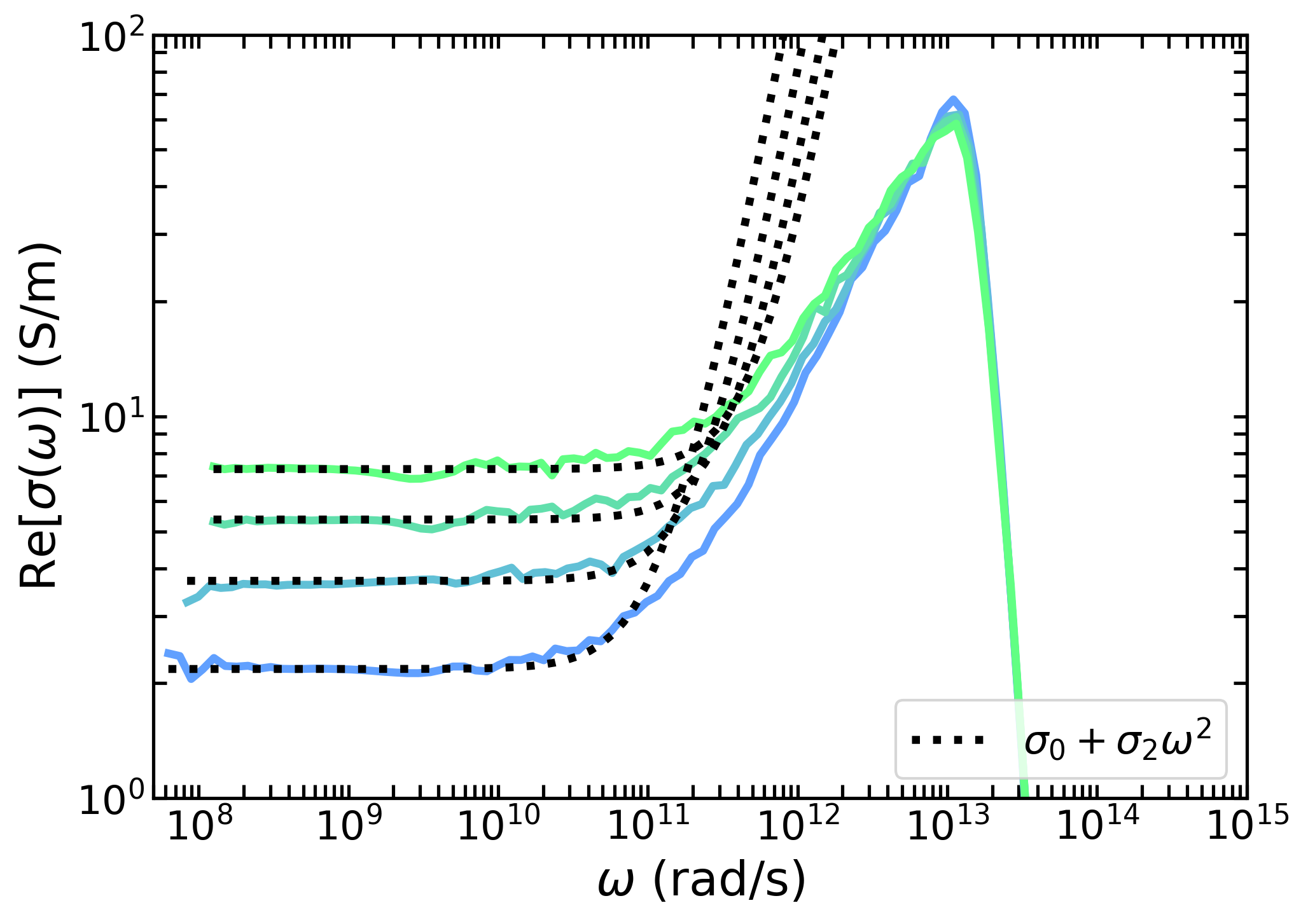}
         \caption{}
     \end{subfigure}
        \caption{\label{fig:EH}\tbf{(a)} Motivated by the relationship in Eq.~\ref{eqn:final EH}, a linear fit of these timeseries within the region highlighted in black produces the $\sigma_0$ and $\sigma_1$ values in Table~\ref{table:sigma}. \tbf{(b)} Clearly the agreement between $\sigma_0$ found from long-time simulation and the Einstein-Helfand relation shows the low-frequency limit of $\sigma(\omega)$ has been successfully reached. \tbf{(c)} In order to remove the noise at low-frequency in $\operatorname{Im}[\sigma(\omega)]$ we replace the data with a low-frequency expansion. \tbf{(d)} Finally $\sigma_2$ is found using a fit of the low-frequency approximation to $\operatorname{Re}[\sigma(\omega)]$.}
\end{figure*}

To make the direct connection we can take $\Delta t = \delta t$, amounting to a approximating to the continuous convolution integral that appears in the GLE with a particular treatment of the endpoints. Numerical artifacts that give the free-streaming term $\neq 0$ are subsumed into the value of $T_1$. Since the timestep is so short, we ignore the fact that, formally, $\zeta(t<2\delta t)$ are not described in the TTM framework. We find that when taking into accout a non-zero free-streaming term, both auxilliary kernel and TTM methods give equivalent results at early times before developing a small, visible offset after around 100~fs that is larger for lower temperatures (and seems to become smaller as time goes on). This is the expected difference due to the lack of $\delta t \rightarrow 0$.

In contrast, if the free streaming term is set to zero when constructing the auxilliary kernels, as suggested by Eq.~\ref{eqn:langevin}, the result is wildly wrong. We see a large amplitude, lower frequency oscillation about zero, i.e. the friction kernel is negative for long periods. In light of the known difficulties discussed above, we must prefer the TTM-derived friction kernel. 

\subsection{Finding \texorpdfstring{$\sigma_n$}{sigma n} Using the Einstein-Helfand Relation}\label{app:einstein}
Ionic liquids by their nature are highly correlated, and therefore a large amount of statistics is required to be certain we have converged in the low frequency limit. This is a computationally taxing task, and therefore a second method has been suggested to find $\sigma_0$ ($= \sigma(\omega\rightarrow0)$) \cite{schroder_computation_2008, picalek_molecular_2007}. The Einstein-Helfand relation utilises the mean-squared total ionic dipole displacement, and obtains static parameters of a system without requiring long and costly calculations. The mean-squared total ionic dipole displacement is defined as, 
\begin{equation}
\langle |\Delta \mbf{M}(t)|^{2}\rangle = \langle |\mbf{M}(t) - \mbf{M}(0)|^2\rangle,
\end{equation}
where
\begin{equation}
\mbf{M}(t) = \sum_{i}^{N}{q_{i} \mbf{r}_{i}(t)}.
\end{equation}
This sums over all $N$ atoms with charge $q_{\mathrm{i}}$ and position $\bm{r}_{\mathrm{i}}$, following the ions outside of the simulation box \cite{cox_finite_2019, caillol_comments_1994}. This expression can be rewritten as,
\begin{equation}
\langle|\Delta \mbf{M}(t)|^{2}\rangle = \int_{0}^{t}\dd{t_2}\int_{0}^{t}{\dd{t_1} \langle \mbf{J}(t_1)\cdot\mbf{J}(t_2)\rangle },
\end{equation}
and then further expressed using properties such as origin independence and time reversal symmetry as:
\begin{equation}
\langle|\Delta \mbf{M}(t)|^{2}\rangle = 2 \int^{t}_{0}{\dd{s} (t-s) \langle \mbf{J}(0)\cdot\mbf{J}(s)\rangle}.
\end{equation}
Provided we consider time frames longer than the correlation time of our system,
\begin{equation}\label{eqn:einstein}
\langle|\Delta \mbf{M}(t)|^{2}\rangle = 2 \left(\frac{3V\sigma_0}{\beta}t - \int^{t}_{0}\dd{s}{\langle \mbf{J}(0)\cdot\mbf{J}(s)\rangle s} \right).
\end{equation}
Therefore the static conductivity can be found from the gradient of a straight line fit of the mean-squared total ionic dipole displacement versus time, providing us with a second route to obtain this constant. 

The second term in Eq.~\ref{eqn:einstein} is usually ignored as it does not contain any obvious insight into the system's behaviour, however our definition in Eq.~\ref{eqn:sig odd} clearly shows the Einstein-Helfand relation can be rewritten as,
\begin{equation}\label{eqn:final EH}
\langle\Delta \mbf{M}^{2}(t)\rangle = \frac{6V}{\beta}\left(\sigma_{0}t + \sigma_1\right),
\end{equation}
and therefore this method can be used to find values for both $\sigma_0$ and $\sigma_1$.

$\langle\Delta \mbf{M}^{2}(t)\rangle$ is plotted in Fig.~\ref{fig:EH}(a) and a fit of this to Eq.~\ref{eqn:final EH} gives the values for $\sigma_0$ and $\sigma_1$ quoted in Table~\ref{table:sigma}. Fig.~\ref{fig:EH}(b) shows the good agreement between $\sigma_0$ values, highlighting that we have indeed reached the low frequency limit with simulation for all temperatures. In order to remove the noise seen at low-frequency in Fig.~\ref{fig:bulkcond}(d) we can replace the data in this region ($\omega \lessapprox 10^{10}$) with a low-frequency approximation ($\operatorname{Im}[\sigma(\omega\rightarrow 0)]=\sigma_1\omega$).

A value for $\sigma_2$ is then found through a fit of the real-part of the bulk conductivity spectra to second-order as shown in Fig.~\ref{fig:EH}(d), where $\sigma_0$ is found using the Einstein-Helfand method. Therefore at no point in our procedure do we find our $\sigma_n$ parameters through a fit to the impedance spectrum. 

\section*{References}
\bibliography{references-2}

\begin{thebibliography}{100}%
\makeatletter
\providecommand \@ifxundefined [1]{%
 \@ifx{#1\undefined}
}%
\providecommand \@ifnum [1]{%
 \ifnum #1\expandafter \@firstoftwo
 \else \expandafter \@secondoftwo
 \fi
}%
\providecommand \@ifx [1]{%
 \ifx #1\expandafter \@firstoftwo
 \else \expandafter \@secondoftwo
 \fi
}%
\providecommand \natexlab [1]{#1}%
\providecommand \enquote  [1]{``#1''}%
\providecommand \bibnamefont  [1]{#1}%
\providecommand \bibfnamefont [1]{#1}%
\providecommand \citenamefont [1]{#1}%
\providecommand \href@noop [0]{\@secondoftwo}%
\providecommand \href [0]{\begingroup \@sanitize@url \@href}%
\providecommand \@href[1]{\@@startlink{#1}\@@href}%
\providecommand \@@href[1]{\endgroup#1\@@endlink}%
\providecommand \@sanitize@url [0]{\catcode `\\12\catcode `\$12\catcode `\&12\catcode `\#12\catcode `\^12\catcode `\_12\catcode `\%12\relax}%
\providecommand \@@startlink[1]{}%
\providecommand \@@endlink[0]{}%
\providecommand \url  [0]{\begingroup\@sanitize@url \@url }%
\providecommand \@url [1]{\endgroup\@href {#1}{\urlprefix }}%
\providecommand \urlprefix  [0]{URL }%
\providecommand \Eprint [0]{\href }%
\providecommand \doibase [0]{https://doi.org/}%
\providecommand \selectlanguage [0]{\@gobble}%
\providecommand \bibinfo  [0]{\@secondoftwo}%
\providecommand \bibfield  [0]{\@secondoftwo}%
\providecommand \translation [1]{[#1]}%
\providecommand \BibitemOpen [0]{}%
\providecommand \bibitemStop [0]{}%
\providecommand \bibitemNoStop [0]{.\EOS\space}%
\providecommand \EOS [0]{\spacefactor3000\relax}%
\providecommand \BibitemShut  [1]{\csname bibitem#1\endcsname}%
\let\auto@bib@innerbib\@empty
\bibitem [{\citenamefont {Perkin}\ \emph {et~al.}(2013)\citenamefont {Perkin}, \citenamefont {Salanne}, \citenamefont {Madden},\ and\ \citenamefont {Lynden-Bell}}]{perkin_is_2013}%
  \BibitemOpen
  \bibfield  {author} {\bibinfo {author} {\bibfnamefont {S.}~\bibnamefont {Perkin}}, \bibinfo {author} {\bibfnamefont {M.}~\bibnamefont {Salanne}}, \bibinfo {author} {\bibfnamefont {P.}~\bibnamefont {Madden}},\ and\ \bibinfo {author} {\bibfnamefont {R.}~\bibnamefont {Lynden-Bell}},\ }\href {https://doi.org/10.1073/pnas.1314188110} {\bibfield  {journal} {\bibinfo  {journal} {Proceedings of the National Academy of Sciences}\ }\textbf {\bibinfo {volume} {110}},\ \bibinfo {pages} {4121} (\bibinfo {year} {2013})}\BibitemShut {NoStop}%
\bibitem [{\citenamefont {Gebbie}\ \emph {et~al.}(2013)\citenamefont {Gebbie}, \citenamefont {Valtiner}, \citenamefont {Banquy}, \citenamefont {Fox}, \citenamefont {Henderson},\ and\ \citenamefont {Israelachvili}}]{gebbie_ionic_2013}%
  \BibitemOpen
  \bibfield  {author} {\bibinfo {author} {\bibfnamefont {M.~A.}\ \bibnamefont {Gebbie}}, \bibinfo {author} {\bibfnamefont {M.}~\bibnamefont {Valtiner}}, \bibinfo {author} {\bibfnamefont {X.}~\bibnamefont {Banquy}}, \bibinfo {author} {\bibfnamefont {E.~T.}\ \bibnamefont {Fox}}, \bibinfo {author} {\bibfnamefont {W.~A.}\ \bibnamefont {Henderson}},\ and\ \bibinfo {author} {\bibfnamefont {J.~N.}\ \bibnamefont {Israelachvili}},\ }\href {https://doi.org/10.1073/pnas.1307871110} {\bibfield  {journal} {\bibinfo  {journal} {Proceedings of the National Academy of Sciences}\ }\textbf {\bibinfo {volume} {110}},\ \bibinfo {pages} {9674} (\bibinfo {year} {2013})}\BibitemShut {NoStop}%
\bibitem [{\citenamefont {Lee}\ \emph {et~al.}(2015)\citenamefont {Lee}, \citenamefont {Vella}, \citenamefont {Perkin},\ and\ \citenamefont {Goriely}}]{lee_are_2015}%
  \BibitemOpen
  \bibfield  {author} {\bibinfo {author} {\bibfnamefont {A.~A.}\ \bibnamefont {Lee}}, \bibinfo {author} {\bibfnamefont {D.}~\bibnamefont {Vella}}, \bibinfo {author} {\bibfnamefont {S.}~\bibnamefont {Perkin}},\ and\ \bibinfo {author} {\bibfnamefont {A.}~\bibnamefont {Goriely}},\ }\href {https://doi.org/10.1021/jz502250z} {\bibfield  {journal} {\bibinfo  {journal} {The Journal of Physical Chemistry Letters}\ }\textbf {\bibinfo {volume} {6}},\ \bibinfo {pages} {159} (\bibinfo {year} {2015})}\BibitemShut {NoStop}%
\bibitem [{\citenamefont {Weingärtner}(2008)}]{weingartner_understanding_2008}%
  \BibitemOpen
  \bibfield  {author} {\bibinfo {author} {\bibfnamefont {H.}~\bibnamefont {Weingärtner}},\ }\href {https://doi.org/10.1002/anie.200604951} {\bibfield  {journal} {\bibinfo  {journal} {Angewandte Chemie International Edition}\ }\textbf {\bibinfo {volume} {47}},\ \bibinfo {pages} {654} (\bibinfo {year} {2008})}\BibitemShut {NoStop}%
\bibitem [{\citenamefont {Nordness}\ and\ \citenamefont {Brennecke}(2020)}]{nordness_ion_2020}%
  \BibitemOpen
  \bibfield  {author} {\bibinfo {author} {\bibfnamefont {O.}~\bibnamefont {Nordness}}\ and\ \bibinfo {author} {\bibfnamefont {J.}~\bibnamefont {Brennecke}},\ }\href {https://pubs.acs.org/doi/full/10.1021/acs.chemrev.0c00373?casa_token=LYnLw3t53_cAAAAA%3AyBGG85bBhElSi1yHaTyCyM2pDqDGXC3uk6TH6yp5pUYEEYjglp47DjxBEQmDSnvVDc2CVyNHqphP5Q} {\bibfield  {journal} {\bibinfo  {journal} {Chemical Reviews}\ }\textbf {\bibinfo {volume} {120}},\ \bibinfo {pages} {12873} (\bibinfo {year} {2020})}\BibitemShut {NoStop}%
\bibitem [{\citenamefont {Gebbie}\ \emph {et~al.}(2017)\citenamefont {Gebbie}, \citenamefont {Smith}, \citenamefont {Dobbs}, \citenamefont {Lee}, \citenamefont {Warr}, \citenamefont {Banquy}, \citenamefont {Valtiner}, \citenamefont {Rutland}, \citenamefont {Israelachvili}, \citenamefont {Perkin},\ and\ \citenamefont {Atkin}}]{gebbie_long_2017}%
  \BibitemOpen
  \bibfield  {author} {\bibinfo {author} {\bibfnamefont {M.~A.}\ \bibnamefont {Gebbie}}, \bibinfo {author} {\bibfnamefont {A.~M.}\ \bibnamefont {Smith}}, \bibinfo {author} {\bibfnamefont {H.~A.}\ \bibnamefont {Dobbs}}, \bibinfo {author} {\bibfnamefont {A.~A.}\ \bibnamefont {Lee}}, \bibinfo {author} {\bibfnamefont {G.~G.}\ \bibnamefont {Warr}}, \bibinfo {author} {\bibfnamefont {X.}~\bibnamefont {Banquy}}, \bibinfo {author} {\bibfnamefont {M.}~\bibnamefont {Valtiner}}, \bibinfo {author} {\bibfnamefont {M.~W.}\ \bibnamefont {Rutland}}, \bibinfo {author} {\bibfnamefont {J.~N.}\ \bibnamefont {Israelachvili}}, \bibinfo {author} {\bibfnamefont {S.}~\bibnamefont {Perkin}},\ and\ \bibinfo {author} {\bibfnamefont {R.}~\bibnamefont {Atkin}},\ }\href {https://doi.org/10.1039/C6CC08820A} {\bibfield  {journal} {\bibinfo  {journal} {Chemical Communications}\ }\textbf {\bibinfo {volume} {53}},\ \bibinfo {pages} {1214} (\bibinfo {year} {2017})}\BibitemShut {NoStop}%
\bibitem [{\citenamefont {Kornyshev}(2007)}]{kornyshev_double-layer_2007}%
  \BibitemOpen
  \bibfield  {author} {\bibinfo {author} {\bibfnamefont {A.~A.}\ \bibnamefont {Kornyshev}},\ }\href {https://doi.org/10.1021/jp067857o} {\bibfield  {journal} {\bibinfo  {journal} {The Journal of Physical Chemistry B}\ }\textbf {\bibinfo {volume} {111}},\ \bibinfo {pages} {5545} (\bibinfo {year} {2007})}\BibitemShut {NoStop}%
\bibitem [{\citenamefont {Bazant}\ \emph {et~al.}(2011)\citenamefont {Bazant}, \citenamefont {Storey},\ and\ \citenamefont {Kornyshev}}]{bazant_double_2011}%
  \BibitemOpen
  \bibfield  {author} {\bibinfo {author} {\bibfnamefont {M.~Z.}\ \bibnamefont {Bazant}}, \bibinfo {author} {\bibfnamefont {B.~D.}\ \bibnamefont {Storey}},\ and\ \bibinfo {author} {\bibfnamefont {A.~A.}\ \bibnamefont {Kornyshev}},\ }\href {https://doi.org/10.1103/PhysRevLett.106.046102} {\bibfield  {journal} {\bibinfo  {journal} {Physical Review Letters}\ }\textbf {\bibinfo {volume} {106}},\ \bibinfo {pages} {046102} (\bibinfo {year} {2011})}\BibitemShut {NoStop}%
\bibitem [{\citenamefont {Wang}\ \emph {et~al.}(2021)\citenamefont {Wang}, \citenamefont {Zhang}, \citenamefont {Gharbi}, \citenamefont {Vivier}, \citenamefont {Gao},\ and\ \citenamefont {Orazem}}]{wang_electrochemical_2021}%
  \BibitemOpen
  \bibfield  {author} {\bibinfo {author} {\bibfnamefont {S.}~\bibnamefont {Wang}}, \bibinfo {author} {\bibfnamefont {J.}~\bibnamefont {Zhang}}, \bibinfo {author} {\bibfnamefont {O.}~\bibnamefont {Gharbi}}, \bibinfo {author} {\bibfnamefont {V.}~\bibnamefont {Vivier}}, \bibinfo {author} {\bibfnamefont {M.}~\bibnamefont {Gao}},\ and\ \bibinfo {author} {\bibfnamefont {M.~E.}\ \bibnamefont {Orazem}},\ }\href {https://doi.org/10.1038/s43586-021-00039-w} {\bibfield  {journal} {\bibinfo  {journal} {Nature Reviews Methods Primers}\ }\textbf {\bibinfo {volume} {1}},\ \bibinfo {pages} {1} (\bibinfo {year} {2021})}\BibitemShut {NoStop}%
\bibitem [{\citenamefont {Usler}\ \emph {et~al.}(2026)\citenamefont {Usler}, \citenamefont {Fertig},\ and\ \citenamefont {Janssen}}]{usler_impedance_2026}%
  \BibitemOpen
  \bibfield  {author} {\bibinfo {author} {\bibfnamefont {A.~L.}\ \bibnamefont {Usler}}, \bibinfo {author} {\bibfnamefont {D.}~\bibnamefont {Fertig}},\ and\ \bibinfo {author} {\bibfnamefont {M.}~\bibnamefont {Janssen}},\ }\href {http://arxiv.org/abs/2606.21980} {\bibfield  {journal} {\bibinfo  {journal} {arXiv:2606.21980 [physics.chem-ph]}\ } (\bibinfo {year} {2026})}\BibitemShut {NoStop}%
\bibitem [{\citenamefont {Cole}\ and\ \citenamefont {Cole}(1941)}]{cole_dispersion_1941}%
  \BibitemOpen
  \bibfield  {author} {\bibinfo {author} {\bibfnamefont {K.~S.}\ \bibnamefont {Cole}}\ and\ \bibinfo {author} {\bibfnamefont {R.~H.}\ \bibnamefont {Cole}},\ }\href {https://doi.org/10.1063/1.1750906} {\bibfield  {journal} {\bibinfo  {journal} {The Journal of Chemical Physics}\ }\textbf {\bibinfo {volume} {9}},\ \bibinfo {pages} {341} (\bibinfo {year} {1941})}\BibitemShut {NoStop}%
\bibitem [{\citenamefont {Davidson}\ and\ \citenamefont {Cole}(1950)}]{davidson_dielectric_1950}%
  \BibitemOpen
  \bibfield  {author} {\bibinfo {author} {\bibfnamefont {D.~W.}\ \bibnamefont {Davidson}}\ and\ \bibinfo {author} {\bibfnamefont {R.~H.}\ \bibnamefont {Cole}},\ }\href {https://doi.org/10.1063/1.1747496} {\bibfield  {journal} {\bibinfo  {journal} {The Journal of Chemical Physics}\ }\textbf {\bibinfo {volume} {18}},\ \bibinfo {pages} {1417} (\bibinfo {year} {1950})}\BibitemShut {NoStop}%
\bibitem [{\citenamefont {Davidson}\ and\ \citenamefont {Cole}(1951)}]{davidson_dielectric_1951}%
  \BibitemOpen
  \bibfield  {author} {\bibinfo {author} {\bibfnamefont {D.~W.}\ \bibnamefont {Davidson}}\ and\ \bibinfo {author} {\bibfnamefont {R.~H.}\ \bibnamefont {Cole}},\ }\href {https://doi.org/10.1063/1.1748105} {\bibfield  {journal} {\bibinfo  {journal} {The Journal of Chemical Physics}\ }\textbf {\bibinfo {volume} {19}},\ \bibinfo {pages} {1484} (\bibinfo {year} {1951})}\BibitemShut {NoStop}%
\bibitem [{\citenamefont {Tee}\ and\ \citenamefont {Searles}(2022)}]{tee_fully_2022}%
  \BibitemOpen
  \bibfield  {author} {\bibinfo {author} {\bibfnamefont {S.~R.}\ \bibnamefont {Tee}}\ and\ \bibinfo {author} {\bibfnamefont {D.~J.}\ \bibnamefont {Searles}},\ }\href {https://doi.org/10.1063/5.0086986} {\bibfield  {journal} {\bibinfo  {journal} {The Journal of Chemical Physics}\ }\textbf {\bibinfo {volume} {156}},\ \bibinfo {pages} {184101} (\bibinfo {year} {2022})}\BibitemShut {NoStop}%
\bibitem [{\citenamefont {Dufils}\ \emph {et~al.}(2019)\citenamefont {Dufils}, \citenamefont {Jeanmairet}, \citenamefont {Rotenberg}, \citenamefont {Sprik},\ and\ \citenamefont {Salanne}}]{dufils_simulating_2019}%
  \BibitemOpen
  \bibfield  {author} {\bibinfo {author} {\bibfnamefont {T.}~\bibnamefont {Dufils}}, \bibinfo {author} {\bibfnamefont {G.}~\bibnamefont {Jeanmairet}}, \bibinfo {author} {\bibfnamefont {B.}~\bibnamefont {Rotenberg}}, \bibinfo {author} {\bibfnamefont {M.}~\bibnamefont {Sprik}},\ and\ \bibinfo {author} {\bibfnamefont {M.}~\bibnamefont {Salanne}},\ }\href {https://doi.org/10.1103/PhysRevLett.123.195501} {\bibfield  {journal} {\bibinfo  {journal} {Physical Review Letters}\ }\textbf {\bibinfo {volume} {123}},\ \bibinfo {pages} {195501} (\bibinfo {year} {2019})}\BibitemShut {NoStop}%
\bibitem [{\citenamefont {Pireddu}\ and\ \citenamefont {Rotenberg}(2023)}]{pireddu_frequency-dependent_2023}%
  \BibitemOpen
  \bibfield  {author} {\bibinfo {author} {\bibfnamefont {G.}~\bibnamefont {Pireddu}}\ and\ \bibinfo {author} {\bibfnamefont {B.}~\bibnamefont {Rotenberg}},\ }\href {https://journals.aps.org/prl/abstract/10.1103/PhysRevLett.130.098001} {\bibfield  {journal} {\bibinfo  {journal} {Phys. Rev. Lett.}\ }\textbf {\bibinfo {volume} {130}} (\bibinfo {year} {2023})}\BibitemShut {NoStop}%
\bibitem [{\citenamefont {Pireddu}\ \emph {et~al.}(2024)\citenamefont {Pireddu}, \citenamefont {Fairchild}, \citenamefont {Niblett}, \citenamefont {Cox},\ and\ \citenamefont {Rotenberg}}]{pireddu_impedance_2024}%
  \BibitemOpen
  \bibfield  {author} {\bibinfo {author} {\bibfnamefont {G.}~\bibnamefont {Pireddu}}, \bibinfo {author} {\bibfnamefont {C.~J.}\ \bibnamefont {Fairchild}}, \bibinfo {author} {\bibfnamefont {S.~P.}\ \bibnamefont {Niblett}}, \bibinfo {author} {\bibfnamefont {S.~J.}\ \bibnamefont {Cox}},\ and\ \bibinfo {author} {\bibfnamefont {B.}~\bibnamefont {Rotenberg}},\ }\href {https://doi.org/10.1073/pnas.2318157121} {\bibfield  {journal} {\bibinfo  {journal} {Proceedings of the National Academy of Sciences}\ }\textbf {\bibinfo {volume} {121}},\ \bibinfo {pages} {e2318157121} (\bibinfo {year} {2024})}\BibitemShut {NoStop}%
\bibitem [{Note1()}]{Note1}%
  \BibitemOpen
  \bibinfo {note} {As determined from the covariance of the local polarization with the total dipole of the fluid}\BibitemShut {NoStop}%
\bibitem [{\citenamefont {J.~Cox}\ and\ \citenamefont {L.~Geissler}(2022)}]{jcox_dielectric_2022}%
  \BibitemOpen
  \bibfield  {author} {\bibinfo {author} {\bibfnamefont {S.}~\bibnamefont {J.~Cox}}\ and\ \bibinfo {author} {\bibfnamefont {P.}~\bibnamefont {L.~Geissler}},\ }\href {https://doi.org/10.1039/D2SC01243J} {\bibfield  {journal} {\bibinfo  {journal} {Chemical Science}\ }\textbf {\bibinfo {volume} {13}},\ \bibinfo {pages} {9102} (\bibinfo {year} {2022})}\BibitemShut {NoStop}%
\bibitem [{\citenamefont {dos Santos}\ and\ \citenamefont {Netz}(2018)}]{dos_santos_dielectric_2018}%
  \BibitemOpen
  \bibfield  {author} {\bibinfo {author} {\bibfnamefont {A.~P.}\ \bibnamefont {dos Santos}}\ and\ \bibinfo {author} {\bibfnamefont {R.~R.}\ \bibnamefont {Netz}},\ }\href {https://doi.org/10.1063/1.5022226} {\bibfield  {journal} {\bibinfo  {journal} {The Journal of Chemical Physics}\ }\textbf {\bibinfo {volume} {148}},\ \bibinfo {pages} {164103} (\bibinfo {year} {2018})}\BibitemShut {NoStop}%
\bibitem [{\citenamefont {Fletcher}\ \emph {et~al.}(2014)\citenamefont {Fletcher}, \citenamefont {Black},\ and\ \citenamefont {Kirkpatrick}}]{fletcher_universal_2014}%
  \BibitemOpen
  \bibfield  {author} {\bibinfo {author} {\bibfnamefont {S.}~\bibnamefont {Fletcher}}, \bibinfo {author} {\bibfnamefont {V.~J.}\ \bibnamefont {Black}},\ and\ \bibinfo {author} {\bibfnamefont {I.}~\bibnamefont {Kirkpatrick}},\ }\href {https://doi.org/10.1007/s10008-013-2328-4} {\bibfield  {journal} {\bibinfo  {journal} {Journal of Solid State Electrochemistry}\ }\textbf {\bibinfo {volume} {18}},\ \bibinfo {pages} {1377} (\bibinfo {year} {2014})}\BibitemShut {NoStop}%
\bibitem [{\citenamefont {Zhang}\ \emph {et~al.}(2004)\citenamefont {Zhang}, \citenamefont {Xu},\ and\ \citenamefont {Jow}}]{zhang_electrochemical_2004}%
  \BibitemOpen
  \bibfield  {author} {\bibinfo {author} {\bibfnamefont {S.~S.}\ \bibnamefont {Zhang}}, \bibinfo {author} {\bibfnamefont {K.}~\bibnamefont {Xu}},\ and\ \bibinfo {author} {\bibfnamefont {T.~R.}\ \bibnamefont {Jow}},\ }\href {https://doi.org/10.1016/j.electacta.2003.10.016} {\bibfield  {journal} {\bibinfo  {journal} {Electrochimica Acta}\ }\textbf {\bibinfo {volume} {49}},\ \bibinfo {pages} {1057} (\bibinfo {year} {2004})}\BibitemShut {NoStop}%
\bibitem [{\citenamefont {Irvine}\ \emph {et~al.}(1990)\citenamefont {Irvine}, \citenamefont {Sinclair},\ and\ \citenamefont {West}}]{irvine_electroceramics_1990}%
  \BibitemOpen
  \bibfield  {author} {\bibinfo {author} {\bibfnamefont {J.~T.~S.}\ \bibnamefont {Irvine}}, \bibinfo {author} {\bibfnamefont {D.~C.}\ \bibnamefont {Sinclair}},\ and\ \bibinfo {author} {\bibfnamefont {A.~R.}\ \bibnamefont {West}},\ }\href {https://doi.org/10.1002/adma.19900020304} {\bibfield  {journal} {\bibinfo  {journal} {Advanced Materials}\ }\textbf {\bibinfo {volume} {2}},\ \bibinfo {pages} {132} (\bibinfo {year} {1990})}\BibitemShut {NoStop}%
\bibitem [{\citenamefont {Hernández}\ \emph {et~al.}(2016)\citenamefont {Hernández}, \citenamefont {Masó},\ and\ \citenamefont {West}}]{hernandez_correct_2016}%
  \BibitemOpen
  \bibfield  {author} {\bibinfo {author} {\bibfnamefont {M.~A.}\ \bibnamefont {Hernández}}, \bibinfo {author} {\bibfnamefont {N.}~\bibnamefont {Masó}},\ and\ \bibinfo {author} {\bibfnamefont {A.~R.}\ \bibnamefont {West}},\ }\href {https://doi.org/10.1063/1.4946008} {\bibfield  {journal} {\bibinfo  {journal} {Applied Physics Letters}\ }\textbf {\bibinfo {volume} {108}},\ \bibinfo {pages} {152901} (\bibinfo {year} {2016})}\BibitemShut {NoStop}%
\bibitem [{\citenamefont {Schwan}(1992)}]{schwan_linear_1992}%
  \BibitemOpen
  \bibfield  {author} {\bibinfo {author} {\bibfnamefont {H.~P.}\ \bibnamefont {Schwan}},\ }\href {https://doi.org/10.1007/BF02368531} {\bibfield  {journal} {\bibinfo  {journal} {Annals of Biomedical Engineering}\ }\textbf {\bibinfo {volume} {20}},\ \bibinfo {pages} {269} (\bibinfo {year} {1992})}\BibitemShut {NoStop}%
\bibitem [{\citenamefont {Chassagne}\ \emph {et~al.}(2016)\citenamefont {Chassagne}, \citenamefont {Dubois}, \citenamefont {Jiménez}, \citenamefont {van~der Ploeg},\ and\ \citenamefont {van Turnhout}}]{chassagne_compensating_2016}%
  \BibitemOpen
  \bibfield  {author} {\bibinfo {author} {\bibfnamefont {C.}~\bibnamefont {Chassagne}}, \bibinfo {author} {\bibfnamefont {E.}~\bibnamefont {Dubois}}, \bibinfo {author} {\bibfnamefont {M.~L.}\ \bibnamefont {Jiménez}}, \bibinfo {author} {\bibfnamefont {J.~P.~M.}\ \bibnamefont {van~der Ploeg}},\ and\ \bibinfo {author} {\bibfnamefont {J.}~\bibnamefont {van Turnhout}},\ }\bibfield  {journal} {\bibinfo  {journal} {Frontiers in Chemistry}\ }\textbf {\bibinfo {volume} {4}},\ \href {https://doi.org/10.3389/fchem.2016.00030} {10.3389/fchem.2016.00030} (\bibinfo {year} {2016})\BibitemShut {NoStop}%
\bibitem [{\citenamefont {Schröder}\ \emph {et~al.}(2008)\citenamefont {Schröder}, \citenamefont {Haberler},\ and\ \citenamefont {Steinhauser}}]{schroder_computation_2008}%
  \BibitemOpen
  \bibfield  {author} {\bibinfo {author} {\bibfnamefont {C.}~\bibnamefont {Schröder}}, \bibinfo {author} {\bibfnamefont {M.}~\bibnamefont {Haberler}},\ and\ \bibinfo {author} {\bibfnamefont {O.}~\bibnamefont {Steinhauser}},\ }\href {https://doi.org/10.1063/1.2868752} {\bibfield  {journal} {\bibinfo  {journal} {The Journal of Chemical Physics}\ }\textbf {\bibinfo {volume} {128}},\ \bibinfo {pages} {134501} (\bibinfo {year} {2008})}\BibitemShut {NoStop}%
\bibitem [{\citenamefont {Picálek}\ and\ \citenamefont {Kolafa}(2007)}]{picalek_molecular_2007}%
  \BibitemOpen
  \bibfield  {author} {\bibinfo {author} {\bibfnamefont {J.}~\bibnamefont {Picálek}}\ and\ \bibinfo {author} {\bibfnamefont {J.}~\bibnamefont {Kolafa}},\ }\href {https://doi.org/10.1016/j.molliq.2006.12.015} {\bibfield  {journal} {\bibinfo  {journal} {Journal of Molecular Liquids}\ }\bibinfo {series} {{EMLG}/{JMLG} 2005 {Special} {Issue}},\ \textbf {\bibinfo {volume} {134}},\ \bibinfo {pages} {29} (\bibinfo {year} {2007})}\BibitemShut {NoStop}%
\bibitem [{\citenamefont {Friesen}\ \emph {et~al.}(2000)\citenamefont {Friesen}, \citenamefont {Özsar},\ and\ \citenamefont {Dunlop}}]{friesen_impedance_2000}%
  \BibitemOpen
  \bibfield  {author} {\bibinfo {author} {\bibfnamefont {G.}~\bibnamefont {Friesen}}, \bibinfo {author} {\bibfnamefont {M.~E.}\ \bibnamefont {Özsar}},\ and\ \bibinfo {author} {\bibfnamefont {E.~D.}\ \bibnamefont {Dunlop}},\ }\href {https://doi.org/10.1016/S0040-6090(99)00764-6} {\bibfield  {journal} {\bibinfo  {journal} {Thin Solid Films}\ }\textbf {\bibinfo {volume} {361-362}},\ \bibinfo {pages} {303} (\bibinfo {year} {2000})}\BibitemShut {NoStop}%
\bibitem [{\citenamefont {Stoppa}\ \emph {et~al.}(2008)\citenamefont {Stoppa}, \citenamefont {Hunger}, \citenamefont {Buchner}, \citenamefont {Hefter}, \citenamefont {Thoman},\ and\ \citenamefont {Helm}}]{stoppa_interactions_2008}%
  \BibitemOpen
  \bibfield  {author} {\bibinfo {author} {\bibfnamefont {A.}~\bibnamefont {Stoppa}}, \bibinfo {author} {\bibfnamefont {J.}~\bibnamefont {Hunger}}, \bibinfo {author} {\bibfnamefont {R.}~\bibnamefont {Buchner}}, \bibinfo {author} {\bibfnamefont {G.}~\bibnamefont {Hefter}}, \bibinfo {author} {\bibfnamefont {A.}~\bibnamefont {Thoman}},\ and\ \bibinfo {author} {\bibfnamefont {H.}~\bibnamefont {Helm}},\ }\href {https://doi.org/10.1021/jp800852z} {\bibfield  {journal} {\bibinfo  {journal} {The Journal of Physical Chemistry B}\ }\textbf {\bibinfo {volume} {112}},\ \bibinfo {pages} {4854} (\bibinfo {year} {2008})}\BibitemShut {NoStop}%
\bibitem [{\citenamefont {Williams}\ and\ \citenamefont {Watts}(1970)}]{williams_non-symmetrical_1970}%
  \BibitemOpen
  \bibfield  {author} {\bibinfo {author} {\bibfnamefont {G.}~\bibnamefont {Williams}}\ and\ \bibinfo {author} {\bibfnamefont {D.~C.}\ \bibnamefont {Watts}},\ }\href {https://doi.org/10.1039/TF9706600080} {\bibfield  {journal} {\bibinfo  {journal} {Transactions of the Faraday Society}\ }\textbf {\bibinfo {volume} {66}},\ \bibinfo {pages} {80} (\bibinfo {year} {1970})}\BibitemShut {NoStop}%
\bibitem [{\citenamefont {Lazanas}\ and\ \citenamefont {Prodromidis}(2023)}]{lazanas_electrochemical_2023}%
  \BibitemOpen
  \bibfield  {author} {\bibinfo {author} {\bibfnamefont {A.~C.}\ \bibnamefont {Lazanas}}\ and\ \bibinfo {author} {\bibfnamefont {M.~I.}\ \bibnamefont {Prodromidis}},\ }\href {https://doi.org/10.1021/acsmeasuresciau.2c00070} {\bibfield  {journal} {\bibinfo  {journal} {ACS Measurement Science Au}\ }\textbf {\bibinfo {volume} {3}},\ \bibinfo {pages} {162} (\bibinfo {year} {2023})}\BibitemShut {NoStop}%
\bibitem [{\citenamefont {Beckmann}(1988)}]{beckmann_spectral_1988}%
  \BibitemOpen
  \bibfield  {author} {\bibinfo {author} {\bibfnamefont {P.~A.}\ \bibnamefont {Beckmann}},\ }\href {https://doi.org/10.1016/0370-1573(88)90073-7} {\bibfield  {journal} {\bibinfo  {journal} {Physics Reports}\ }\textbf {\bibinfo {volume} {171}},\ \bibinfo {pages} {85} (\bibinfo {year} {1988})}\BibitemShut {NoStop}%
\bibitem [{\citenamefont {Iglesias}\ \emph {et~al.}(2017)\citenamefont {Iglesias}, \citenamefont {Vilão},\ and\ \citenamefont {Reis}}]{iglesias_approach_2017}%
  \BibitemOpen
  \bibfield  {author} {\bibinfo {author} {\bibfnamefont {T.~P.}\ \bibnamefont {Iglesias}}, \bibinfo {author} {\bibfnamefont {G.}~\bibnamefont {Vilão}},\ and\ \bibinfo {author} {\bibfnamefont {J.~C.~R.}\ \bibnamefont {Reis}},\ }\href {https://doi.org/10.1063/1.4985839} {\bibfield  {journal} {\bibinfo  {journal} {Journal of Applied Physics}\ }\textbf {\bibinfo {volume} {122}},\ \bibinfo {pages} {074102} (\bibinfo {year} {2017})}\BibitemShut {NoStop}%
\bibitem [{\citenamefont {Powles}(1951)}]{powles_interpretation_1951}%
  \BibitemOpen
  \bibfield  {author} {\bibinfo {author} {\bibfnamefont {J.~G.}\ \bibnamefont {Powles}},\ }\href {https://doi.org/10.1088/0370-1301/64/1/109} {\bibfield  {journal} {\bibinfo  {journal} {Proceedings of the Physical Society. Section B}\ }\textbf {\bibinfo {volume} {64}},\ \bibinfo {pages} {81} (\bibinfo {year} {1951})}\BibitemShut {NoStop}%
\bibitem [{\citenamefont {Shoar~Abouzari}\ \emph {et~al.}(2009)\citenamefont {Shoar~Abouzari}, \citenamefont {Berkemeier}, \citenamefont {Schmitz},\ and\ \citenamefont {Wilmer}}]{shoar_abouzari_physical_2009}%
  \BibitemOpen
  \bibfield  {author} {\bibinfo {author} {\bibfnamefont {M.~R.}\ \bibnamefont {Shoar~Abouzari}}, \bibinfo {author} {\bibfnamefont {F.}~\bibnamefont {Berkemeier}}, \bibinfo {author} {\bibfnamefont {G.}~\bibnamefont {Schmitz}},\ and\ \bibinfo {author} {\bibfnamefont {D.}~\bibnamefont {Wilmer}},\ }\href {https://doi.org/10.1016/j.ssi.2009.04.002} {\bibfield  {journal} {\bibinfo  {journal} {Solid State Ionics}\ }\textbf {\bibinfo {volume} {180}},\ \bibinfo {pages} {922} (\bibinfo {year} {2009})}\BibitemShut {NoStop}%
\bibitem [{\citenamefont {Rezaei~Niya}\ and\ \citenamefont {Hoorfar}(2016)}]{rezaei_niya_possible_2016}%
  \BibitemOpen
  \bibfield  {author} {\bibinfo {author} {\bibfnamefont {S.~M.}\ \bibnamefont {Rezaei~Niya}}\ and\ \bibinfo {author} {\bibfnamefont {M.}~\bibnamefont {Hoorfar}},\ }\href {https://doi.org/10.1016/j.electacta.2015.11.142} {\bibfield  {journal} {\bibinfo  {journal} {Electrochimica Acta}\ }\textbf {\bibinfo {volume} {188}},\ \bibinfo {pages} {98} (\bibinfo {year} {2016})}\BibitemShut {NoStop}%
\bibitem [{\citenamefont {Córdoba-Torres}\ \emph {et~al.}(2015)\citenamefont {Córdoba-Torres}, \citenamefont {Mesquita},\ and\ \citenamefont {Nogueira}}]{cordoba-torres_relationship_2015}%
  \BibitemOpen
  \bibfield  {author} {\bibinfo {author} {\bibfnamefont {P.}~\bibnamefont {Córdoba-Torres}}, \bibinfo {author} {\bibfnamefont {T.~J.}\ \bibnamefont {Mesquita}},\ and\ \bibinfo {author} {\bibfnamefont {R.~P.}\ \bibnamefont {Nogueira}},\ }\href {https://doi.org/10.1021/jp512063f} {\bibfield  {journal} {\bibinfo  {journal} {The Journal of Physical Chemistry C}\ }\textbf {\bibinfo {volume} {119}},\ \bibinfo {pages} {4136} (\bibinfo {year} {2015})}\BibitemShut {NoStop}%
\bibitem [{\citenamefont {Wei}\ and\ \citenamefont {Patey}(1991)}]{wei_dielectric_1991}%
  \BibitemOpen
  \bibfield  {author} {\bibinfo {author} {\bibfnamefont {D.}~\bibnamefont {Wei}}\ and\ \bibinfo {author} {\bibfnamefont {G.~N.}\ \bibnamefont {Patey}},\ }\href {https://doi.org/10.1063/1.460257} {\bibfield  {journal} {\bibinfo  {journal} {The Journal of Chemical Physics}\ }\textbf {\bibinfo {volume} {94}},\ \bibinfo {pages} {6795} (\bibinfo {year} {1991})}\BibitemShut {NoStop}%
\bibitem [{\citenamefont {Chandra}\ \emph {et~al.}(1993)\citenamefont {Chandra}, \citenamefont {Wei},\ and\ \citenamefont {Patey}}]{chandra_frequency_1993}%
  \BibitemOpen
  \bibfield  {author} {\bibinfo {author} {\bibfnamefont {A.}~\bibnamefont {Chandra}}, \bibinfo {author} {\bibfnamefont {D.}~\bibnamefont {Wei}},\ and\ \bibinfo {author} {\bibfnamefont {G.~N.}\ \bibnamefont {Patey}},\ }\href {https://doi.org/10.1063/1.465274} {\bibfield  {journal} {\bibinfo  {journal} {The Journal of Chemical Physics}\ }\textbf {\bibinfo {volume} {99}},\ \bibinfo {pages} {2083} (\bibinfo {year} {1993})}\BibitemShut {NoStop}%
\bibitem [{\citenamefont {Chandra}\ and\ \citenamefont {Bagchi}(2000{\natexlab{a}})}]{chandra_frequency_2000}%
  \BibitemOpen
  \bibfield  {author} {\bibinfo {author} {\bibfnamefont {A.}~\bibnamefont {Chandra}}\ and\ \bibinfo {author} {\bibfnamefont {B.}~\bibnamefont {Bagchi}},\ }\href {https://doi.org/10.1063/1.480751} {\bibfield  {journal} {\bibinfo  {journal} {The Journal of Chemical Physics}\ }\textbf {\bibinfo {volume} {112}},\ \bibinfo {pages} {1876} (\bibinfo {year} {2000}{\natexlab{a}})}\BibitemShut {NoStop}%
\bibitem [{\citenamefont {Chandra}\ and\ \citenamefont {Bagchi}(2000{\natexlab{b}})}]{chandra_beyond_2000}%
  \BibitemOpen
  \bibfield  {author} {\bibinfo {author} {\bibfnamefont {A.}~\bibnamefont {Chandra}}\ and\ \bibinfo {author} {\bibfnamefont {B.}~\bibnamefont {Bagchi}},\ }\href {https://doi.org/10.1021/jp001052d} {\bibfield  {journal} {\bibinfo  {journal} {The Journal of Physical Chemistry B}\ }\textbf {\bibinfo {volume} {104}},\ \bibinfo {pages} {9067} (\bibinfo {year} {2000}{\natexlab{b}})}\BibitemShut {NoStop}%
\bibitem [{\citenamefont {Dufrêche}\ \emph {et~al.}(2002)\citenamefont {Dufrêche}, \citenamefont {Bernard}, \citenamefont {Turq}, \citenamefont {Mukherjee},\ and\ \citenamefont {Bagchi}}]{dufreche_ionic_2002}%
  \BibitemOpen
  \bibfield  {author} {\bibinfo {author} {\bibfnamefont {J.-F.}\ \bibnamefont {Dufrêche}}, \bibinfo {author} {\bibfnamefont {O.}~\bibnamefont {Bernard}}, \bibinfo {author} {\bibfnamefont {P.}~\bibnamefont {Turq}}, \bibinfo {author} {\bibfnamefont {A.}~\bibnamefont {Mukherjee}},\ and\ \bibinfo {author} {\bibfnamefont {B.}~\bibnamefont {Bagchi}},\ }\href {https://doi.org/10.1103/PhysRevLett.88.095902} {\bibfield  {journal} {\bibinfo  {journal} {Physical Review Letters}\ }\textbf {\bibinfo {volume} {88}},\ \bibinfo {pages} {095902} (\bibinfo {year} {2002})}\BibitemShut {NoStop}%
\bibitem [{\citenamefont {Zwanzig}(1960)}]{zwanzig_ensemble_1960}%
  \BibitemOpen
  \bibfield  {author} {\bibinfo {author} {\bibfnamefont {R.}~\bibnamefont {Zwanzig}},\ }\href {https://doi.org/10.1063/1.1731409} {\bibfield  {journal} {\bibinfo  {journal} {The Journal of Chemical Physics}\ }\textbf {\bibinfo {volume} {33}},\ \bibinfo {pages} {1338} (\bibinfo {year} {1960})}\BibitemShut {NoStop}%
\bibitem [{\citenamefont {Fujisaka}\ and\ \citenamefont {Inoue}(1987)}]{fujisaka_continued_1987}%
  \BibitemOpen
  \bibfield  {author} {\bibinfo {author} {\bibfnamefont {H.}~\bibnamefont {Fujisaka}}\ and\ \bibinfo {author} {\bibfnamefont {M.}~\bibnamefont {Inoue}},\ }\href {https://doi.org/10.1143/PTP.78.1203} {\bibfield  {journal} {\bibinfo  {journal} {Progress of Theoretical Physics}\ }\textbf {\bibinfo {volume} {78}},\ \bibinfo {pages} {1203} (\bibinfo {year} {1987})}\BibitemShut {NoStop}%
\bibitem [{\citenamefont {Boon}\ and\ \citenamefont {Yip}(1991)}]{boon_molecular_1991}%
  \BibitemOpen
  \bibfield  {author} {\bibinfo {author} {\bibfnamefont {J.~P.}\ \bibnamefont {Boon}}\ and\ \bibinfo {author} {\bibfnamefont {S.}~\bibnamefont {Yip}},\ }\href@noop {} {\emph {\bibinfo {title} {Molecular {Hydrodynamics}}}}\ (\bibinfo  {publisher} {Courier Corporation},\ \bibinfo {year} {1991})\BibitemShut {NoStop}%
\bibitem [{\citenamefont {Stoppa}\ \emph {et~al.}(2010)\citenamefont {Stoppa}, \citenamefont {Zech}, \citenamefont {Kunz},\ and\ \citenamefont {Buchner}}]{stoppa_conductivity_2010}%
  \BibitemOpen
  \bibfield  {author} {\bibinfo {author} {\bibfnamefont {A.}~\bibnamefont {Stoppa}}, \bibinfo {author} {\bibfnamefont {O.}~\bibnamefont {Zech}}, \bibinfo {author} {\bibfnamefont {W.}~\bibnamefont {Kunz}},\ and\ \bibinfo {author} {\bibfnamefont {R.}~\bibnamefont {Buchner}},\ }\href {https://doi.org/10.1021/je900789j} {\bibfield  {journal} {\bibinfo  {journal} {Journal of Chemical \& Engineering Data}\ }\textbf {\bibinfo {volume} {55}},\ \bibinfo {pages} {1768} (\bibinfo {year} {2010})}\BibitemShut {NoStop}%
\bibitem [{\citenamefont {Hayes}\ \emph {et~al.}(2015)\citenamefont {Hayes}, \citenamefont {Warr},\ and\ \citenamefont {Atkin}}]{hayes_structure_2015}%
  \BibitemOpen
  \bibfield  {author} {\bibinfo {author} {\bibfnamefont {R.}~\bibnamefont {Hayes}}, \bibinfo {author} {\bibfnamefont {G.~G.}\ \bibnamefont {Warr}},\ and\ \bibinfo {author} {\bibfnamefont {R.}~\bibnamefont {Atkin}},\ }\href {https://doi.org/10.1021/cr500411q} {\bibfield  {journal} {\bibinfo  {journal} {Chemical Reviews}\ }\textbf {\bibinfo {volume} {115}},\ \bibinfo {pages} {6357} (\bibinfo {year} {2015})}\BibitemShut {NoStop}%
\bibitem [{\citenamefont {Pei}\ \emph {et~al.}(2022)\citenamefont {Pei}, \citenamefont {Zhang}, \citenamefont {Ma}, \citenamefont {Fan}, \citenamefont {Zhang},\ and\ \citenamefont {Wang}}]{pei_ionic_2022}%
  \BibitemOpen
  \bibfield  {author} {\bibinfo {author} {\bibfnamefont {Y.}~\bibnamefont {Pei}}, \bibinfo {author} {\bibfnamefont {Y.}~\bibnamefont {Zhang}}, \bibinfo {author} {\bibfnamefont {J.}~\bibnamefont {Ma}}, \bibinfo {author} {\bibfnamefont {M.}~\bibnamefont {Fan}}, \bibinfo {author} {\bibfnamefont {S.}~\bibnamefont {Zhang}},\ and\ \bibinfo {author} {\bibfnamefont {J.}~\bibnamefont {Wang}},\ }\href {https://doi.org/10.1016/j.mtnano.2021.100159} {\bibfield  {journal} {\bibinfo  {journal} {Materials Today Nano}\ }\textbf {\bibinfo {volume} {17}},\ \bibinfo {pages} {100159} (\bibinfo {year} {2022})}\BibitemShut {NoStop}%
\bibitem [{\citenamefont {Ngo}\ \emph {et~al.}(2000)\citenamefont {Ngo}, \citenamefont {LeCompte}, \citenamefont {Hargens},\ and\ \citenamefont {McEwen}}]{ngo_thermal_2000}%
  \BibitemOpen
  \bibfield  {author} {\bibinfo {author} {\bibfnamefont {H.~L.}\ \bibnamefont {Ngo}}, \bibinfo {author} {\bibfnamefont {K.}~\bibnamefont {LeCompte}}, \bibinfo {author} {\bibfnamefont {L.}~\bibnamefont {Hargens}},\ and\ \bibinfo {author} {\bibfnamefont {A.~B.}\ \bibnamefont {McEwen}},\ }\href {https://doi.org/10.1016/S0040-6031(00)00373-7} {\bibfield  {journal} {\bibinfo  {journal} {Thermochimica Acta}\ }\textbf {\bibinfo {volume} {357-358}},\ \bibinfo {pages} {97} (\bibinfo {year} {2000})}\BibitemShut {NoStop}%
\bibitem [{\citenamefont {Eftekhari}(2017)}]{eftekhari_supercapacitors_2017}%
  \BibitemOpen
  \bibfield  {author} {\bibinfo {author} {\bibfnamefont {A.}~\bibnamefont {Eftekhari}},\ }\href {https://doi.org/10.1016/j.ensm.2017.06.009} {\bibfield  {journal} {\bibinfo  {journal} {Energy Storage Materials}\ }\textbf {\bibinfo {volume} {9}},\ \bibinfo {pages} {47} (\bibinfo {year} {2017})}\BibitemShut {NoStop}%
\bibitem [{\citenamefont {Miao}\ \emph {et~al.}(2021)\citenamefont {Miao}, \citenamefont {Song}, \citenamefont {Zhu}, \citenamefont {Li}, \citenamefont {Gan},\ and\ \citenamefont {Liu}}]{miao_ionic_2021}%
  \BibitemOpen
  \bibfield  {author} {\bibinfo {author} {\bibfnamefont {L.}~\bibnamefont {Miao}}, \bibinfo {author} {\bibfnamefont {Z.}~\bibnamefont {Song}}, \bibinfo {author} {\bibfnamefont {D.}~\bibnamefont {Zhu}}, \bibinfo {author} {\bibfnamefont {L.}~\bibnamefont {Li}}, \bibinfo {author} {\bibfnamefont {L.}~\bibnamefont {Gan}},\ and\ \bibinfo {author} {\bibfnamefont {M.}~\bibnamefont {Liu}},\ }\href {https://doi.org/10.1021/acs.energyfuels.1c00321} {\bibfield  {journal} {\bibinfo  {journal} {Energy \& Fuels}\ }\textbf {\bibinfo {volume} {35}},\ \bibinfo {pages} {8443} (\bibinfo {year} {2021})}\BibitemShut {NoStop}%
\bibitem [{\citenamefont {Salanne}(2018)}]{salanne_ionic_2018}%
  \BibitemOpen
  \bibfield  {author} {\bibinfo {author} {\bibfnamefont {M.}~\bibnamefont {Salanne}},\ }in\ \href {https://doi.org/10.1007/978-3-319-89794-3_2} {\emph {\bibinfo {booktitle} {Ionic {Liquids} {II}}}},\ \bibinfo {editor} {edited by\ \bibinfo {editor} {\bibfnamefont {B.}~\bibnamefont {Kirchner}}\ and\ \bibinfo {editor} {\bibfnamefont {E.}~\bibnamefont {Perlt}}}\ (\bibinfo  {publisher} {Springer International Publishing},\ \bibinfo {address} {Cham},\ \bibinfo {year} {2018})\ pp.\ \bibinfo {pages} {29--53}\BibitemShut {NoStop}%
\bibitem [{\citenamefont {Ray}\ and\ \citenamefont {Saruhan}(2021)}]{ray_application_2021}%
  \BibitemOpen
  \bibfield  {author} {\bibinfo {author} {\bibfnamefont {A.}~\bibnamefont {Ray}}\ and\ \bibinfo {author} {\bibfnamefont {B.}~\bibnamefont {Saruhan}},\ }\href {https://doi.org/10.3390/ma14112942} {\bibfield  {journal} {\bibinfo  {journal} {Materials}\ }\textbf {\bibinfo {volume} {14}},\ \bibinfo {pages} {2942} (\bibinfo {year} {2021})}\BibitemShut {NoStop}%
\bibitem [{\citenamefont {Merlet}\ \emph {et~al.}(2012)\citenamefont {Merlet}, \citenamefont {Salanne},\ and\ \citenamefont {Rotenberg}}]{merlet_new_2012}%
  \BibitemOpen
  \bibfield  {author} {\bibinfo {author} {\bibfnamefont {C.}~\bibnamefont {Merlet}}, \bibinfo {author} {\bibfnamefont {M.}~\bibnamefont {Salanne}},\ and\ \bibinfo {author} {\bibfnamefont {B.}~\bibnamefont {Rotenberg}},\ }\href {https://doi.org/10.1021/jp3008877} {\bibfield  {journal} {\bibinfo  {journal} {The Journal of Physical Chemistry C}\ }\textbf {\bibinfo {volume} {116}},\ \bibinfo {pages} {7687} (\bibinfo {year} {2012})}\BibitemShut {NoStop}%
\bibitem [{\citenamefont {Shamim}\ and\ \citenamefont {McKenna}(2010)}]{shamim_glass_2010}%
  \BibitemOpen
  \bibfield  {author} {\bibinfo {author} {\bibfnamefont {N.}~\bibnamefont {Shamim}}\ and\ \bibinfo {author} {\bibfnamefont {G.~B.}\ \bibnamefont {McKenna}},\ }\href {https://doi.org/10.1021/jp1044089} {\bibfield  {journal} {\bibinfo  {journal} {The Journal of Physical Chemistry B}\ }\textbf {\bibinfo {volume} {114}},\ \bibinfo {pages} {15742} (\bibinfo {year} {2010})}\BibitemShut {NoStop}%
\bibitem [{\citenamefont {Cerrillo}\ and\ \citenamefont {Cao}(2014)}]{cerrillo_non-markovian_2014}%
  \BibitemOpen
  \bibfield  {author} {\bibinfo {author} {\bibfnamefont {J.}~\bibnamefont {Cerrillo}}\ and\ \bibinfo {author} {\bibfnamefont {J.}~\bibnamefont {Cao}},\ }\href {https://doi.org/10.1103/PhysRevLett.112.110401} {\bibfield  {journal} {\bibinfo  {journal} {Physical Review Letters}\ }\textbf {\bibinfo {volume} {112}},\ \bibinfo {pages} {110401} (\bibinfo {year} {2014})}\BibitemShut {NoStop}%
\bibitem [{\citenamefont {Fong}\ \emph {et~al.}(2020)\citenamefont {Fong}, \citenamefont {Self}, \citenamefont {McCloskey},\ and\ \citenamefont {Persson}}]{fong_onsager_2020}%
  \BibitemOpen
  \bibfield  {author} {\bibinfo {author} {\bibfnamefont {K.~D.}\ \bibnamefont {Fong}}, \bibinfo {author} {\bibfnamefont {J.}~\bibnamefont {Self}}, \bibinfo {author} {\bibfnamefont {B.~D.}\ \bibnamefont {McCloskey}},\ and\ \bibinfo {author} {\bibfnamefont {K.~A.}\ \bibnamefont {Persson}},\ }\href {https://doi.org/10.1021/acs.macromol.0c02001} {\bibfield  {journal} {\bibinfo  {journal} {Macromolecules}\ }\textbf {\bibinfo {volume} {53}},\ \bibinfo {pages} {9503} (\bibinfo {year} {2020})}\BibitemShut {NoStop}%
\bibitem [{\citenamefont {Rusciano}\ \emph {et~al.}(2022)\citenamefont {Rusciano}, \citenamefont {Pastore},\ and\ \citenamefont {Greco}}]{rusciano_fickian_2022}%
  \BibitemOpen
  \bibfield  {author} {\bibinfo {author} {\bibfnamefont {F.}~\bibnamefont {Rusciano}}, \bibinfo {author} {\bibfnamefont {R.}~\bibnamefont {Pastore}},\ and\ \bibinfo {author} {\bibfnamefont {F.}~\bibnamefont {Greco}},\ }\href {https://doi.org/10.1103/PhysRevLett.128.168001} {\bibfield  {journal} {\bibinfo  {journal} {Physical Review Letters}\ }\textbf {\bibinfo {volume} {128}},\ \bibinfo {pages} {168001} (\bibinfo {year} {2022})}\BibitemShut {NoStop}%
\bibitem [{\citenamefont {de~Souza}\ and\ \citenamefont {Wales}(2008)}]{de_souza_energy_2008}%
  \BibitemOpen
  \bibfield  {author} {\bibinfo {author} {\bibfnamefont {V.~K.}\ \bibnamefont {de~Souza}}\ and\ \bibinfo {author} {\bibfnamefont {D.~J.}\ \bibnamefont {Wales}},\ }\href {https://doi.org/10.1063/1.2992128} {\bibfield  {journal} {\bibinfo  {journal} {The Journal of Chemical Physics}\ }\textbf {\bibinfo {volume} {129}},\ \bibinfo {pages} {164507} (\bibinfo {year} {2008})}\BibitemShut {NoStop}%
\bibitem [{\citenamefont {Roy}\ \emph {et~al.}(2010)\citenamefont {Roy}, \citenamefont {Patel}, \citenamefont {Conte},\ and\ \citenamefont {Maroncelli}}]{roy_dynamics_2010}%
  \BibitemOpen
  \bibfield  {author} {\bibinfo {author} {\bibfnamefont {D.}~\bibnamefont {Roy}}, \bibinfo {author} {\bibfnamefont {N.}~\bibnamefont {Patel}}, \bibinfo {author} {\bibfnamefont {S.}~\bibnamefont {Conte}},\ and\ \bibinfo {author} {\bibfnamefont {M.}~\bibnamefont {Maroncelli}},\ }\href {https://doi.org/10.1021/jp1004709} {\bibfield  {journal} {\bibinfo  {journal} {The Journal of Physical Chemistry B}\ }\textbf {\bibinfo {volume} {114}},\ \bibinfo {pages} {8410} (\bibinfo {year} {2010})}\BibitemShut {NoStop}%
\bibitem [{\citenamefont {Knorr}\ \emph {et~al.}(2016)\citenamefont {Knorr}, \citenamefont {Stange}, \citenamefont {Fumino}, \citenamefont {Weinhold},\ and\ \citenamefont {Ludwig}}]{knorr_spectroscopic_2016}%
  \BibitemOpen
  \bibfield  {author} {\bibinfo {author} {\bibfnamefont {A.}~\bibnamefont {Knorr}}, \bibinfo {author} {\bibfnamefont {P.}~\bibnamefont {Stange}}, \bibinfo {author} {\bibfnamefont {K.}~\bibnamefont {Fumino}}, \bibinfo {author} {\bibfnamefont {F.}~\bibnamefont {Weinhold}},\ and\ \bibinfo {author} {\bibfnamefont {R.}~\bibnamefont {Ludwig}},\ }\href {https://doi.org/10.1002/cphc.201501134} {\bibfield  {journal} {\bibinfo  {journal} {ChemPhysChem}\ }\textbf {\bibinfo {volume} {17}},\ \bibinfo {pages} {458} (\bibinfo {year} {2016})}\BibitemShut {NoStop}%
\bibitem [{\citenamefont {Shiraishi}\ \emph {et~al.}(2023)\citenamefont {Shiraishi}, \citenamefont {Mizuno},\ and\ \citenamefont {Ikeda}}]{shiraishi_joharigoldstein_2023}%
  \BibitemOpen
  \bibfield  {author} {\bibinfo {author} {\bibfnamefont {K.}~\bibnamefont {Shiraishi}}, \bibinfo {author} {\bibfnamefont {H.}~\bibnamefont {Mizuno}},\ and\ \bibinfo {author} {\bibfnamefont {A.}~\bibnamefont {Ikeda}},\ }\href {https://doi.org/10.1073/pnas.2215153120} {\bibfield  {journal} {\bibinfo  {journal} {Proceedings of the National Academy of Sciences}\ }\textbf {\bibinfo {volume} {120}},\ \bibinfo {pages} {e2215153120} (\bibinfo {year} {2023})}\BibitemShut {NoStop}%
\bibitem [{\citenamefont {Karmakar}(2016)}]{karmakar_overview_2016}%
  \BibitemOpen
  \bibfield  {author} {\bibinfo {author} {\bibfnamefont {S.}~\bibnamefont {Karmakar}},\ }\href {https://doi.org/10.1088/1742-6596/759/1/012008} {\bibfield  {journal} {\bibinfo  {journal} {Journal of Physics: Conference Series}\ }\textbf {\bibinfo {volume} {759}},\ \bibinfo {pages} {012008} (\bibinfo {year} {2016})}\BibitemShut {NoStop}%
\bibitem [{\citenamefont {Sha}\ \emph {et~al.}(2019)\citenamefont {Sha}, \citenamefont {Ma}, \citenamefont {Li}, \citenamefont {Luo}, \citenamefont {Zhu},\ and\ \citenamefont {Fayer}}]{sha_dynamical_2019}%
  \BibitemOpen
  \bibfield  {author} {\bibinfo {author} {\bibfnamefont {M.}~\bibnamefont {Sha}}, \bibinfo {author} {\bibfnamefont {X.}~\bibnamefont {Ma}}, \bibinfo {author} {\bibfnamefont {N.}~\bibnamefont {Li}}, \bibinfo {author} {\bibfnamefont {F.}~\bibnamefont {Luo}}, \bibinfo {author} {\bibfnamefont {G.}~\bibnamefont {Zhu}},\ and\ \bibinfo {author} {\bibfnamefont {M.~D.}\ \bibnamefont {Fayer}},\ }\href {https://doi.org/10.1063/1.5126231} {\bibfield  {journal} {\bibinfo  {journal} {The Journal of Chemical Physics}\ }\textbf {\bibinfo {volume} {151}},\ \bibinfo {pages} {154502} (\bibinfo {year} {2019})}\BibitemShut {NoStop}%
\bibitem [{\citenamefont {Debye}\ and\ \citenamefont {Falkenhagen}(1928)}]{debye_dispersion_1928}%
  \BibitemOpen
  \bibfield  {author} {\bibinfo {author} {\bibfnamefont {P.}~\bibnamefont {Debye}}\ and\ \bibinfo {author} {\bibfnamefont {H.}~\bibnamefont {Falkenhagen}},\ }\href@noop {} {\bibfield  {journal} {\bibinfo  {journal} {Phys. Z}\ }\textbf {\bibinfo {volume} {29}},\ \bibinfo {pages} {121} (\bibinfo {year} {1928})}\BibitemShut {NoStop}%
\bibitem [{\citenamefont {Lesikar}\ \emph {et~al.}(1980)\citenamefont {Lesikar}, \citenamefont {Simmons},\ and\ \citenamefont {Moynihan}}]{lesikar_debye-falkenhagen_1980}%
  \BibitemOpen
  \bibfield  {author} {\bibinfo {author} {\bibfnamefont {A.~V.}\ \bibnamefont {Lesikar}}, \bibinfo {author} {\bibfnamefont {C.~J.}\ \bibnamefont {Simmons}},\ and\ \bibinfo {author} {\bibfnamefont {C.~T.}\ \bibnamefont {Moynihan}},\ }\href {https://doi.org/10.1016/0022-3093(80)90101-5} {\bibfield  {journal} {\bibinfo  {journal} {Journal of Non-Crystalline Solids}\ }\bibinfo {series} {Proceedings of the {Fifth} {University} {Conference} on {Glass} {Science}},\ \textbf {\bibinfo {volume} {40}},\ \bibinfo {pages} {171} (\bibinfo {year} {1980})}\BibitemShut {NoStop}%
\bibitem [{\citenamefont {Anderson}(1994)}]{anderson_debye-falkenhagen_1994}%
  \BibitemOpen
  \bibfield  {author} {\bibinfo {author} {\bibfnamefont {J.~E.}\ \bibnamefont {Anderson}},\ }\href {https://doi.org/10.1016/0022-3093(94)90642-4} {\bibfield  {journal} {\bibinfo  {journal} {Journal of Non-Crystalline Solids}\ }\bibinfo {series} {Proceedings of the {Second} {Internatinal} {Discussion} {Meeting} on {Relaxations} in {Complex} {Systems}},\ \textbf {\bibinfo {volume} {172-174}},\ \bibinfo {pages} {1190} (\bibinfo {year} {1994})}\BibitemShut {NoStop}%
\bibitem [{\citenamefont {Ibuki}\ and\ \citenamefont {Nakahara}(1990)}]{ibuki_effect_1990}%
  \BibitemOpen
  \bibfield  {author} {\bibinfo {author} {\bibfnamefont {K.}~\bibnamefont {Ibuki}}\ and\ \bibinfo {author} {\bibfnamefont {M.}~\bibnamefont {Nakahara}},\ }\href {https://doi.org/10.1063/1.458217} {\bibfield  {journal} {\bibinfo  {journal} {The Journal of Chemical Physics}\ }\textbf {\bibinfo {volume} {92}},\ \bibinfo {pages} {7323} (\bibinfo {year} {1990})}\BibitemShut {NoStop}%
\bibitem [{\citenamefont {Bonneau}\ \emph {et~al.}(2024)\citenamefont {Bonneau}, \citenamefont {Avni}, \citenamefont {Andelman},\ and\ \citenamefont {Orland}}]{bonneau_frequency-dependent_2024}%
  \BibitemOpen
  \bibfield  {author} {\bibinfo {author} {\bibfnamefont {H.}~\bibnamefont {Bonneau}}, \bibinfo {author} {\bibfnamefont {Y.}~\bibnamefont {Avni}}, \bibinfo {author} {\bibfnamefont {D.}~\bibnamefont {Andelman}},\ and\ \bibinfo {author} {\bibfnamefont {H.}~\bibnamefont {Orland}},\ }\href {https://doi.org/10.1063/5.0236073} {\bibfield  {journal} {\bibinfo  {journal} {The Journal of Chemical Physics}\ }\textbf {\bibinfo {volume} {161}},\ \bibinfo {pages} {244501} (\bibinfo {year} {2024})}\BibitemShut {NoStop}%
\bibitem [{\citenamefont {Janssen}(2018)}]{janssen_mode-coupling_2018}%
  \BibitemOpen
  \bibfield  {author} {\bibinfo {author} {\bibfnamefont {L.~M.~C.}\ \bibnamefont {Janssen}},\ }\bibfield  {journal} {\bibinfo  {journal} {Frontiers in Physics}\ }\textbf {\bibinfo {volume} {6}},\ \href {https://doi.org/10.3389/fphy.2018.00097} {10.3389/fphy.2018.00097} (\bibinfo {year} {2018})\BibitemShut {NoStop}%
\bibitem [{\citenamefont {Canales}\ and\ \citenamefont {Sesé}(1998)}]{canales_generalized_1998}%
  \BibitemOpen
  \bibfield  {author} {\bibinfo {author} {\bibfnamefont {M.}~\bibnamefont {Canales}}\ and\ \bibinfo {author} {\bibfnamefont {G.}~\bibnamefont {Sesé}},\ }\href {https://doi.org/10.1063/1.477226} {\bibfield  {journal} {\bibinfo  {journal} {The Journal of Chemical Physics}\ }\textbf {\bibinfo {volume} {109}},\ \bibinfo {pages} {6004} (\bibinfo {year} {1998})}\BibitemShut {NoStop}%
\bibitem [{Note2()}]{Note2}%
  \BibitemOpen
  \bibinfo {note} {This is because the random force remains in the complementary subspace of $\protect \bm {\protect \mathit {v}}_i$ at all times, while the velocities of all the other ions $\protect \bm {\protect \mathit {v}}_j$ start in the complementary subspace and only partial enter the projected space under time evolution.}\BibitemShut {Stop}%
\bibitem [{\citenamefont {Carof}\ \emph {et~al.}(2014)\citenamefont {Carof}, \citenamefont {Vuilleumier},\ and\ \citenamefont {Rotenberg}}]{carof_two_2014}%
  \BibitemOpen
  \bibfield  {author} {\bibinfo {author} {\bibfnamefont {A.}~\bibnamefont {Carof}}, \bibinfo {author} {\bibfnamefont {R.}~\bibnamefont {Vuilleumier}},\ and\ \bibinfo {author} {\bibfnamefont {B.}~\bibnamefont {Rotenberg}},\ }\href {https://doi.org/10.1063/1.4868653} {\bibfield  {journal} {\bibinfo  {journal} {The Journal of Chemical Physics}\ }\textbf {\bibinfo {volume} {140}},\ \bibinfo {pages} {124103} (\bibinfo {year} {2014})}\BibitemShut {NoStop}%
\bibitem [{\citenamefont {Jung}\ \emph {et~al.}(2017)\citenamefont {Jung}, \citenamefont {Hanke},\ and\ \citenamefont {Schmid}}]{jung_iterative_2017}%
  \BibitemOpen
  \bibfield  {author} {\bibinfo {author} {\bibfnamefont {G.}~\bibnamefont {Jung}}, \bibinfo {author} {\bibfnamefont {M.}~\bibnamefont {Hanke}},\ and\ \bibinfo {author} {\bibfnamefont {F.}~\bibnamefont {Schmid}},\ }\href {https://doi.org/10.1021/acs.jctc.7b00274} {\bibfield  {journal} {\bibinfo  {journal} {Journal of Chemical Theory and Computation}\ }\textbf {\bibinfo {volume} {13}},\ \bibinfo {pages} {2481} (\bibinfo {year} {2017})}\BibitemShut {NoStop}%
\bibitem [{\citenamefont {Daldrop}\ \emph {et~al.}(2018)\citenamefont {Daldrop}, \citenamefont {Kappler}, \citenamefont {Brünig},\ and\ \citenamefont {Netz}}]{daldrop_butane_2018}%
  \BibitemOpen
  \bibfield  {author} {\bibinfo {author} {\bibfnamefont {J.~O.}\ \bibnamefont {Daldrop}}, \bibinfo {author} {\bibfnamefont {J.}~\bibnamefont {Kappler}}, \bibinfo {author} {\bibfnamefont {F.~N.}\ \bibnamefont {Brünig}},\ and\ \bibinfo {author} {\bibfnamefont {R.~R.}\ \bibnamefont {Netz}},\ }\href {https://doi.org/10.1073/pnas.1722327115} {\bibfield  {journal} {\bibinfo  {journal} {Proceedings of the National Academy of Sciences}\ }\textbf {\bibinfo {volume} {115}},\ \bibinfo {pages} {5169} (\bibinfo {year} {2018})}\BibitemShut {NoStop}%
\bibitem [{\citenamefont {Sears}(1965)}]{sears_itinerant_1965}%
  \BibitemOpen
  \bibfield  {author} {\bibinfo {author} {\bibfnamefont {V.~F.}\ \bibnamefont {Sears}},\ }\href {https://doi.org/10.1088/0370-1328/86/5/306} {\bibfield  {journal} {\bibinfo  {journal} {Proceedings of the Physical Society}\ }\textbf {\bibinfo {volume} {86}},\ \bibinfo {pages} {953} (\bibinfo {year} {1965})}\BibitemShut {NoStop}%
\bibitem [{\citenamefont {Damle}\ \emph {et~al.}(1968)\citenamefont {Damle}, \citenamefont {Sjölander},\ and\ \citenamefont {Singwi}}]{damle_itinerant-oscillator_1968-1}%
  \BibitemOpen
  \bibfield  {author} {\bibinfo {author} {\bibfnamefont {P.~S.}\ \bibnamefont {Damle}}, \bibinfo {author} {\bibfnamefont {A.}~\bibnamefont {Sjölander}},\ and\ \bibinfo {author} {\bibfnamefont {K.~S.}\ \bibnamefont {Singwi}},\ }\href {https://doi.org/10.1103/PhysRev.165.277} {\bibfield  {journal} {\bibinfo  {journal} {Physical Review}\ }\textbf {\bibinfo {volume} {165}},\ \bibinfo {pages} {277} (\bibinfo {year} {1968})}\BibitemShut {NoStop}%
\bibitem [{\citenamefont {Netz}(2026)}]{netz_barrier_crossing_2026}%
  \BibitemOpen
  \bibfield  {author} {\bibinfo {author} {\bibfnamefont {R.~R.}\ \bibnamefont {Netz}},\ }\href {http://arxiv.org/abs/2601.09861} {\bibfield  {journal} {\bibinfo  {journal} {arXiv:2601.09861 [cond-mat]}\ } (\bibinfo {year} {2026})}\BibitemShut {NoStop}%
\bibitem [{\citenamefont {Kudlik}\ \emph {et~al.}(1999)\citenamefont {Kudlik}, \citenamefont {Benkhof}, \citenamefont {Blochowicz}, \citenamefont {Tschirwitz},\ and\ \citenamefont {Rössler}}]{kudlik_dielectric_1999}%
  \BibitemOpen
  \bibfield  {author} {\bibinfo {author} {\bibfnamefont {A.}~\bibnamefont {Kudlik}}, \bibinfo {author} {\bibfnamefont {S.}~\bibnamefont {Benkhof}}, \bibinfo {author} {\bibfnamefont {T.}~\bibnamefont {Blochowicz}}, \bibinfo {author} {\bibfnamefont {C.}~\bibnamefont {Tschirwitz}},\ and\ \bibinfo {author} {\bibfnamefont {E.}~\bibnamefont {Rössler}},\ }\href {https://doi.org/10.1016/S0022-2860(98)00871-0} {\bibfield  {journal} {\bibinfo  {journal} {Journal of Molecular Structure}\ }\textbf {\bibinfo {volume} {479}},\ \bibinfo {pages} {201} (\bibinfo {year} {1999})}\BibitemShut {NoStop}%
\bibitem [{\citenamefont {Lasia}(2022)}]{lasia_origin_2022}%
  \BibitemOpen
  \bibfield  {author} {\bibinfo {author} {\bibfnamefont {A.}~\bibnamefont {Lasia}},\ }\href {https://doi.org/10.1021/acs.jpclett.1c03782} {\bibfield  {journal} {\bibinfo  {journal} {The Journal of Physical Chemistry Letters}\ }\textbf {\bibinfo {volume} {13}},\ \bibinfo {pages} {580} (\bibinfo {year} {2022})}\BibitemShut {NoStop}%
\bibitem [{\citenamefont {Illien}(2025)}]{illien_deankawasaki_2025}%
  \BibitemOpen
  \bibfield  {author} {\bibinfo {author} {\bibfnamefont {P.}~\bibnamefont {Illien}},\ }\href {https://doi.org/10.1088/1361-6633/adee2e} {\bibfield  {journal} {\bibinfo  {journal} {Reports on Progress in Physics}\ }\textbf {\bibinfo {volume} {88}},\ \bibinfo {pages} {086601} (\bibinfo {year} {2025})}\BibitemShut {NoStop}%
\bibitem [{\citenamefont {Illien}\ \emph {et~al.}(2024)\citenamefont {Illien}, \citenamefont {Carof},\ and\ \citenamefont {Rotenberg}}]{illien_stochastic_2024}%
  \BibitemOpen
  \bibfield  {author} {\bibinfo {author} {\bibfnamefont {P.}~\bibnamefont {Illien}}, \bibinfo {author} {\bibfnamefont {A.}~\bibnamefont {Carof}},\ and\ \bibinfo {author} {\bibfnamefont {B.}~\bibnamefont {Rotenberg}},\ }\href {https://doi.org/10.1103/PhysRevLett.133.268002} {\bibfield  {journal} {\bibinfo  {journal} {Physical Review Letters}\ }\textbf {\bibinfo {volume} {133}},\ \bibinfo {pages} {268002} (\bibinfo {year} {2024})}\BibitemShut {NoStop}%
\bibitem [{\citenamefont {Varghese}\ \emph {et~al.}(2025)\citenamefont {Varghese}, \citenamefont {Illien},\ and\ \citenamefont {Rotenberg}}]{varghese_dynamic_2025}%
  \BibitemOpen
  \bibfield  {author} {\bibinfo {author} {\bibfnamefont {S.}~\bibnamefont {Varghese}}, \bibinfo {author} {\bibfnamefont {P.}~\bibnamefont {Illien}},\ and\ \bibinfo {author} {\bibfnamefont {B.}~\bibnamefont {Rotenberg}},\ }\href {https://doi.org/10.1063/5.0292306} {\bibfield  {journal} {\bibinfo  {journal} {The Journal of Chemical Physics}\ }\textbf {\bibinfo {volume} {163}},\ \bibinfo {pages} {124107} (\bibinfo {year} {2025})}\BibitemShut {NoStop}%
\bibitem [{\citenamefont {Varghese}\ \emph {et~al.}(2026)\citenamefont {Varghese}, \citenamefont {Rotenberg},\ and\ \citenamefont {Illien}}]{varghese_solvent-induced_2026}%
  \BibitemOpen
  \bibfield  {author} {\bibinfo {author} {\bibfnamefont {S.}~\bibnamefont {Varghese}}, \bibinfo {author} {\bibfnamefont {B.}~\bibnamefont {Rotenberg}},\ and\ \bibinfo {author} {\bibfnamefont {P.}~\bibnamefont {Illien}},\ }\href {http://arxiv.org/abs/2605.06293} {\bibfield  {journal} {\bibinfo  {journal} {arXiv:2605.06293 [cond-mat.soft]}\ } (\bibinfo {year} {2026})}\BibitemShut {NoStop}%
\bibitem [{\citenamefont {Fedorov}\ and\ \citenamefont {Kornyshev}(2014)}]{fedorov_ionic_2014}%
  \BibitemOpen
  \bibfield  {author} {\bibinfo {author} {\bibfnamefont {M.~V.}\ \bibnamefont {Fedorov}}\ and\ \bibinfo {author} {\bibfnamefont {A.~A.}\ \bibnamefont {Kornyshev}},\ }\href {https://doi.org/10.1021/cr400374x} {\bibfield  {journal} {\bibinfo  {journal} {Chemical Reviews}\ }\textbf {\bibinfo {volume} {114}},\ \bibinfo {pages} {2978} (\bibinfo {year} {2014})}\BibitemShut {NoStop}%
\bibitem [{\citenamefont {Borghi}\ and\ \citenamefont {Podestà}(2020)}]{borghi_ionic_2020}%
  \BibitemOpen
  \bibfield  {author} {\bibinfo {author} {\bibfnamefont {F.}~\bibnamefont {Borghi}}\ and\ \bibinfo {author} {\bibfnamefont {A.}~\bibnamefont {Podestà}},\ }\href {https://doi.org/10.1080/23746149.2020.1736949} {\bibfield  {journal} {\bibinfo  {journal} {Advances in Physics: X}\ }\textbf {\bibinfo {volume} {5}},\ \bibinfo {pages} {1736949} (\bibinfo {year} {2020})}\BibitemShut {NoStop}%
\bibitem [{\citenamefont {Kritikos}\ \emph {et~al.}(2016)\citenamefont {Kritikos}, \citenamefont {Vergadou},\ and\ \citenamefont {Economou}}]{kritikos_molecular_2016}%
  \BibitemOpen
  \bibfield  {author} {\bibinfo {author} {\bibfnamefont {G.}~\bibnamefont {Kritikos}}, \bibinfo {author} {\bibfnamefont {N.}~\bibnamefont {Vergadou}},\ and\ \bibinfo {author} {\bibfnamefont {I.~G.}\ \bibnamefont {Economou}},\ }\href {https://doi.org/10.1021/acs.jpcc.5b09947} {\bibfield  {journal} {\bibinfo  {journal} {The Journal of Physical Chemistry C}\ }\textbf {\bibinfo {volume} {120}},\ \bibinfo {pages} {1013} (\bibinfo {year} {2016})}\BibitemShut {NoStop}%
\bibitem [{\citenamefont {Hayes}\ \emph {et~al.}(2010)\citenamefont {Hayes}, \citenamefont {Warr},\ and\ \citenamefont {Atkin}}]{hayes_at_2010}%
  \BibitemOpen
  \bibfield  {author} {\bibinfo {author} {\bibfnamefont {R.}~\bibnamefont {Hayes}}, \bibinfo {author} {\bibfnamefont {G.~G.}\ \bibnamefont {Warr}},\ and\ \bibinfo {author} {\bibfnamefont {R.}~\bibnamefont {Atkin}},\ }\href {https://doi.org/10.1039/B920393A} {\bibfield  {journal} {\bibinfo  {journal} {Physical Chemistry Chemical Physics}\ }\textbf {\bibinfo {volume} {12}},\ \bibinfo {pages} {1709} (\bibinfo {year} {2010})}\BibitemShut {NoStop}%
\bibitem [{\citenamefont {Merlet}\ \emph {et~al.}(2014)\citenamefont {Merlet}, \citenamefont {Limmer}, \citenamefont {Salanne}, \citenamefont {van Roij}, \citenamefont {Madden}, \citenamefont {Chandler},\ and\ \citenamefont {Rotenberg}}]{merlet_electric_2014}%
  \BibitemOpen
  \bibfield  {author} {\bibinfo {author} {\bibfnamefont {C.}~\bibnamefont {Merlet}}, \bibinfo {author} {\bibfnamefont {D.~T.}\ \bibnamefont {Limmer}}, \bibinfo {author} {\bibfnamefont {M.}~\bibnamefont {Salanne}}, \bibinfo {author} {\bibfnamefont {R.}~\bibnamefont {van Roij}}, \bibinfo {author} {\bibfnamefont {P.~A.}\ \bibnamefont {Madden}}, \bibinfo {author} {\bibfnamefont {D.}~\bibnamefont {Chandler}},\ and\ \bibinfo {author} {\bibfnamefont {B.}~\bibnamefont {Rotenberg}},\ }\href {https://doi.org/10.1021/jp503224w} {\bibfield  {journal} {\bibinfo  {journal} {The Journal of Physical Chemistry C}\ }\textbf {\bibinfo {volume} {118}},\ \bibinfo {pages} {18291} (\bibinfo {year} {2014})}\BibitemShut {NoStop}%
\bibitem [{\citenamefont {Ntim}\ and\ \citenamefont {Sulpizi}(2020)}]{ntim_role_2020}%
  \BibitemOpen
  \bibfield  {author} {\bibinfo {author} {\bibfnamefont {S.}~\bibnamefont {Ntim}}\ and\ \bibinfo {author} {\bibfnamefont {M.}~\bibnamefont {Sulpizi}},\ }\href {https://doi.org/10.1039/D0CP00409J} {\bibfield  {journal} {\bibinfo  {journal} {Physical Chemistry Chemical Physics}\ }\textbf {\bibinfo {volume} {22}},\ \bibinfo {pages} {10786} (\bibinfo {year} {2020})}\BibitemShut {NoStop}%
\bibitem [{\citenamefont {Lhermerout}\ and\ \citenamefont {Perkin}(2018)}]{lhermerout_nanoconfined_2018}%
  \BibitemOpen
  \bibfield  {author} {\bibinfo {author} {\bibfnamefont {R.}~\bibnamefont {Lhermerout}}\ and\ \bibinfo {author} {\bibfnamefont {S.}~\bibnamefont {Perkin}},\ }\href {https://doi.org/10.1103/PhysRevFluids.3.014201} {\bibfield  {journal} {\bibinfo  {journal} {Physical Review Fluids}\ }\textbf {\bibinfo {volume} {3}},\ \bibinfo {pages} {014201} (\bibinfo {year} {2018})}\BibitemShut {NoStop}%
\bibitem [{\citenamefont {Dufils}\ \emph {et~al.}(2021)\citenamefont {Dufils}, \citenamefont {Sprik},\ and\ \citenamefont {Salanne}}]{dufils_computational_2021}%
  \BibitemOpen
  \bibfield  {author} {\bibinfo {author} {\bibfnamefont {T.}~\bibnamefont {Dufils}}, \bibinfo {author} {\bibfnamefont {M.}~\bibnamefont {Sprik}},\ and\ \bibinfo {author} {\bibfnamefont {M.}~\bibnamefont {Salanne}},\ }\href {https://doi.org/10.1021/acs.jpclett.1c01131} {\bibfield  {journal} {\bibinfo  {journal} {The Journal of Physical Chemistry Letters}\ }\textbf {\bibinfo {volume} {12}},\ \bibinfo {pages} {4357} (\bibinfo {year} {2021})}\BibitemShut {NoStop}%
\bibitem [{\citenamefont {Cross}\ \emph {et~al.}(2026)\citenamefont {Cross}, \citenamefont {Garcia}, \citenamefont {Charlaix},\ and\ \citenamefont {Kékicheff}}]{cross_short-range_2026}%
  \BibitemOpen
  \bibfield  {author} {\bibinfo {author} {\bibfnamefont {B.}~\bibnamefont {Cross}}, \bibinfo {author} {\bibfnamefont {L.}~\bibnamefont {Garcia}}, \bibinfo {author} {\bibfnamefont {E.}~\bibnamefont {Charlaix}},\ and\ \bibinfo {author} {\bibfnamefont {P.}~\bibnamefont {Kékicheff}},\ }\href {https://doi.org/10.1073/pnas.2517939123} {\bibfield  {journal} {\bibinfo  {journal} {Proceedings of the National Academy of Sciences}\ }\textbf {\bibinfo {volume} {123}},\ \bibinfo {pages} {e2517939123} (\bibinfo {year} {2026})}\BibitemShut {NoStop}%
\bibitem [{\citenamefont {Andersen}(1983)}]{andersen_rattle_1983}%
  \BibitemOpen
  \bibfield  {author} {\bibinfo {author} {\bibfnamefont {H.~C.}\ \bibnamefont {Andersen}},\ }\href {https://doi.org/10.1016/0021-9991(83)90014-1} {\bibfield  {journal} {\bibinfo  {journal} {Journal of Computational Physics}\ }\textbf {\bibinfo {volume} {52}},\ \bibinfo {pages} {24} (\bibinfo {year} {1983})}\BibitemShut {NoStop}%
\bibitem [{\citenamefont {Thompson}\ \emph {et~al.}(2022)\citenamefont {Thompson}, \citenamefont {Aktulga}, \citenamefont {Berger}, \citenamefont {Bolintineanu}, \citenamefont {Brown}, \citenamefont {Crozier}, \citenamefont {in~'t Veld}, \citenamefont {Kohlmeyer}, \citenamefont {Moore}, \citenamefont {Nguyen}, \citenamefont {Shan}, \citenamefont {Stevens}, \citenamefont {Tranchida}, \citenamefont {Trott},\ and\ \citenamefont {Plimpton}}]{thompson_lammps_2022}%
  \BibitemOpen
  \bibfield  {author} {\bibinfo {author} {\bibfnamefont {A.~P.}\ \bibnamefont {Thompson}}, \bibinfo {author} {\bibfnamefont {H.~M.}\ \bibnamefont {Aktulga}}, \bibinfo {author} {\bibfnamefont {R.}~\bibnamefont {Berger}}, \bibinfo {author} {\bibfnamefont {D.~S.}\ \bibnamefont {Bolintineanu}}, \bibinfo {author} {\bibfnamefont {W.~M.}\ \bibnamefont {Brown}}, \bibinfo {author} {\bibfnamefont {P.~S.}\ \bibnamefont {Crozier}}, \bibinfo {author} {\bibfnamefont {P.~J.}\ \bibnamefont {in~'t Veld}}, \bibinfo {author} {\bibfnamefont {A.}~\bibnamefont {Kohlmeyer}}, \bibinfo {author} {\bibfnamefont {S.~G.}\ \bibnamefont {Moore}}, \bibinfo {author} {\bibfnamefont {T.~D.}\ \bibnamefont {Nguyen}}, \bibinfo {author} {\bibfnamefont {R.}~\bibnamefont {Shan}}, \bibinfo {author} {\bibfnamefont {M.~J.}\ \bibnamefont {Stevens}}, \bibinfo {author} {\bibfnamefont {J.}~\bibnamefont {Tranchida}}, \bibinfo {author} {\bibfnamefont {C.}~\bibnamefont {Trott}},\ and\ \bibinfo {author} {\bibfnamefont {S.~J.}\ \bibnamefont {Plimpton}},\ }\href
  {https://doi.org/10.1016/j.cpc.2021.108171} {\bibfield  {journal} {\bibinfo  {journal} {Computer Physics Communications}\ }\textbf {\bibinfo {volume} {271}},\ \bibinfo {pages} {108171} (\bibinfo {year} {2022})}\BibitemShut {NoStop}%
\bibitem [{\citenamefont {Cox}\ and\ \citenamefont {Sprik}(2019)}]{cox_finite_2019}%
  \BibitemOpen
  \bibfield  {author} {\bibinfo {author} {\bibfnamefont {S.~J.}\ \bibnamefont {Cox}}\ and\ \bibinfo {author} {\bibfnamefont {M.}~\bibnamefont {Sprik}},\ }\href {https://doi.org/10.1063/1.5099207} {\bibfield  {journal} {\bibinfo  {journal} {The Journal of Chemical Physics}\ }\textbf {\bibinfo {volume} {151}},\ \bibinfo {pages} {064506} (\bibinfo {year} {2019})}\BibitemShut {NoStop}%
\bibitem [{\citenamefont {Caillol}(1994)}]{caillol_comments_1994}%
  \BibitemOpen
  \bibfield  {author} {\bibinfo {author} {\bibfnamefont {J.}~\bibnamefont {Caillol}},\ }\href {https://doi.org/10.1063/1.468422} {\bibfield  {journal} {\bibinfo  {journal} {The Journal of Chemical Physics}\ }\textbf {\bibinfo {volume} {101}},\ \bibinfo {pages} {6080} (\bibinfo {year} {1994})}\BibitemShut {NoStop}%
\bibitem [{\citenamefont {Wan}\ \emph {et~al.}(2015)\citenamefont {Wan}, \citenamefont {Saccoccio}, \citenamefont {Chen},\ and\ \citenamefont {Ciucci}}]{wan_influence_2015}%
  \BibitemOpen
  \bibfield  {author} {\bibinfo {author} {\bibfnamefont {T.~H.}\ \bibnamefont {Wan}}, \bibinfo {author} {\bibfnamefont {M.}~\bibnamefont {Saccoccio}}, \bibinfo {author} {\bibfnamefont {C.}~\bibnamefont {Chen}},\ and\ \bibinfo {author} {\bibfnamefont {F.}~\bibnamefont {Ciucci}},\ }\href {https://doi.org/10.1016/j.electacta.2015.09.097} {\bibfield  {journal} {\bibinfo  {journal} {Electrochimica Acta}\ }\textbf {\bibinfo {volume} {184}},\ \bibinfo {pages} {483} (\bibinfo {year} {2015})}\BibitemShut {NoStop}%
\bibitem [{\citenamefont {Wang}\ \emph {et~al.}(2025)\citenamefont {Wang}, \citenamefont {Wang}, \citenamefont {Py}, \citenamefont {Maradesa}, \citenamefont {Liu}, \citenamefont {Wan}, \citenamefont {Saccoccio},\ and\ \citenamefont {Ciucci}}]{wang_drttools_2025}%
  \BibitemOpen
  \bibfield  {author} {\bibinfo {author} {\bibfnamefont {Z.}~\bibnamefont {Wang}}, \bibinfo {author} {\bibfnamefont {Y.}~\bibnamefont {Wang}}, \bibinfo {author} {\bibfnamefont {B.}~\bibnamefont {Py}}, \bibinfo {author} {\bibfnamefont {A.}~\bibnamefont {Maradesa}}, \bibinfo {author} {\bibfnamefont {J.}~\bibnamefont {Liu}}, \bibinfo {author} {\bibfnamefont {T.~H.}\ \bibnamefont {Wan}}, \bibinfo {author} {\bibfnamefont {M.}~\bibnamefont {Saccoccio}},\ and\ \bibinfo {author} {\bibfnamefont {F.}~\bibnamefont {Ciucci}},\ }\href {https://doi.org/10.1021/acselectrochem.5c00334} {\bibfield  {journal} {\bibinfo  {journal} {ACS Electrochemistry}\ }\textbf {\bibinfo {volume} {1}},\ \bibinfo {pages} {2680} (\bibinfo {year} {2025})}\BibitemShut {NoStop}%
\end{thebibliography}%

\appendix
\setcounter{figure}{0}
\section{Fitting to an RC model}\label{app:RCfit}
\vspace{-8pt}

\renewcommand{\thefigure}{\thesection\arabic{figure}}
\begin{figure*}[!ht]
     \centering
     \begin{subfigure}[b]{0.49\textwidth}
         \centering
         \includegraphics[width=\textwidth]{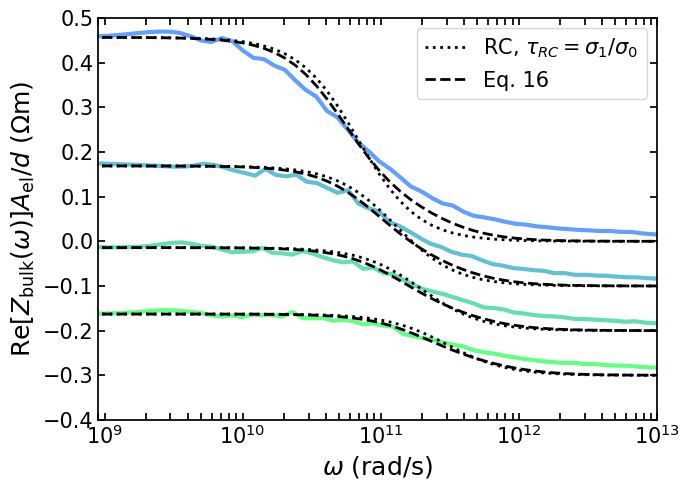}
         \caption{}
     \end{subfigure}
     \hfill
     \begin{subfigure}[b]{0.49\textwidth}
         \centering
         \includegraphics[width=\textwidth]{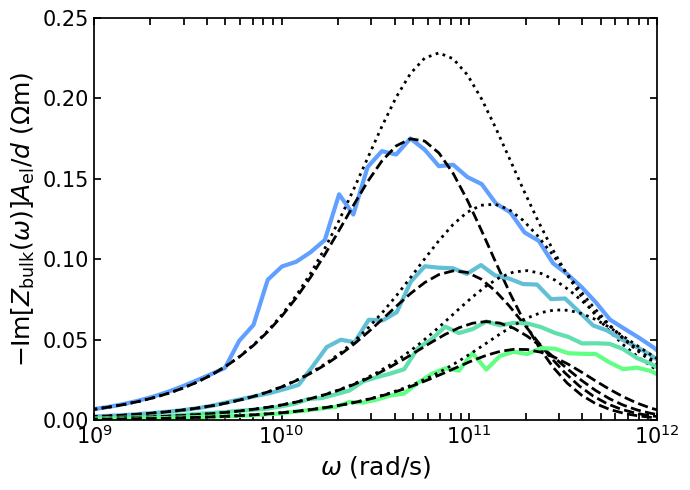}
         \caption{}
     \end{subfigure}
        \caption{\label{fig:ReImZ}Independent \textbf{(a)} real and \textbf{(b)} imaginary parts of the impedance data presented together in Fig.~\ref{fig:Nyquist}. From Eq.~\ref{eqn:ZRC}, the RC model's timescale, $\tau_{RC}$, is defined by the position of the maximum in the imaginary part. As the maxima from defining $\tau_{RC}=\sigma_1/\sigma_0$ are too high in frequency compared to the simulation data (worsening with increasing temperature), the quotient as obtained from a fit would be wrong. The resistance prefactor in Eq.~\ref{eqn:ZRC} determines the height of both real and imaginary parts, which means the good agreement in panel (a)---which gives $\sigma_0$---is inconsistent due to the poor agreement of the same quantity in panel (b). That is, since truncation of $\sigma(\omega)$ at first order is a poor approximation, the RC fit cannot extract faithful values of $\sigma_1$ and $\sigma_0$.}
\end{figure*}

Figure~\ref{fig:ReImZ} shows both the real and imaginary parts of the impedance that have been previously plotted in Fig.~\ref{fig:Nyquist} but now resolved in frequency-space, alongside the results for an RC representation, and our new model (Eq.~\ref{eqn:Zbulk 2order}). Figure~\ref{fig:ReImZ}(b) highlights that the maximum in $\operatorname{Im}[Z_{\mrm{bulk}}(\omega)]$ found from simulation, and the resulting maximum in $\operatorname{Im}[Z_{\mrm{RC}}(\omega)]$ when $\tau_{RC}=\sigma_1/\sigma_0$ do not align, therefore we can state that a value of $\tau_{RC}$ found from a fit to the data would not be equal to $\sigma_1/\sigma_0$.

\section{Distribution of Relaxation Times (DRT) Analysis}\label{app:DRT}
\vspace{-8pt}
As is often common practice, we have gone on to convert our impedance data into a DRT spectrum. DRT is an analysis method by which the impedance as a function of frequency of an electrochemical device is converted into
a function of the distribution of time constants within the system (Eq.~\ref{eqn:DRT}). The result is a spectrum of peaks, each peak occurring at a particular $\tau$, corresponding to a timescale of a chemical or physical process occurring in the system.

\begin{figure*}[!ht]
     \centering
     \begin{subfigure}[b]{0.49\textwidth}
         \centering
         \includegraphics[width=\textwidth]{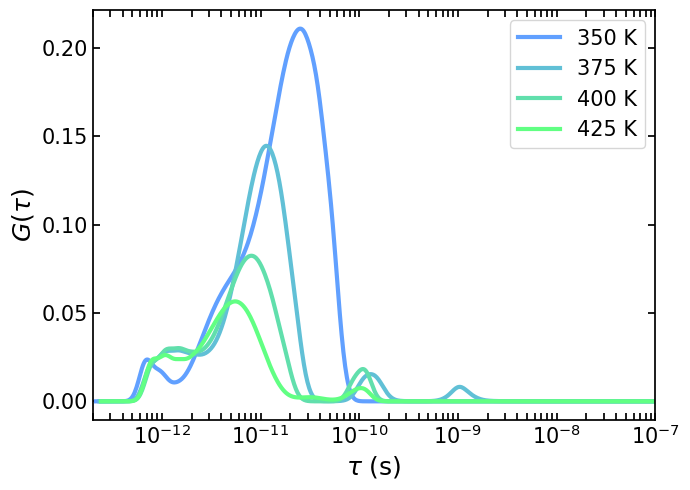}
         \caption{}
     \end{subfigure}
     \hfill
     \begin{subfigure}[b]{0.49\textwidth}
         \centering
         \includegraphics[width=\textwidth]{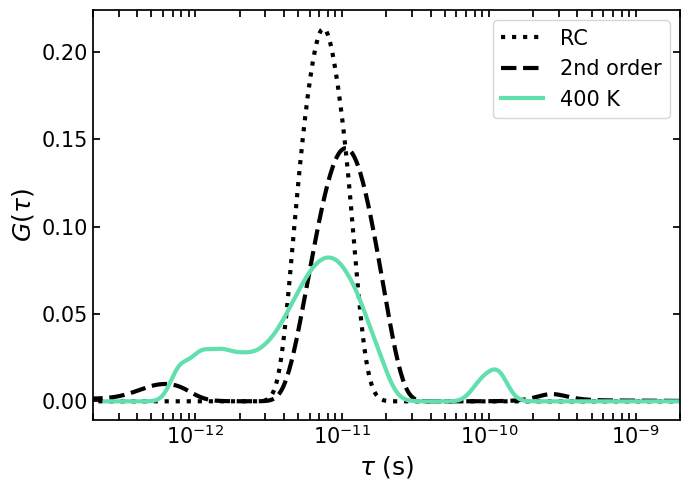}
         \caption{}
     \end{subfigure}
        \caption{\label{fig:DRT}\textbf{(a)} DRT deconvolution as described in Eq.~\ref{eqn:DRT} for the $Z_{\mathrm{bulk}}(\omega)$ simulation data in Fig.~\ref{fig:Nyquist}. This plot clearly shows that at all temperatures this procedure produces a minimum of 3 peaks, corresponding to at least 3 different RC-like timescales. Crucially a Cole-Cole or Cole-Davidson model, which would be the standard choice for modelling these data, would only produce a single, broad peak. However \textbf{(b)} shows that our new impedance model (Eq.~\ref{eqn:Zbulk 2order}) also contains multiple timescales which themselves agree fairly well with the simulation result despite the very simple model used to produce them. This calculation was done using pyDRTtools using the optimal regularization parameter \cite{wan_influence_2015, wang_drttools_2025}.}
\end{figure*}

We note that the standard RC kernel used in Eq.~\ref{eqn:DRT} enforces the assumption that our impedance data is composed of multiple independent RC-like processes. While in the Main Text we have argued this not to be true, as this is the response of a bulk electrolyte with no spatial heterogeneity, we use this method simply to reinforce this point. While one can model these data using a superposition of RC circuits and produce a pleasing fit, the timescales found in Fig.~\ref{fig:DRT}(a) when compared to those in Fig.~\ref{fig:timescales}(b)~and~(c) do not align. Therefore DRT does not effectively report on the microscopic timescales found in this work.

Figure~\ref{fig:DRT}(b) shows how closely a second-order impedance model and our simulation data resemble each other when analysed using DRT. However using a single second-order model to fit our data will produce fitting parameters that directly report on the underlying transport mechanisms, whereas when using a superposition of 3 RC models, as this DRT calculation suggests, we lose all microscopic interpretability. 
\section{Determination of single-ion memory functions}\label{app:cross_terms}
\vspace{-8pt}

\begin{figure*}[!ht]
     \centering
     \begin{subfigure}[b]{0.49\textwidth}
         \centering
         \includegraphics[width=\textwidth]{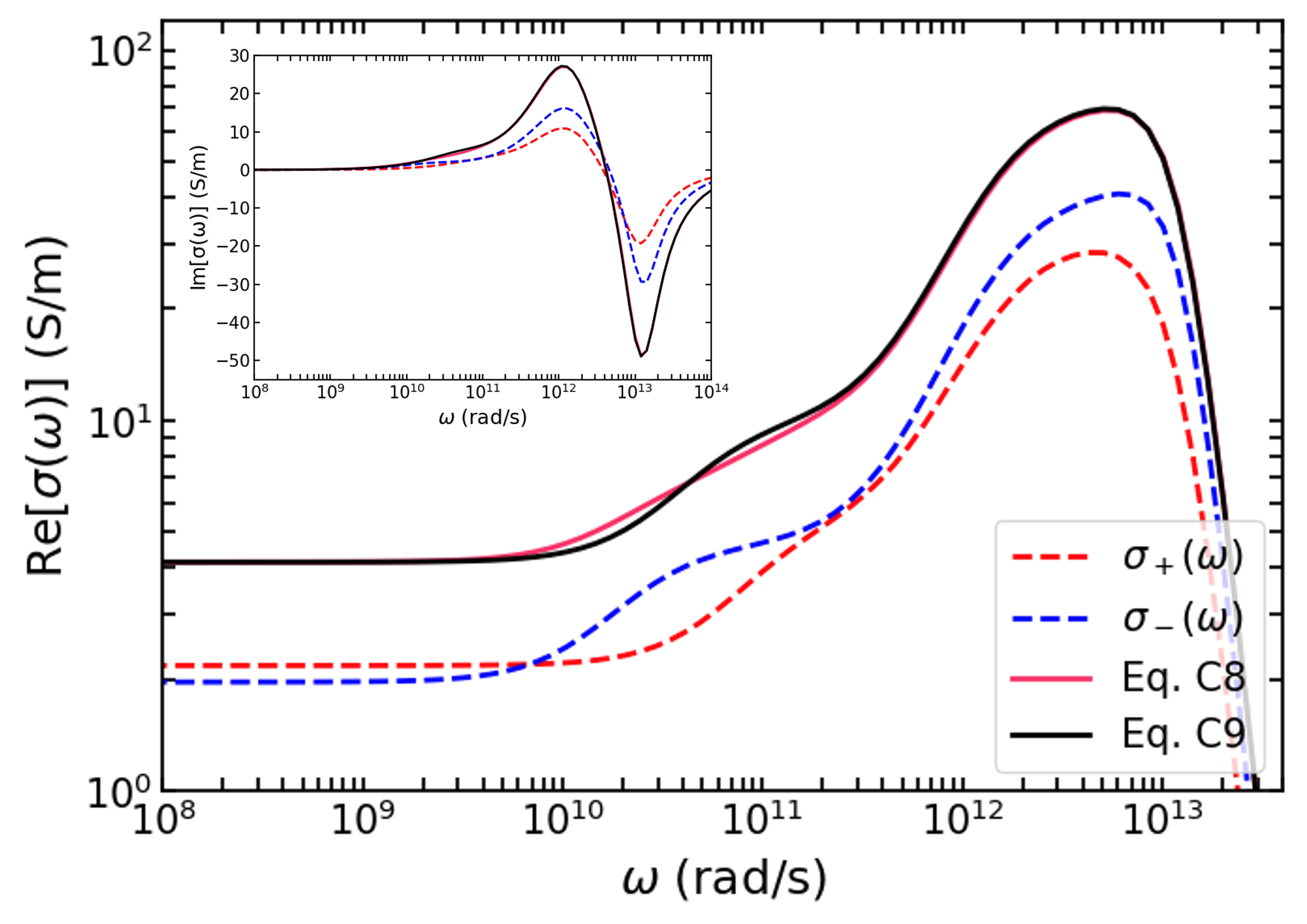}
         \caption{}
     \end{subfigure}
     \hfill
     \begin{subfigure}[b]{0.49\textwidth}
         \centering
         \includegraphics[width=\textwidth]{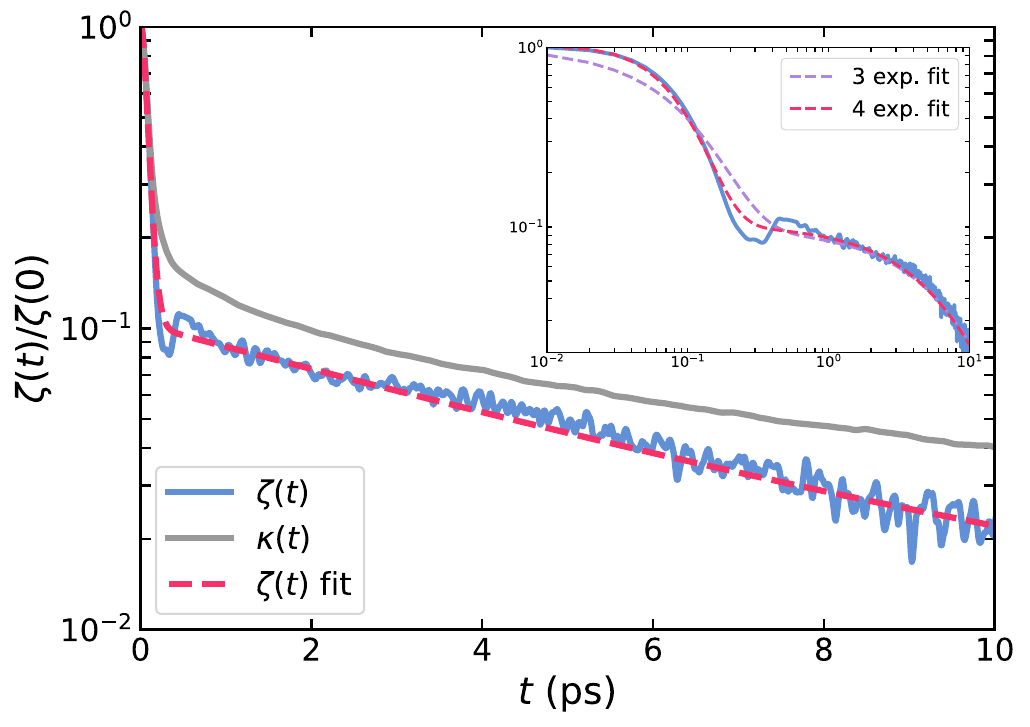}
         \caption{}
     \end{subfigure}
     \hfill
     \begin{subfigure}[b]{0.49\textwidth}
         \centering
         \includegraphics[width=\textwidth]{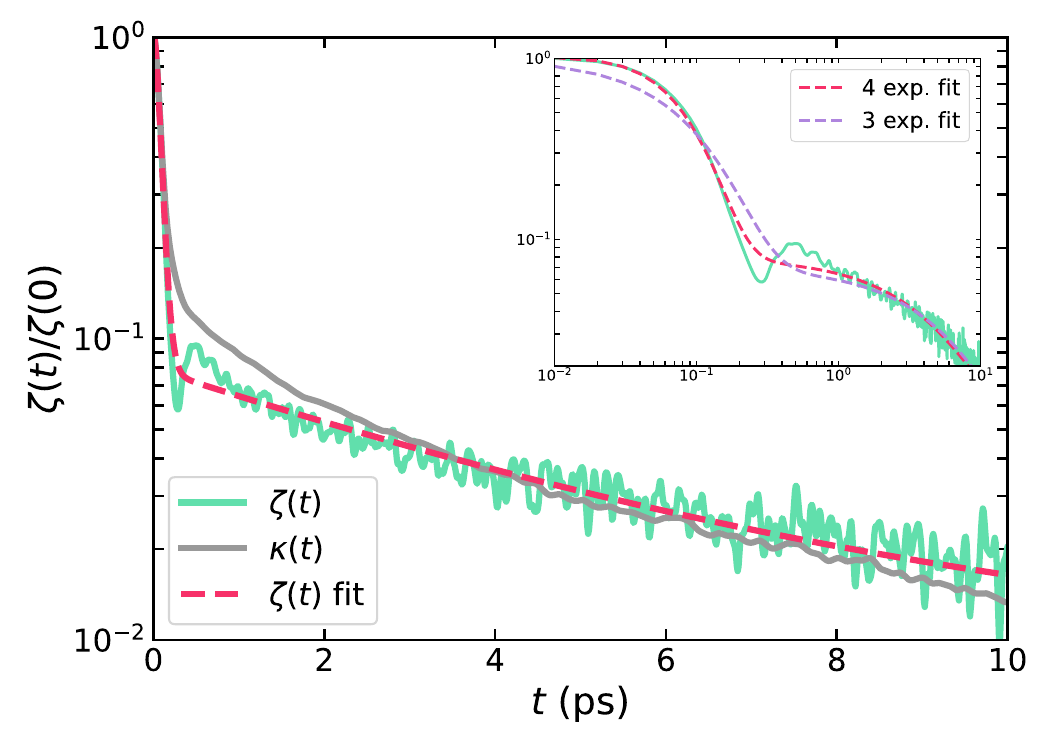}
         \caption{}
     \end{subfigure}
     \hfill
     \begin{subfigure}[b]{0.49\textwidth}
         \centering
         \includegraphics[width=\textwidth]{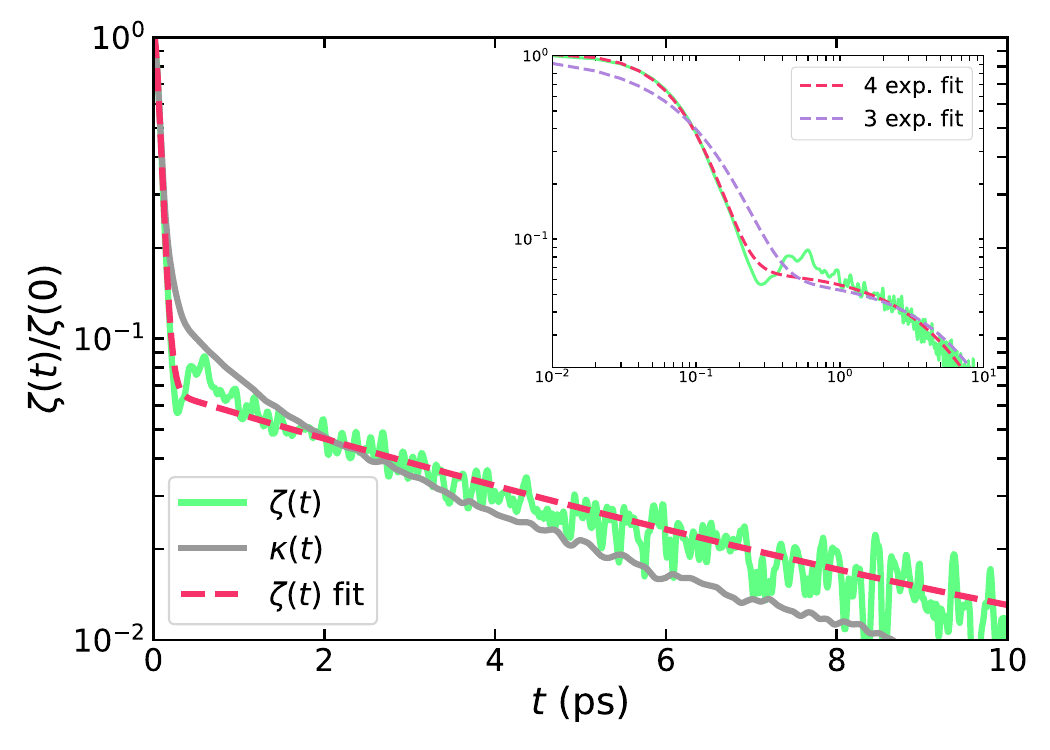}
         \caption{}
     \end{subfigure}
        \caption{\label{fig:singleion} Data justifying the use of the single particle kernel $\zeta(t)$ \textbf{(a)} Check for the validity of $\zeta_+(t)\approx \zeta_-(t)$ by comparing Eq.~\ref{eqn:equalzeta} (Model) and Eq.~\ref{eqn:fullmodel} (Corrected Model). \textbf{(b)}, \textbf{(c)} and \textbf{(d)} show the single-particle memory function, alongside the four exponential fit (outlined in Section~\ref{sec:DF}) at 350~K, 400~K and 425~K respectively. The 325~K plot is shown in the Main Text Fig.~\ref{fig:zeta_data}(b).}
\end{figure*}

In our two component ionic liquid, we suggest we can write the total current correlation function as:
\begin{equation}\label{eqn:JACFaprx}
    \langle\mbf{J}(t)\cdot\mbf{J}(0)\rangle \simeq \sum_iq_i^2\langle\mbf{v}_i(t-t')\cdot\mbf{v}_i(0)\rangle,
\end{equation}
taking the time derivative of this gives:
\begin{equation}\label{eqn:asymzeta}
\begin{split}
     \frac{\partial}{\partial t}\langle\mbf{J}(t)\cdot\mbf{J}(0)\rangle \simeq -\int_0^t\dd{t'} \Big(\zeta_+(t')N_+q_+^2\langle\mbf{v}_+(t-t')\cdot\mbf{v}_+(0)\rangle \\
     + \zeta_-(t')N_-q_-^2\langle\mbf{v}_-(t-t')\cdot\mbf{v}_-(0)\rangle\Big).
\end{split}
\end{equation}
We must note that is only by assuming $\zeta_+(t) \approx \zeta_-(t)$ that we can write:
\begin{align}
     \frac{\partial}{\partial t}\langle\mbf{J}(t)\cdot\mbf{J}(0)\rangle &\simeq -\int_0^t\dd{t'}\zeta(t')\sum_iq_i^2\langle\mbf{v}_i(t-t')\cdot\mbf{v}_i(0)\rangle \\
     &\simeq -\int_0^t\dd{t'}\zeta(t')\langle\mbf{J}(t-t')\cdot\mbf{J}(0)\rangle
\end{align}
However our electrolyte is not symmetric and so we must question how significant this approximation is. 
Taking a Laplace transform of the left-hand side of Eq.~\ref{eqn:asymzeta}:
\begin{align}
    \mathcal{L}\left[\frac{\partial}{\partial t}\langle\mbf{J}(t)\cdot\mbf{J}(0)\rangle\right] &= i\omega\mathcal{L}\left[\langle\mbf{J}(t)\cdot\mbf{J}(0)\rangle\right] - \langle\mbf{J}^2\rangle \\
    &= \frac{3\Omega}{\beta}i\omega \sigma(\omega) - \langle\mbf{J}^2\rangle \\
    &\simeq \frac{3\Omega}{\beta}i\omega(\sigma_+(\omega) + \sigma_-(\omega))
 - \langle\mbf{J}^2\rangle
\end{align}
where the last line follows from Eq.~\ref{eqn:JACFaprx}.
Now taking the Laplace transform of the right-hand side:
\begin{align*}
    \mathcal{L}\bigg[-&\int_0^t\dd{t'} (\zeta_+(t')N_+q_+^2\langle\mbf{v}_+(t-t')\cdot\mbf{v}_+(0)\rangle \\ 
    + &\zeta_-(t')N_-q_-^2\langle\mbf{v}_-(t-t')\cdot\mbf{v}_-(0)\rangle)\bigg] 
    \\
    &= -\zeta_+(t')N_+q_+^2\mathcal{L}[\langle\mbf{v}_+(t-t')\cdot\mbf{v}_+(0)\rangle] \\
    &\hspace{20pt}- \zeta_-(t')N_-q_-^2\mathcal{L}[\langle\mbf{v}_-(t-t')\cdot\mbf{v}_-(0)\rangle]\\
    &=-\frac{3\Omega}{\beta}\left(\zeta_+(\omega)\sigma_+(\omega) + \zeta_-(\omega)\sigma_-(\omega)\right),
\end{align*}
and equating these we can write
\begin{equation}\label{eqn:fullmodel}
    \sigma_+(\omega) + \sigma_-(\omega) = \frac{\frac{3\Omega}{\beta}\langle\mbf{J}^2\rangle + (\zeta_+(\omega)-\zeta_-(\omega))\sigma_-(\omega)}{i\omega + \zeta_+(\omega)}.
\end{equation}
Using our approximation that $\zeta_+(t) \approx \zeta_-(t)$ we arrive at Eq.~\ref{eqn:sigma_from_zeta},
\begin{equation}\label{eqn:equalzeta}
    \sigma_+(\omega) + \sigma_-(\omega) \simeq \frac{3\Omega}{\beta}\langle\mbf{J}^2\rangle  \frac{1}{i\omega + \zeta(\omega)}
\end{equation}
and we can therefore identify the 'correction term' for this approximation as $(\zeta_+(\omega)-\zeta_-(\omega))\sigma_-(\omega)/i\omega + \zeta_+(\omega)$. \\
We can quantify this correction with our model to understand the significance of this approximation and the result is shown in Fig.~\ref{fig:singleion}(a), highlighting that this approximation at 375~K is good. Figure~\ref{fig:singleion}(b), (c) and (d) shows the fit to the single-particle memory at 350, 400 and 425~K respectively. This follows the discussion in Section~\ref{sec:analytical memory}.
\onecolumngrid
\clearpage
\twocolumngrid
\section{Single-Particle Conductivity}\label{app:singleioncond}
\begin{figure*}[!ht]
     \centering
     \begin{subfigure}[b]{0.48\textwidth}
         \centering
         \includegraphics[width=\textwidth]{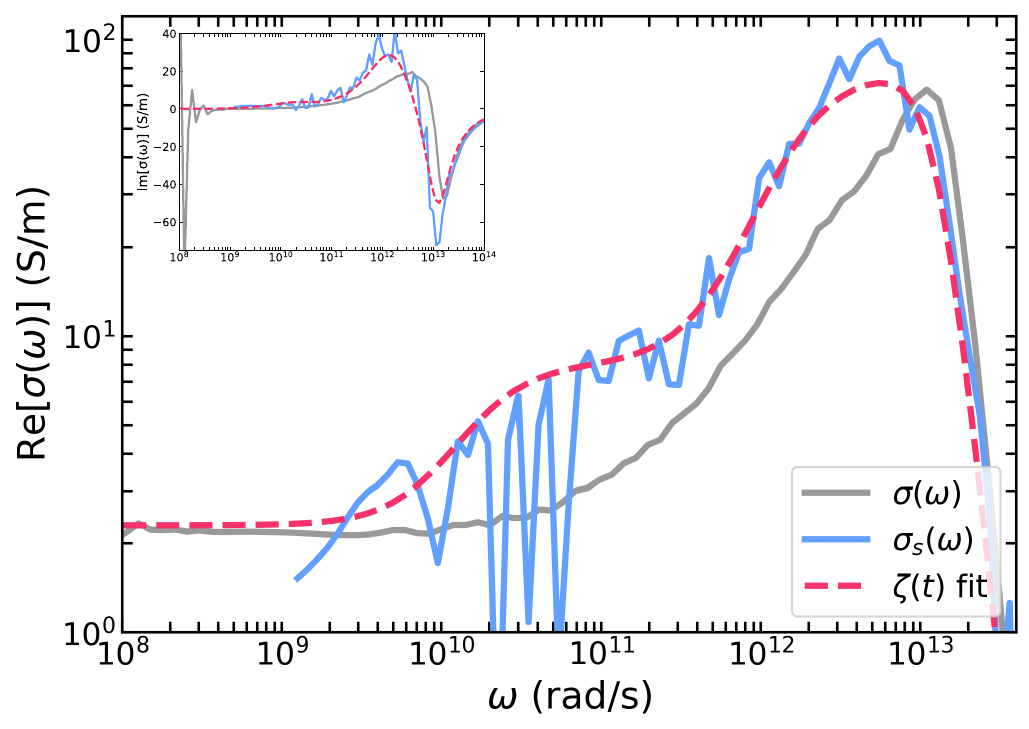}
         \caption{}
     \end{subfigure}
     \hfill
     \begin{subfigure}[b]{0.48\textwidth}
         \centering
         \includegraphics[width=\textwidth]{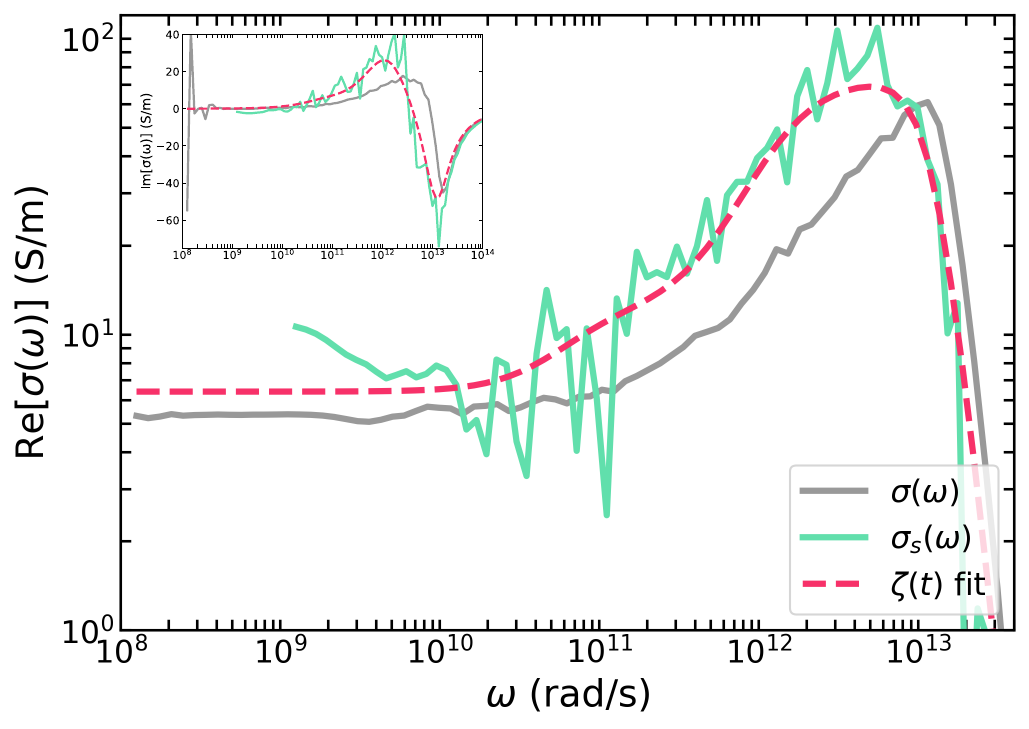}
         \caption{}
     \end{subfigure}
     \hfill
     \begin{subfigure}[b]{0.48\textwidth}
         \centering
         \includegraphics[width=\textwidth]{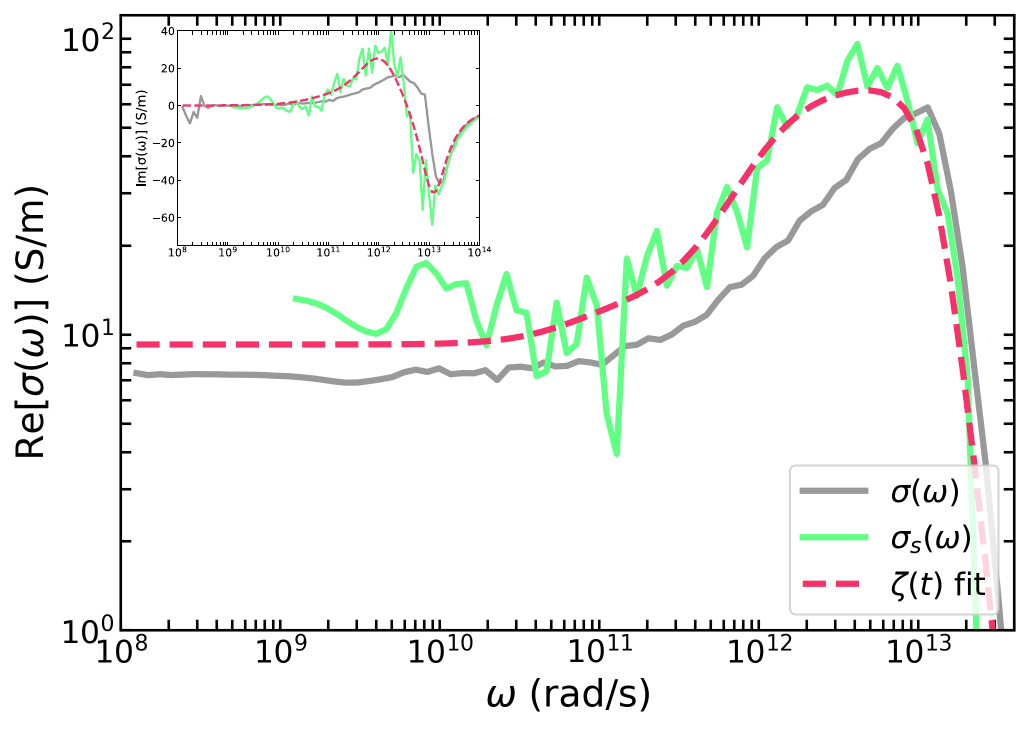}
         \caption{}
     \end{subfigure}
     \vspace{-8pt}
        \caption{\label{fig:singleion_cond} Real-part of the single-particle conductivity found from simulation (using Eq.~\ref{eqn:sigma-FDT-single}) alongside that for the model found using Eq.~\ref{eqn:sigma_from_zeta} and the fit to $\zeta(t)$ shown in Fig.~\ref{fig:singleion} at \textbf{(a)} 350~K, \textbf{(b)} 400~K and \textbf{(c)} 425~K. Insets: Imaginary-part. The real and imaginary parts for 375~K are shown in Main Text Figs.~\ref{fig:zeta_data}(c)~and~(d) respectively.}
        \vspace{-4pt}
\end{figure*}

Figure~\ref{fig:singleion_cond} shows the simulation data in Fig.~\ref{fig:singleion}(b), (c) and (d) transformed using Eq.~\ref{eqn:sigma-FDT-single} into the single-particle conductivity. The four exponential fit of the memory function is also transformed in the same way and plotted as the dashed line, such that the good agreement between the two demonstrates the validity of the fit.
\vspace{6pt}
\section{Itinerant Oscillator Kernel Derivation}\label{app:itinerant}
\vspace{-8pt}

First taking the Laplace transform of Eqs.~\ref{eqn:itinerant1}~and~\ref{eqn:itinerant0}, using the convolution theorem this gives,
\begin{equation}
    \mathcal{F}[\bm{\Dot{v}}_1(t)](s)+\gamma_1(s)\bm{v}_1(s) + \frac{\omega_1^2}{s}[\bm{v}_1(s)-\bm{v}_0(s)] = \frac{\bm{F}_1(s)}{m_1},
\end{equation}
\begin{equation}
    \mathcal{F}[\bm{\Dot{v}}_0(t)](s)+\gamma_0(s)\bm{v}_0(s) + \frac{\omega_0^2}{s}[\bm{v}_0(s)-\bm{v}_1(s)] = \frac{\bm{F}_0(s)}{m_0}.
\end{equation}
We can then close these expressions with $\bm{v}_1(0)$ to produce correlation functions.
\begin{equation}\label{motion1}
\begin{split}
    \mathcal{F}[\langle\bm{v}_1(0)\cdot\bm{\Dot{v}}_1(t)\rangle](s)+\gamma_1(s)\langle\bm{v}_1(0)\cdot\bm{v}_1(s)\rangle + \\ \frac{\omega_1^2}{s}[\langle\bm{v}_1(0)\cdot\bm{v}_1(s)\rangle-\langle\bm{v}_1(0)\cdot\bm{v}_0(s)\rangle] = 0,
\end{split}
\end{equation}
\begin{equation}\label{motion2}
\begin{split}
\mathcal{F}[\langle\bm{v}_1(0)\cdot\bm{\Dot{v}}_0(t)\rangle](s)+\gamma_0(s)\langle\bm{v}_1(0)\cdot\bm{v}_0(s)\rangle + \\
\frac{\omega_0^2}{s}[\langle\bm{v}_1(0)\cdot\bm{v}_0(s)\rangle-\langle\bm{v}_1(0)\cdot\bm{v}_1(s)\rangle] = 0.
\end{split}
\end{equation}
Rewriting equation \ref{motion2} to substitute for $\langle\bm{v}_1(0)\cdot\bm{v}_0(s)\rangle$ in equation \ref{motion1}, noting that $\mathcal{F}[\langle\bm{v}_1(0)\cdot\bm{\Dot{v}}_0(t)\rangle](s) = s\langle\bm{v}_1(0)\cdot\bm{v}_0(s)\rangle$, we then get the expression
\begin{equation}\label{motion3}
\begin{split}
    \mathcal{F}[\langle\bm{v}_1(0)\cdot\bm{\dot{v}}_1(t)\rangle](s)+\langle\bm{v}_1(0)\cdot\bm{v}_1(s)\rangle\bigg\lbrace\gamma_1(s) + \\
    \frac{\omega_1^2}{s}\left[1-\frac{\omega_0}{s^2 +s\gamma_0 + \omega_0^2}\right]\bigg\rbrace = 0,
\end{split}
\end{equation}
Comparing equation \ref{motion3} to the definition of the memory function,
\begin{equation}\label{eqn:langevin_app}
    \frac{\partial}{\partial t} \langle\mbf{v}_i(t)\cdot\mbf{v}_i(0)\rangle = -\int^{t}_0 \dd{t'} \zeta(t')\langle\mbf{v}_i(t-t')\cdot\mbf{v}_i(0)\rangle,
\end{equation}
gives Eq.~\ref{eqn:IOkernel1}.

\begin{widetext}
\begin{center}
\captionof{table}{$\sigma_n$ values}\label{table:sigma}
\begin{tabular}{ |p{3.5cm}||p{3.5cm}|p{3.5cm}|p{3.5cm}|  }
  \hline
 $T$ (K)&$\sigma_0$ (Sm\textsuperscript{-1})&$\sigma_1\times10^{-11}$ (Ssm\textsuperscript{-1})&$\sigma_2\times10^{-22}$ (Ss\textsuperscript{2}m\textsuperscript{-1})\\
 \hline
 350 & 2.19 $\pm$ 0.001 & 3.20 $\pm$ 0.08 & 1.50 $\pm$ 0.09\\
 375 & 3.72 $\pm$ 0.001 & 2.88 $\pm$ 0.13 & 1.0 $\pm$ 0.22\\
 400 & 5.38 $\pm$ 0.003 & 2.72 $\pm$ 0.17 & 0.80 $\pm$ 0.12\\
 425 & 7.29 $\pm$ 0.004 & 2.40 $\pm$ 0.28 & 0.50 $\pm$ 0.08\\
 \hline
\end{tabular}
\end{center}
\end{widetext}

\section{Two Exponential Memory Function}\label{app:IO2exp}
In order to produce an analytical expression for a double exponential memory function of the form:
\begin{equation}\label{eqn:exp_fit}
    \zeta(t)=A\mathrm{e}^{-at} + B\mathrm{e}^{-bt}
\end{equation}
we must make several approximations to the form of Eq.~\ref{eqn:IOkernel1}:
\begin{equation}\label{eqn:IOkernel1app}
\zeta(\omega)=\gamma_1(\omega)+\frac{\omega_1^2}{\imi\omega+\frac{\omega_0^2}{\imi\omega+\gamma_0(\omega)}}.
\end{equation}
Assuming the two friction kernels have a finite lifetime ($\gamma_0(\omega) = e^{-dt}$ and $\gamma_1(\omega) = e^{-ct}$) we can rewrite this as:
\begin{equation}
    \zeta(\omega) = \frac{\gamma_1}{c+\imi\omega} + \frac{\omega_1^2}{\imi\omega+\frac{\omega_0^2}{\imi\omega +\frac{\gamma_0}{b+\imi\omega}}}.
\end{equation}
Provided that the cage-environment friction relaxes sufficiently fast ($b\gg\omega$) we can ignore this timescale in the second term,
\begin{equation}
    \zeta(\omega) = \frac{\gamma_1}{c+\imi\omega} + \frac{\omega_1^2}{\imi\omega+\frac{\omega_0^2}{\imi\omega +\frac{\gamma_0}{b}}}.
\end{equation}
Alongside this, if we also assume the magnitude of the cage-environment friction is sufficiently large ($\gamma_0\gg b\omega$) we arrive at a two exponential form,
\begin{equation}\label{eqn:IOdouble}
    \zeta(\omega) = \frac{\gamma_1}{c+\imi\omega} + \frac{\omega_1^2}{\imi\omega+\frac{b\omega_0^2}{\gamma_0}}.
\end{equation}
The simplifying assumption can be justified if we consider the cage quasiparticle like a large, Brownian particle. It will experience large drag ($\gamma_0$) because of its large size, and there will be good timescale separation as a result of its larger mass compared to the particles of the environment.

While this is a simpler form compared to the four exponential result in the text, when fitting the single-particle memory with a single exponential for long-times it is not possible to recover the Nernst-Einstein limit while still producing a reasonable fit. This clarifies that the long-time behaviour in $\zeta(t)$ is bi-exponential, a result that Eq.~\ref{eqn:IOdouble} cannot capture. However this expression and the resulting conductivity expression presented below may be informative for other systems.

\section{Conductivity Expressions}\label{app:itinerant_cond}
\vspace{-6pt}
We have derived the conductivity expressions, including the first two moments, for the cases of $\kappa(t)\sim\delta(t), ~\exp(-t/\tau),~ ~\gamma\delta(t)+(1-\gamma)\exp(-t/\tau),~\mathrm{and}~A\exp(-at)+B\exp(-bt)$, and present them for reference in Table~\ref{table:formulae}.

For a double exponential memory function (Eq.~\ref{eqn:exp_fit}) we can use Eq.~\ref{eqn:sigma_from_zeta} to find the form of the frequency-dependent conductivity,
\begin{widetext}

\begin{equation}\label{eqn:2exp_cond}
    \sigma(\omega)=\beta\sum^2_{\alpha=1} \rho_\alpha q_\alpha^2 \frac{1}{m\beta}\frac{ab-\omega^2+i\omega(a+b)}{Ab+Ba-\omega^2(a+b)+\mathrm{i}(\omega(A+B+ab)-\omega^3)},
\end{equation}
where the parameters are given a microscopic interpretation in Eq.~\ref{eqn:IOdouble}. The real part is given by 
\begin{equation}\label{eq:Re_2exp_cond}
\operatorname{Re}[\sigma(\omega)]\propto \frac{(ab-\omega^2)(Ab+Ba-\omega^2(a+b))+\omega(a+b)(\omega(A+B+ab)-\omega^3)}{(Ab+Ba-\omega^2(a+b))^2+(\omega(A+B+ab)-\omega^3)^2}.
\end{equation}

\end{widetext}
However since a double exponential form for the single-particle memory function was not found to be appropriate, at least a four exponential memory would be necessary to describe the memory function over the full time domain as shown in Fig.~\ref{fig:zeta_data} of the Main Text. At that point, writing down analytic expressions, especially for the real and imaginary parts, becomes cumbersome.

The analysis therefore is best performed on the timescales in the impedance spectrum itself, which relate the moments of the conductivity. The number of exponentials composing the memory function sets the number of distinct timescales that appear in the hierarchy \cite{boon_molecular_1991}. This fixes the number of conductivity moments required to have full knowledge of the spectrum, with each new timescale entering progressively with each higher moment. For example, $\tau_1^{(\sigma)}$ can be consistently defined from the conductivity moments as the quantity $\frac{\sigma_1}{\sigma_0}+\tau_0^{(\sigma)}$, and compared to any number of exponentials greater than 1. In contrast $\tau_2^{(\sigma)}$ obtains its stable form $\frac{\sigma_2}{\sigma_0}-2\tau_0^{(\sigma)}\tau_1^{(\sigma)}+(\tau_0^{(\sigma)})^2$ with the inclusion of 2 exponentials, and then is invariant to any higher number.

For the three-exponentials employed in the Main Text, Eq.~\ref{eqn:kappaexp}, the frequency-resolved conductivity is given by the straightforward but lengthy extension
\begin{widetext}
\begin{equation}
    \sigma(\omega) = \frac{\beta}{3 \Omega}\langle \bm{J}^2\rangle \left( \frac{-\imi\omega^3 - (a+b+c)\omega^2 +\imi\omega(ab+ac+bc)+abc}{\omega^4 -\imi(a+b+c)\omega^3 -x_2\omega^2 +\imi x_1 \omega +(Abc+Bac+Cab)} \right)
\end{equation}
where $x_1 = abc+A(b+c)+B(a+c)+C(a+b)$ and $x_2 = (ab+ac+bc+A+B+C)$. Eqs.~\ref{eqn:sigma-Re}~and~\ref{eqn:sigma-Im} define the conductivity moments as Taylor coefficients, such that
\begin{equation}
    \sigma_0 = \sigma(\omega\to 0)\propto\frac{1}{\frac{A}{a}+\frac{B}{b}+\frac{C}{c}} \equiv \tau_0^{(3)}
\end{equation}
\begin{equation}
    \frac{\sigma_1}{\sigma_0} = \lim_{\omega\to 0}\left\lbrace -\imi \frac{\partial \sigma(\omega)}{\partial \omega}\right\rbrace / \sigma_0 = \frac{\frac{A}{a^2}+\frac{B}{b^2}+\frac{C}{c^2}}{\frac{A}{a}+\frac{B}{b}+\frac{C}{c}} - \tau_0^{(3)} \equiv\tau_1^{(3)}-\tau_0^{(3)}
\end{equation}
\begin{align}
    \frac{\sigma_2}{\sigma_0} &= \lim_{\omega\to 0}\left\lbrace \frac{1}{2}\frac{\partial^2 \sigma(\omega)}{\partial^2 \omega}\right\rbrace / \sigma_0 \nonumber\\
    &= - \frac{1}{\left(\frac{A}{a}+\frac{B}{b}+\frac{C}{c} \right)^2} + 2\cdot\frac{\frac{A}{a^2}+\frac{B}{b^2}+\frac{C}{c^2}}{\left(\frac{A}{a}+\frac{B}{b}+\frac{C}{c} \right)^2} + \frac{\frac{A}{a^3}+\frac{B}{b^3}+\frac{C}{c^3}}{\frac{A}{a}+\frac{B}{b}+\frac{C}{c}} - \frac{\left(\frac{A}{a^2}+\frac{B}{b^2}+\frac{C}{c^2} \right)^2}{\left(\frac{A}{a}+\frac{B}{b}+\frac{C}{c} \right)^2} \\
    &=  - (\tau_0^{(3)})^2 + 2 \tau_0^{(3)}\tau_1^{(3)} + \frac{\left(\frac{A}{a^3}+\frac{B}{b^3}+\frac{C}{c^3}\right)\left(\frac{A}{a}+\frac{B}{b}+\frac{C}{c}\right)-
    \left(\frac{A}{a^2}+\frac{B}{b^2}+\frac{C}{c^2}\right)^2}{\left(\frac{A}{a}+\frac{B}{b}+\frac{C}{c}\right)^2} \nonumber\\
    &= - (\tau_0^{(3)})^2 + 2 \tau_0^{(3)}\tau_1^{(3)} + \frac{\frac{AB}{ab}\left(\frac{1}{a}-\frac{1}{b}\right)^2 + \frac{AC}{ac}\left(\frac{1}{a}-\frac{1}{c}\right)^2 + \frac{BC}{bc}\left(\frac{1}{b}-\frac{1}{c}\right)^2}{\left(\frac{A}{a}+\frac{B}{b}+\frac{C}{c}\right)^2} \nonumber\\
    &\equiv - (\tau_0^{(3)})^2 + 2 \tau_0^{(3)}\tau_1^{(3)} + (\tau_2^{(3)})^2
\end{align}
\end{widetext}
which are the derivations for Eqs.~\ref{eqn:tau0},~\ref{eqn:tau1},~and~\ref{eqn:tau2} in the Main Text.

\section{Minimal Memory}\label{app:minimalmemory}
In Section~\ref{sec:timescales} we argue the minimal form for the conductivity to be to be a linear combination of three positive exponentials. This is as a result of the form of the memory function found using the itinerant oscillator model in Eq.~\ref{eqn:3exptime}, where we can ignore the exponential with the negative prefactor due to the lack of oscillations in the memory function at short-times (in comparison to those seen in the single-particle memory function at short times).

\begin{figure*}[!ht]
     \centering
     \begin{subfigure}[b]{0.49\textwidth}
         \centering
         \includegraphics[width=\textwidth]{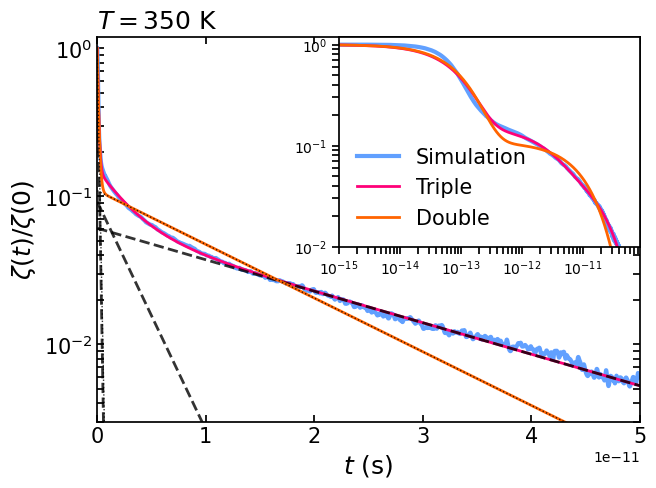}
         \caption{}
     \end{subfigure}
     \hfill
     \begin{subfigure}[b]{0.49\textwidth}
         \centering
         \includegraphics[width=\textwidth]{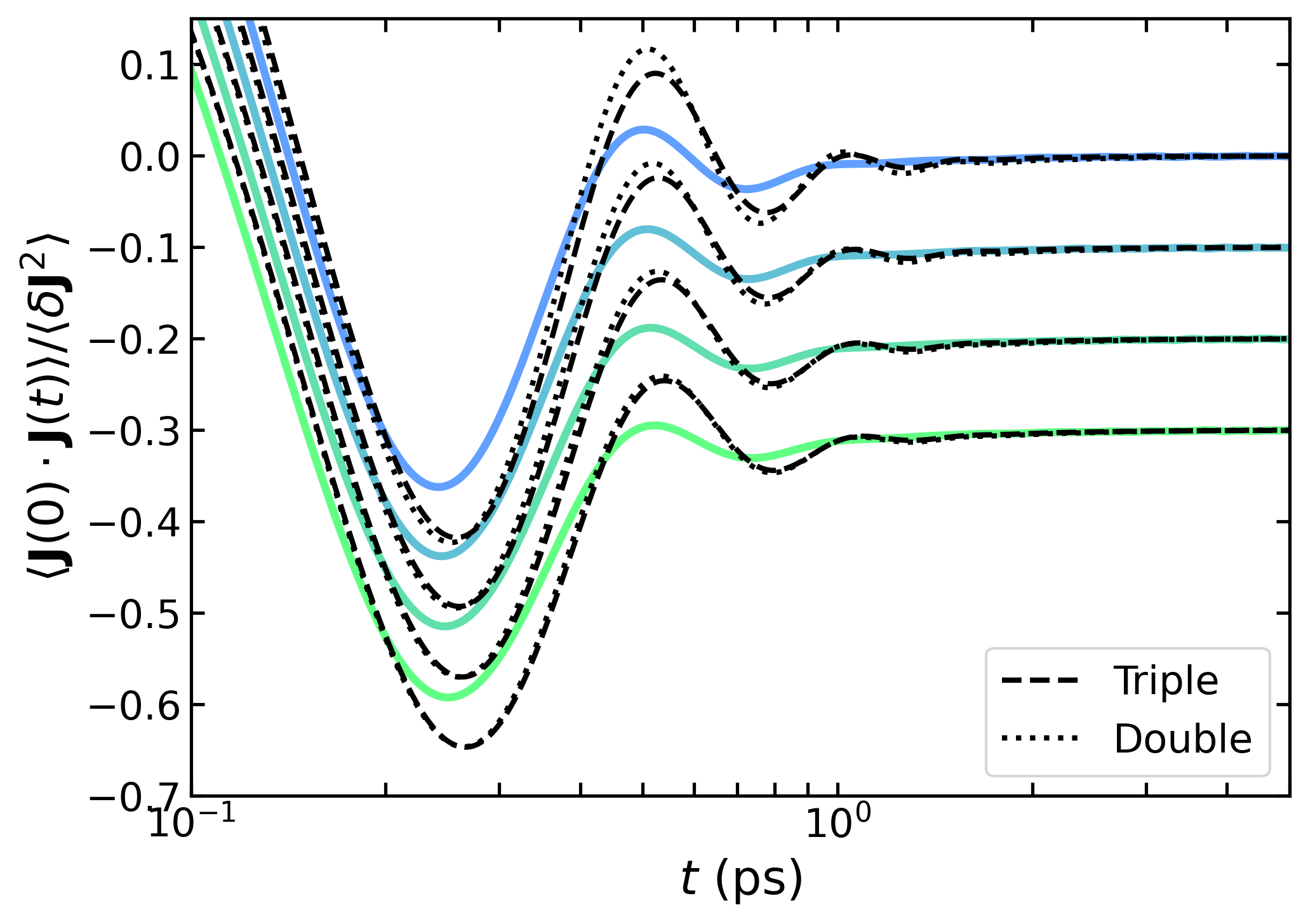}
         \caption{}
     \end{subfigure}
     \hfill
     \begin{subfigure}[b]{0.49\textwidth}
         \centering
         \includegraphics[width=\textwidth]{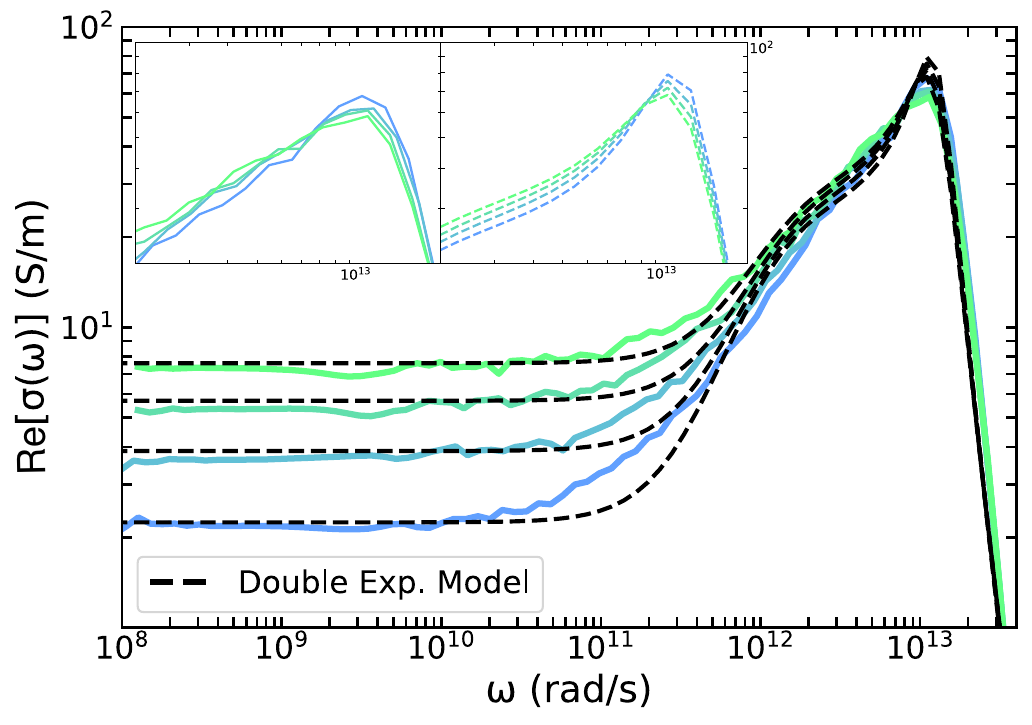}
         \caption{}
     \end{subfigure}
        \caption{\label{fig:minimal} \textbf{(a)} Friction kernel at 350 K, fitted with either a two or three exponential model. Clearly the two exponential model performs worse, and fails to capture the longest behaviour at long times. \textbf{(b)} Despite its poor performance, when transformed to a current autocorrelation function, the two exponential model performs similarly well to the three exponential model and so therefore could be considered minimal. \textbf{(c)} When further transformed to the full conductivity we can see that a double exponential model can capture the low and high frequency behaviour just as well, however in the intermediate frequency range a triple exponential form is best.}
\end{figure*}

Figure~\ref{fig:minimal} shows that a three exponential memory is the minimal acceptable form. While Fig.~\ref{fig:minimal}(b) and (c) may suggest that two exponentials are sufficient, in (a) we can clearly see that a double exponential memory poorly approximates the memory function over the entire trajectory. This is further highlighted in Fig.~\ref{fig:timescales}(c) where the $\tau_2^{(2)}$ value found from the double exponential fit is significantly smaller than the true value ($\tau_2^{(\sigma)}$), whereas $\tau_2^{(3)}$ is a significant improvement.

\begin{sidewaystable*}
  \centering
   \vspace{200pt}
   \caption{\parbox{0.6\textwidth}{Summary of the functional forms of the current-current time correlation function, conductivity spectrum, and its first two moments, for different models of the memory function $\kappa(t)$. \label{table:formulae}}}
    \begin{tabular}{|c|c|c|c|c|}
      \hline &&&&\\
      \textbf{} &
      \textbf{Markovian} &
      \textbf{Exponential Model}&
      \textbf{Exp.+Delta} &
      \textbf{Two Exp.}\\
      \textbf{} &
      \textbf{} &
      \textbf{(DF $\omega\ll\tau_\mathrm{atm}$)} &
      \textbf{} &
      \textbf{}\\
      \hline
      &&&&\\
      \textbf{$\zeta(t)$}
      & $\frac{1}{\tau_0}\delta(t)$
      & $\frac{1}{\tau_{0}\tau_{1}} \mathrm{e}^{-t/\tau_{1}}$
      & $\frac{\gamma}{\tau_0}\delta(t) + \frac{(1-\gamma)}{\tau_0\tau_1}\mathrm{e}^{-t/\tau_1}$
      & $A\e{-at}+B\e{-b t}$
      \\
      &&&&\\
      \hline
      &&&&\\
      \textbf{$\zeta(\omega)$}
      & $\frac{1}{\tau_0}$
      & $\frac{1}{\tau_{0}\tau_{1}}\left(\frac{1}{\frac{1}{\tau_{1}}+i\omega}\right)$
      & $\frac{\gamma}{\tau_0} + \frac{(1-\gamma)}{\tau_{0}\tau_{1}}\left(\frac{1}{\frac{1}{\tau_{1}}+i\omega}\right)$
      & $\frac{A}{s+a}+\frac{B}{s+b}$
      \\
      &&&&\\
      \hline
      &&&&\\
      \textbf{$\langle\bm{J}(0)\cdot\bm{J}(t)\rangle$}
      & $\frac{1}{\tau_0}\mathrm{e}^{-t/\tau_0}$
      & biexponential
      & biexponential
      & triexponential
      \\
      &&&&\\
      \hline
      &&&&\\
      \textbf{$\sigma(\omega)$}
      & $\beta\sum^2_{\alpha=1} \rho_\alpha q_\alpha^2 \frac{1}{m\beta}\frac{1}{\frac{1}{\tau_0}+i\omega}$
      & $\beta\sum^2_{\alpha=1} \rho_\alpha q_\alpha^2 \frac{1}{\beta m}\frac{\frac{1}{\tau_{1}} + i\omega}{\frac{1}{\tau_{0}\tau_{1}} - \omega^2 +i\omega\frac{1}{\tau_{1}}}$
      & $\beta\sum^2_{\alpha=1} \rho_\alpha q_\alpha^2 \frac{1}{\beta m}\frac{\frac{1}{\tau_{1}} + i\omega}{\frac{1}{\tau_{0}\tau_{1}} - \omega^2 +i\omega(\frac{\gamma}{\tau_0} + \frac{1}{\tau_{1}})}$
      & Eq.~\ref{eqn:2exp_cond}
      \\
      &&&&\\
      \hline
      &&&&\\
      \textbf{$\sigma_1/\sigma_0$}
      & $-\tau_0$
      & $\tau_{1}-\tau_{0}$
      & $(1-\gamma)\tau_1 - \tau_0$
      & $\tau_{1}-\tau_{0}$
      \\
      &&&&\\
      \hline
      &&&&\\
      \textbf{$\sigma_2/\sigma_0$}
      & $-\tau_0^2$
      & $2\tau_{0}\tau_{1} - \tau_{0}^2$
      & $\gamma(1-\gamma)\tau_1^2 + 2(1-\gamma)\tau_0\tau_1 -\tau_0^2$
      & 
      $2\tau_{0}\tau_{1}-\tau_{0}^2 + \tau_2^2$
      \\
      &&&&\\
      \hline
    \end{tabular}
    
\end{sidewaystable*}

\end{document}